\documentclass{egpubl}[review]
\usepackage{hpg25DL}

\WsPaper           

\usepackage[T1]{fontenc}
\usepackage{dfadobe}  
\usepackage{listings}
\usepackage{cite}  
\usepackage{multirow}
\usepackage{lineno}
\usepackage{subfig}
\usepackage{amsmath}
\usepackage{bm}
\usepackage{array}
\usepackage[absolute,overlay]{textpos}
\BibtexOrBiblatex
\electronicVersion
\PrintedOrElectronic
\ifpdf \usepackage[pdftex]{graphicx} \pdfcompresslevel=9
\else \usepackage[dvips]{graphicx} \fi
\usepackage{tikz}
\usepackage{egweblnk} 

\usepackage{amssymb}
\usepackage[nolist]{acronym}
\begin{acronym}
\acro{GPU}[GPU]{Graphics Processing Unit}
\acro{DGF}[DGF]{Dense Geometry Format}
\end{acronym}

\newcommand{\weight}{\mathop{}\!\mathrm{w}}
\newcommand{\clamp}{\mathop{}\!\mathrm{clamp}}

\title[GATE: Geometry-Aware Trained Encoding]{GATE: Geometry-Aware Trained Encoding}

\author[Jakub Bok\v{s}ansk\'y \& Daniel Meister \&  Carsten Benthin]
{
\parbox{\textwidth}{\centering Jakub Bok\v{s}ansk\'y\orcid{0000-0003-0087-2645}, Daniel Meister\orcid{0000-0002-3149-1442} and Carsten Benthin\orcid{0000-0003-3337-1636}}\\
{\parbox{\textwidth}{\centering Advanced Micro Devices, Inc.}}
}

\begin{document}

\teaser{

\begin{center}
\begin{minipage}{0.33\textwidth}
    \begin{tikzpicture}
        \node[anchor=south west, inner sep=0] (image) at (0,0) {\includegraphics[width=\linewidth]{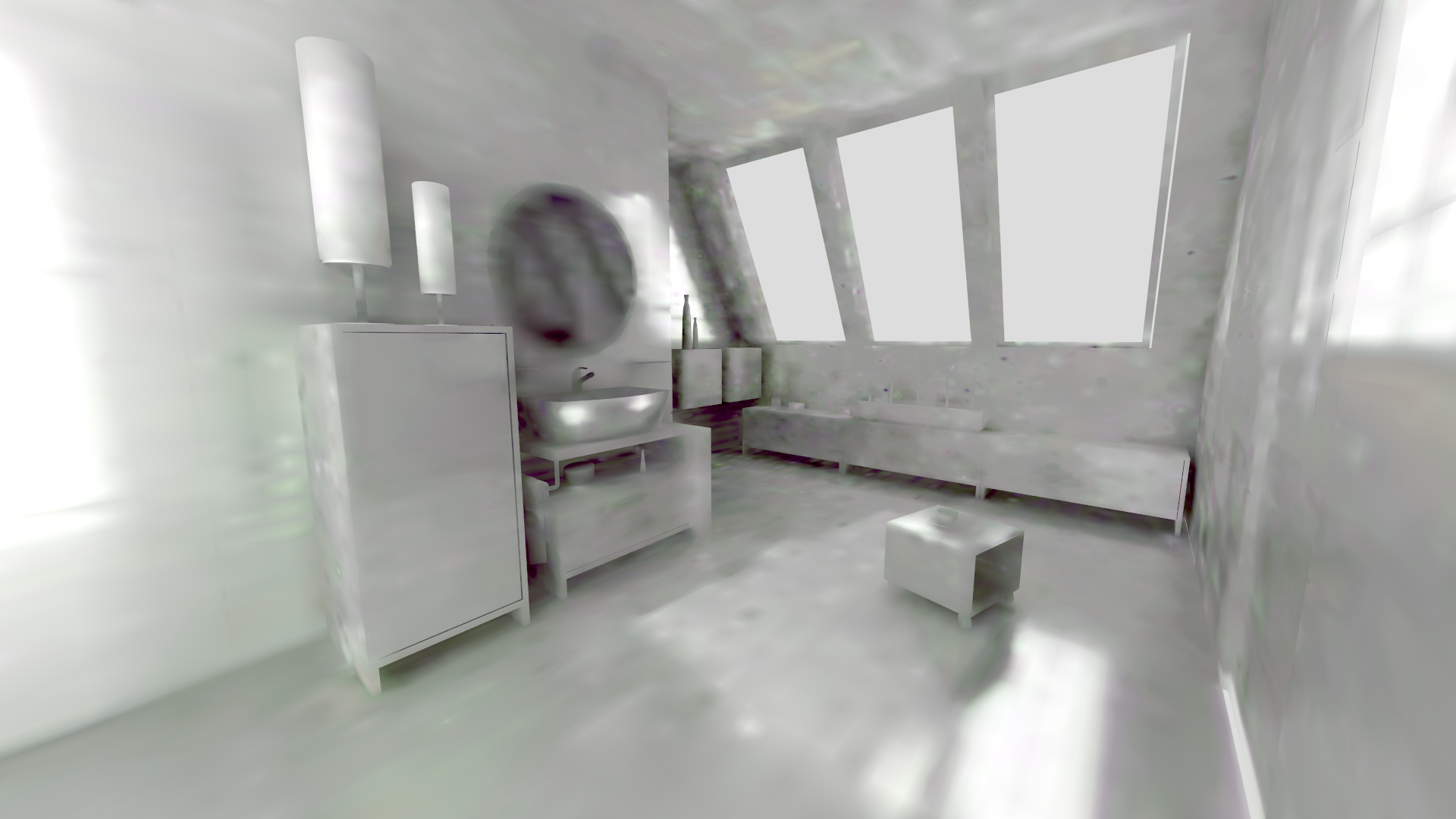}};
        \node[anchor=south east, xshift=-2mm, yshift=2mm, fill=white, opacity=0.7, text opacity=1, font=\scriptsize] 
            at (image.south east) {FLIP 0.149};
       \node[anchor=north west, xshift=2mm, yshift=-2mm, fill=white, opacity=0.7, text opacity=1, font=\scriptsize] 
           at (image.north west) {Hash-Grid};
    \end{tikzpicture}
\end{minipage}%
\begin{minipage}{0.33\textwidth}
    \begin{tikzpicture}
        \node[anchor=south west, inner sep=0] (image) at (0,0) {\includegraphics[width=\linewidth]{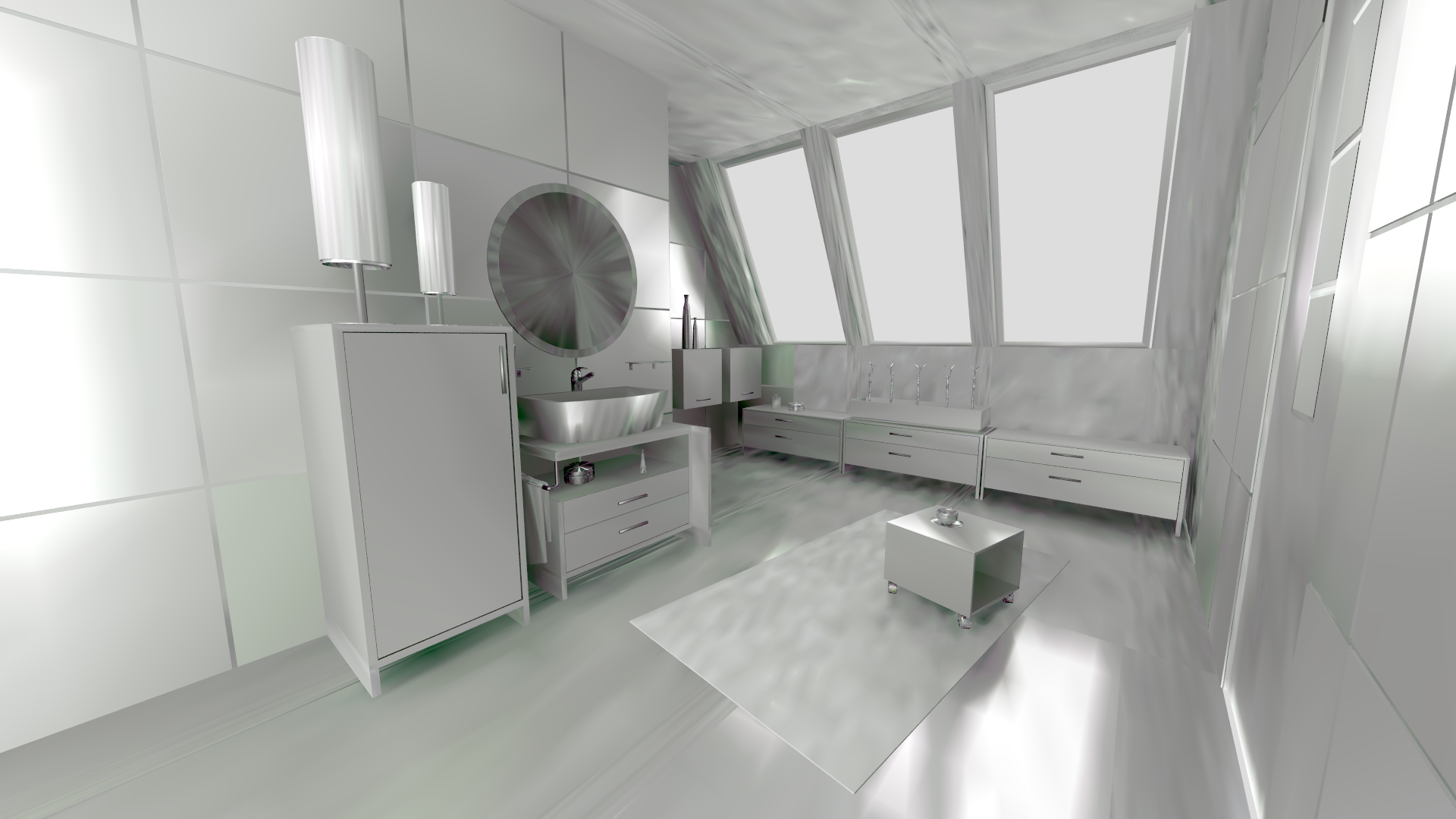}};
        \node[anchor=south east, xshift=-2mm, yshift=2mm, fill=white, opacity=0.7, text opacity=1, font=\scriptsize] 
            at (image.south east) {FLIP 0.130};
       \node[anchor=north west, xshift=2mm, yshift=-2mm, fill=white, opacity=0.7, text opacity=1, font=\scriptsize] 
           at (image.north west) {GATE (ours)};
    \end{tikzpicture}
\end{minipage}%
\begin{minipage}{0.33\textwidth}
    \begin{tikzpicture}
        \node[anchor=south west, inner sep=0] (image) at (0,0) {\includegraphics[width=\linewidth]{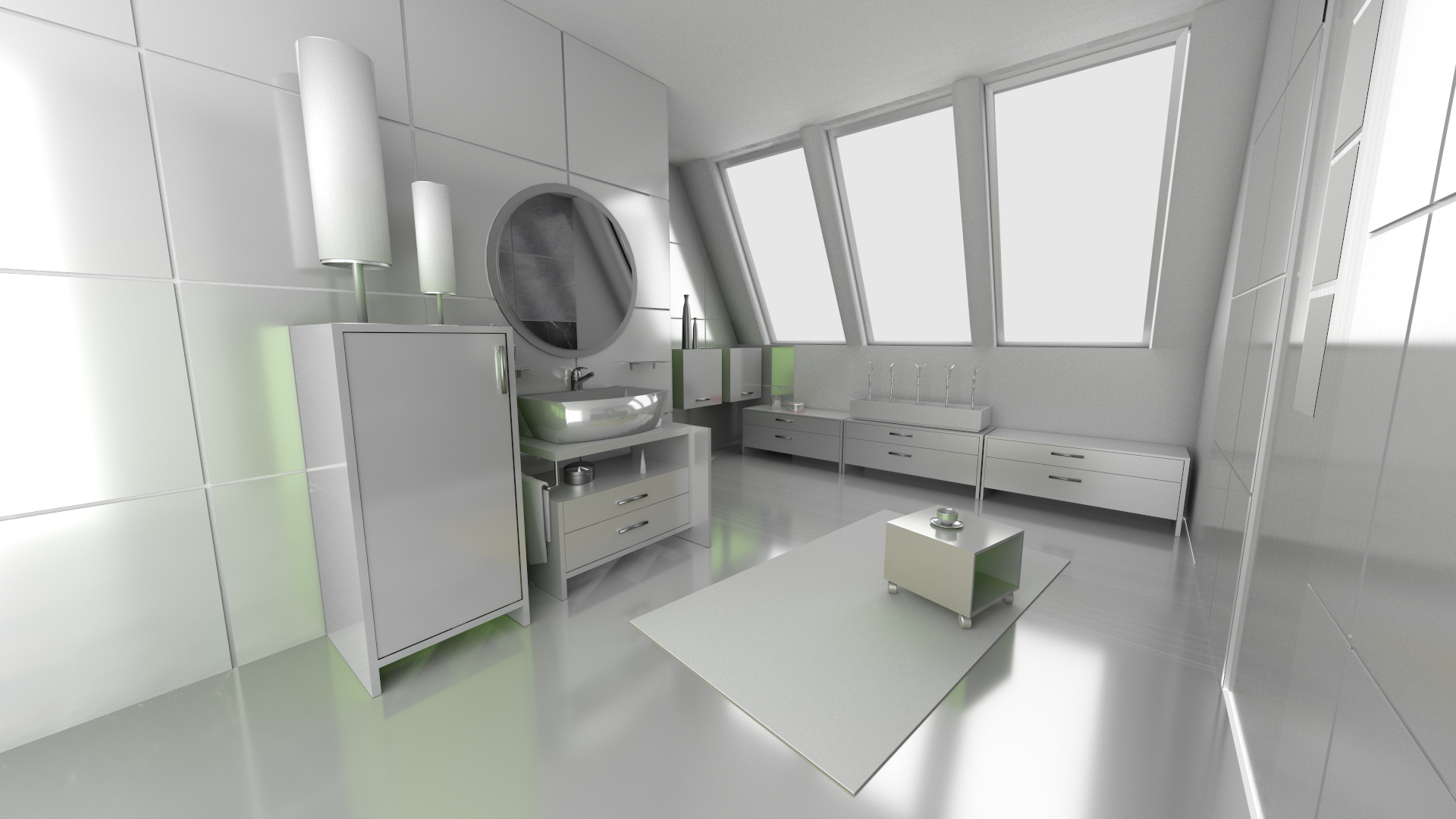}};
       \node[anchor=north west, xshift=2mm, yshift=-2mm, fill=white, opacity=0.7, text opacity=1, font=\scriptsize] 
           at (image.north west) {Ground Truth};
    \end{tikzpicture}
\end{minipage}%

\begin{minipage}{0.33\textwidth}
    \begin{tikzpicture}
        \node[anchor=south west, inner sep=0] (image) at (0,0) {\includegraphics[width=\linewidth]{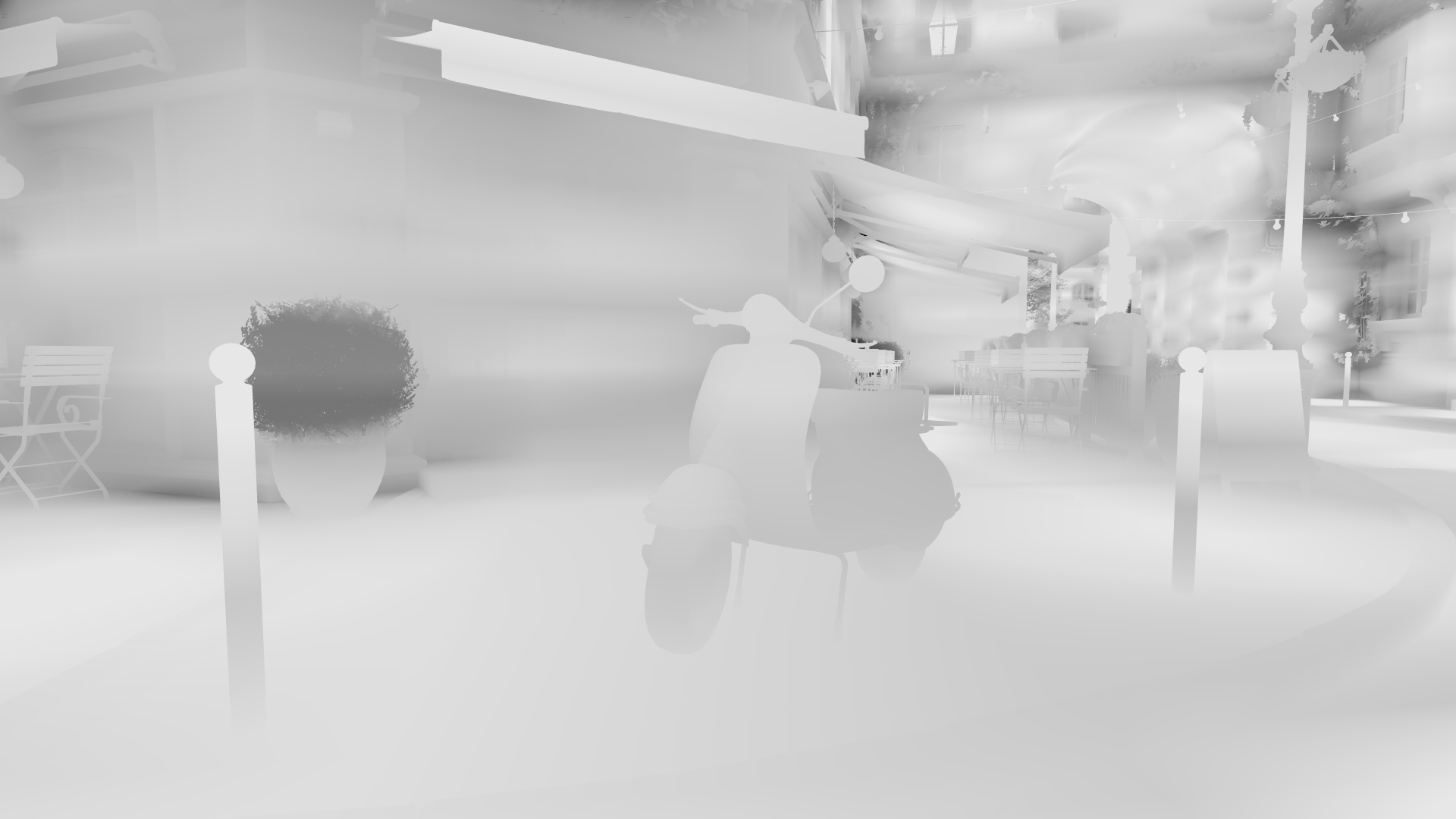}};
        \node[anchor=south east, xshift=-2mm, yshift=2mm, fill=white, opacity=0.7, text opacity=1, font=\scriptsize] 
            at (image.south east) {FLIP 0.121};
       \node[anchor=north west, xshift=2mm, yshift=-2mm, fill=white, opacity=0.7, text opacity=1, font=\scriptsize] 
           at (image.north west) {Hash-Grid};
    \end{tikzpicture}
\end{minipage}%
\begin{minipage}{0.33\textwidth}
    \begin{tikzpicture}
        \node[anchor=south west, inner sep=0] (image) at (0,0) {\includegraphics[width=\linewidth]{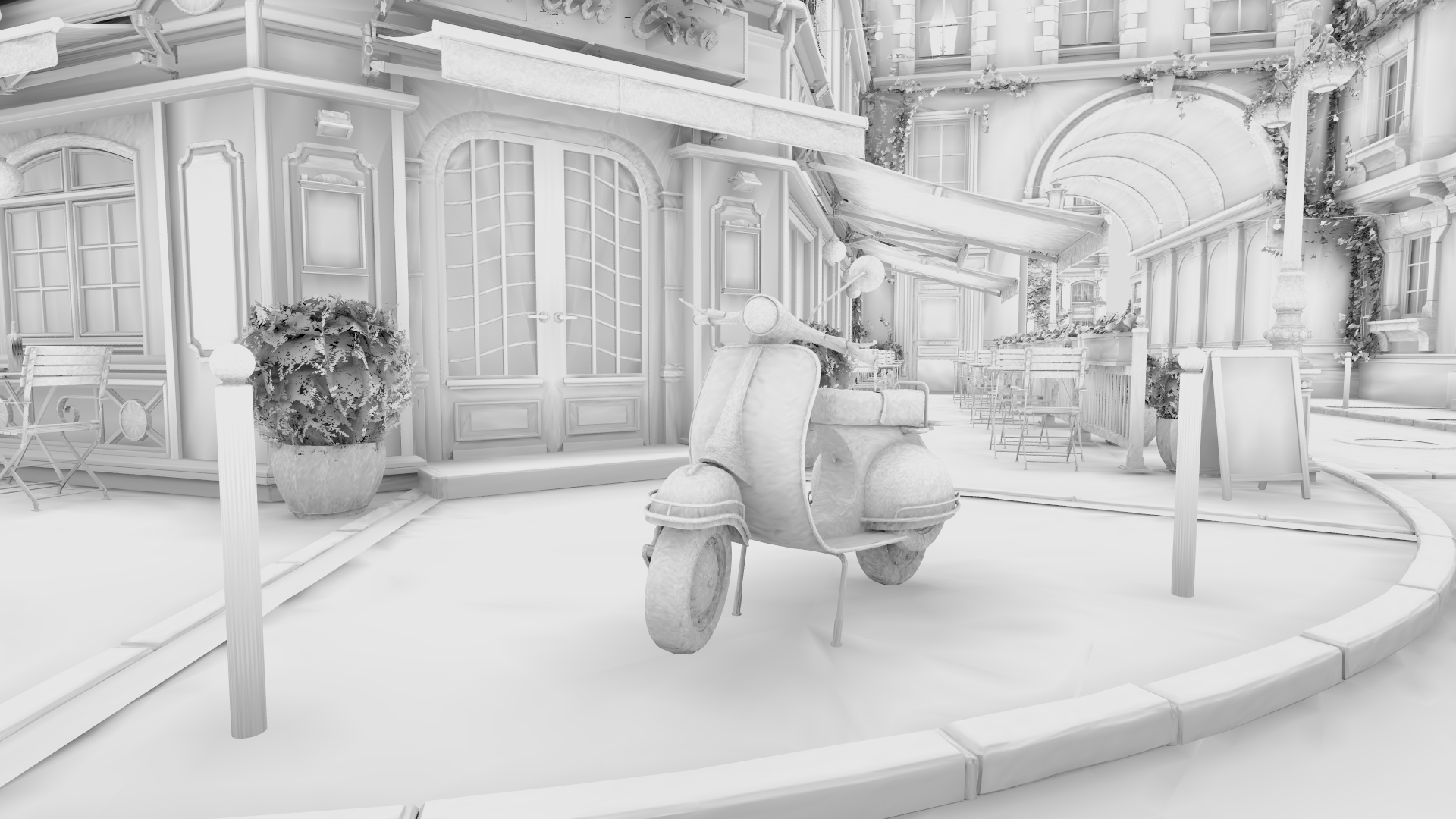}};
        \node[anchor=south east, xshift=-2mm, yshift=2mm, fill=white, opacity=0.7, text opacity=1, font=\scriptsize] 
            at (image.south east) {FLIP 0.038};
       \node[anchor=north west, xshift=2mm, yshift=-2mm, fill=white, opacity=0.7, text opacity=1, font=\scriptsize] 
           at (image.north west) {GATE (ours)};
    \end{tikzpicture}
\end{minipage}%
\begin{minipage}{0.33\textwidth}
    \begin{tikzpicture}
        \node[anchor=south west, inner sep=0] (image) at (0,0) {\includegraphics[width=\linewidth]{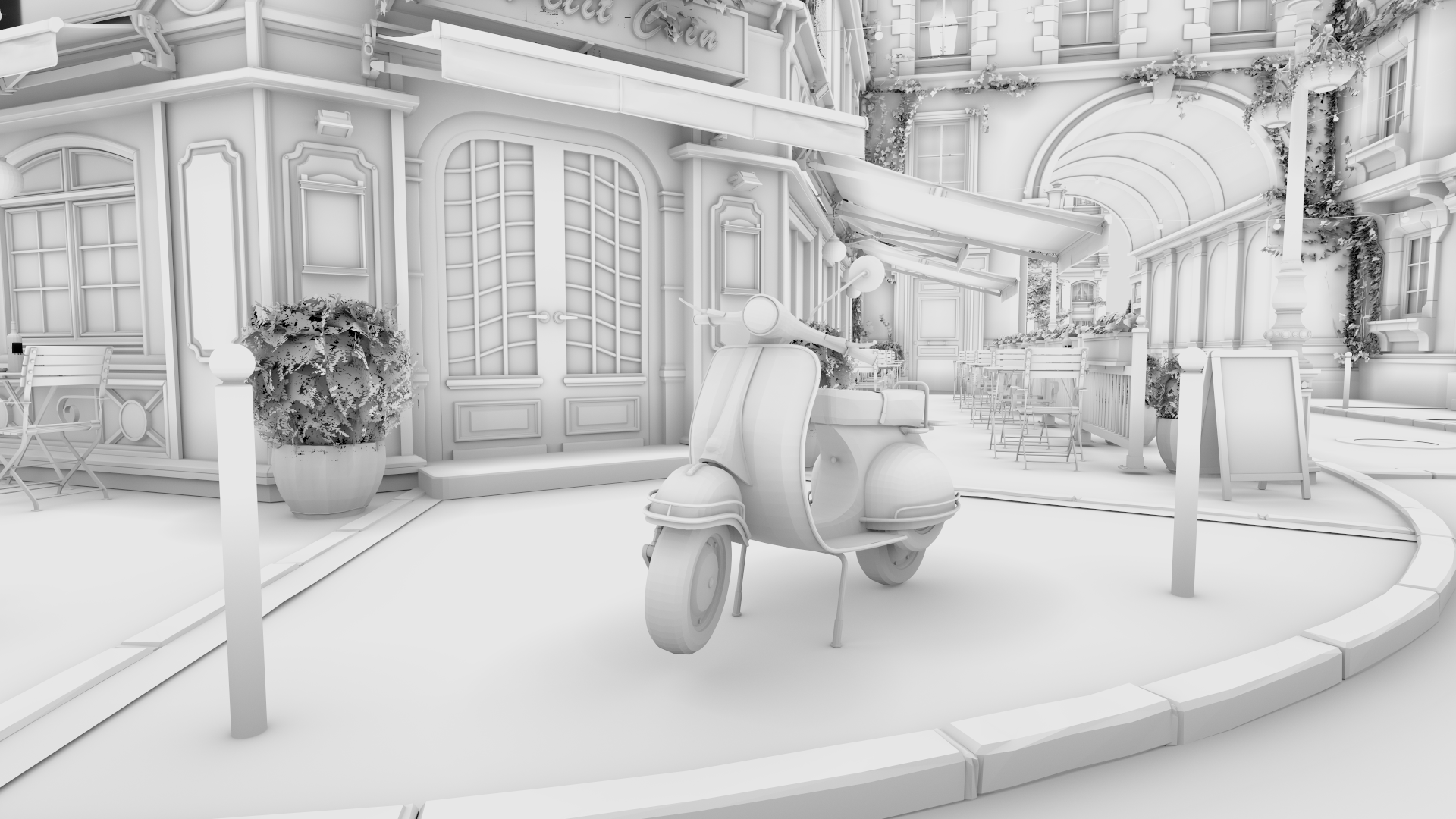}};
       \node[anchor=north west, xshift=2mm, yshift=-2mm, fill=white, opacity=0.7, text opacity=1, font=\scriptsize] 
           at (image.north west) {Ground Truth};
    \end{tikzpicture}
\end{minipage}%
\end{center}
\caption{\label{fig:teaser}
\emph{Bathroom} scene (top) rendered with neural radiance caching (NRC) and \emph{Bistro-Exterior} (bottom) rendered with neural ambient occlusion (NAO) after 1024 (NRC) and 128 (NAO) training steps using multi-dimensional hash-grid encoding~\cite{muller2022instant} (left), our GATE method (middle), and ground truth (right). GATE provides faster training (up to $3.7 \times$) and inference times (up to $2.7 \times$) and at the same time achieves higher quality (lower error vs. ground truth), especially for teapot-in-the-stadium scenarios (e.g., Vespa in \emph{Bistro}).}
}

\maketitle
\begin{abstract}
The encoding of input parameters is one of the fundamental building blocks of neural network algorithms. Its goal is to map the input data to a higher-dimensional space~\cite{rahaman2019}, typically supported by trained feature vectors~\cite{muller2022instant}. The mapping is crucial for the efficiency and approximation quality of neural networks. We propose a novel geometry-aware encoding called GATE that stores feature vectors on the surface of triangular meshes. Our encoding is suitable for neural rendering-related algorithms, for example, neural radiance caching~\cite{muller2021real}. It also avoids limitations of previous hash-based encoding schemes, such as hash collisions, selection of resolution versus scene size, and divergent memory access. Our approach decouples feature vector density from geometry density using mesh colors~\cite{yuksel2010mesh}, while allowing for finer control over neural network training and adaptive level-of-detail.
\end{abstract}  

\section{Introduction}
Deep learning has become ubiquitous in many industrial and research areas. In computer graphics, the multilayer perceptron (MLP), one of the simplest models, has proven to be a powerful tool capable of approximating various complex phenomena (e.g., global illumination~\cite{muller2021real,Dereviannykh2025} or geometry representation~\cite{Takikawa2021}) with appropriate input encoding. Compared to advanced models, the advantage is that MLPs can be trained online in several milliseconds on modern GPUs~\cite{muller2021real,Dereviannykh2025}. Input encoding proved to be a crucial component for capturing high-frequency signals from low-dimensional inputs~\cite{rahaman2019}. The principle is to map low-dimensional signals to a higher-dimensional space to help MLPs capture the high-frequency details. Various encoding strategies have been proposed; among others, those with learnable parameters became popular~\cite{muller2022instant}.

In this paper, we present a novel positional encoding\footnote{Throughout the paper, we use the term \emph{positional encoding} for the encoding of positions and \emph{frequency encoding} for positional encoding based on frequencies~\cite{Vaswani2017, mildenhall2020nerf}.} to store quantities on triangular meshes in the context of ray tracing-based rendering. We exploit the fact that the rays originate from triangle surfaces (except camera rays), hence we store the feature vectors directly on triangles and use barycentric coordinates for interpolation. A positive side effect of using barycentric interpolation is that the input is implicitly normalized, which otherwise needs to be done manually (based on scene bounds). Our approach also adapts to the geometry of the scene, efficiently dealing with the \emph{teapot-in-a-stadium} problem (i.e., a detailed object inside a large scene) compared to uniform grid-based approaches~\cite{muller2022instant}. Our contributions include:
\begin{itemize}
    \item A novel positional encoding method for triangular meshes with trainable parameters that allows real-time online training.
    \item An adaptive algorithm distributing feature vectors on triangles according to the sizes of triangles.
    \item A distribution scheme for training samples that limits training to a subset of triangles.
    \item Application of our encoding in two rendering scenarios: neural ambient occlusion and neural radiance caching.
\end{itemize}

\section{Previous Work} 
Computer graphics typically deals with low-dimensional data (e.g., 3D positions, normal vectors, or RGB colors) compared to high-dimensional data for computer vision or natural language processing (e.g., images, words, or sentences). Feeding such input directly to an MLP blurs out the high-frequency details. Therefore, various encoding methods which map low-dimensional input to a high-dimensional space have been proposed to help MLPs representing high-frequency details.

\emph{One-hot encoding}, one of the first encoding techniques, divides the range of input values into a predefined number of bins, where each bin corresponds to a single output dimension. The bin in which an input value falls is set to one while the other dimensions remain zero. A drawback is that different values assigned to the same bin are mapped to the same value. M\"{u}ller et al.~\cite{muller2019neural} generalized \emph{one-hot encoding} into \emph{one-blob encoding}. The idea is to center a Gaussian kernel around the input value and dice the Gaussian into bins, where the value of each bin is obtained by integration over the corresponding Gaussian segment. In practice, the Gaussian kernel can be replaced with a quartic kernel for higher performance~\cite{muller2021real}.

\emph{Frequency encoding}~\cite{Vaswani2017, mildenhall2020nerf} represents an input value as a sequence of sines and cosines, where the input is multiplied by $2^{i-1}\pi$ for $i$-th frequency and fed to the sine and cosine functions. The number of frequencies is a hyperparameter and should be set according to the highest frequency of the captured input data.

Parametric encodings combine traditional data structures and neural networks. The concept is to store additional learnable parameters (i.e., latent feature vectors) in a data structure spanning across the input domain, where the feature vectors are optimized by stochastic gradient descent together with the network weights. An input value, before it is fed to a neural network, is encoded by interpolating the feature vectors in the proximity of the input value. Most of the learnable parameters are now stored in the data structure, and the neural networks serves only as a lightweight decoder.

Octrees~\cite{Takikawa2021, Martell2021} and uniform grids with single or multiple levels~\cite{Hadadan2021, Kuznetsov2021, Takikawa2022, Weier2023} are widely used to store feature vectors. The input data are encoded by tri-linearly interpolated feature vectors in the grid/octree cell's corners. In the case of multiple levels, the feature vectors from the individual levels are concatenated. The memory footprint of grid-based data structures might be very large. M\"{u}ller et al.~\cite{muller2022instant} proposed to represent a multi-resolution grid in a hash table, significantly reducing memory requirements, and similarly, Takikawa et al.~\cite{Takikawa2021} proposed to represent feature vectors as a sparse voxel octree. Also, Takikawa et al.~\cite{takikawa2023compact} combined multi-resolution hash-grid encoding with a pair of codebooks, while Govindarajan et al.~\cite{govindarajan2024laghashes} used Gaussian interpolation to reduce the size of the hash table. Fujieda et al.~\cite{Fujieda2023} proposed applying frequency encoding on top of grid-based encoding such that the sines and cosines are multiplied by the coefficients of the interpolated feature vector.

The encodings discussed so far assume normalized uniformly distributed data of arbitrary dimension (in theory). Sivaram et al.~\cite{Sivaram2024} proposed an encoding customized to represent 3D meshes. The mesh is decomposed into coarse quadrangular patches, where each patch stores four feature vectors in the corners. Encoded data are obtained by bi-linearly interpolating the feature vectors of a given patch. The interpolated point within the patch is also fed into the frequency encoding and concatenated with the interpolated feature vector. The feature vectors are trained to represent displacements for points on the patches. Thies et al.~\cite{thies2019} proposed a deferred neural rendering framework that stores latent feature vectors as textures mapped on imperfect models reconstructed by photometric methods. The neural renderer is trained together with feature vectors to match photorealistic target images. Similarly to these approaches, our method naturally adapts to the geometry of the scene, unlike general encodings. We store feature vectors directly on the triangular meshes, using (implicitly normalized) barycentric coordinates for interpolation. We train an MLP and the feature vectors to represent various photometric quantities (e.g., radiance or ambient occlusion) computed by a physically-based renderer in contrast to the neural renderer~\cite{thies2019} or the displacement~\cite{Sivaram2024}.

Neural methods for compressing textures~\cite{fujieda2024neural, vaidyanathan2023random} and representing materials~\cite{Kuznetsov2021, zeltner2024real} use encodings tailored for their input domains (e.g., the multi-resolution 2D grid or a hierarchical texture). These methods are aimed at offline training and generally do not support algorithms operating in 3D space, such as the online-trained neural radiance cache.

\section{Geometry-Aware Trained Encoding (GATE)}
\label{sec:OurMethod}

\begin{figure}
\centering
\includegraphics[width=0.45\textwidth]{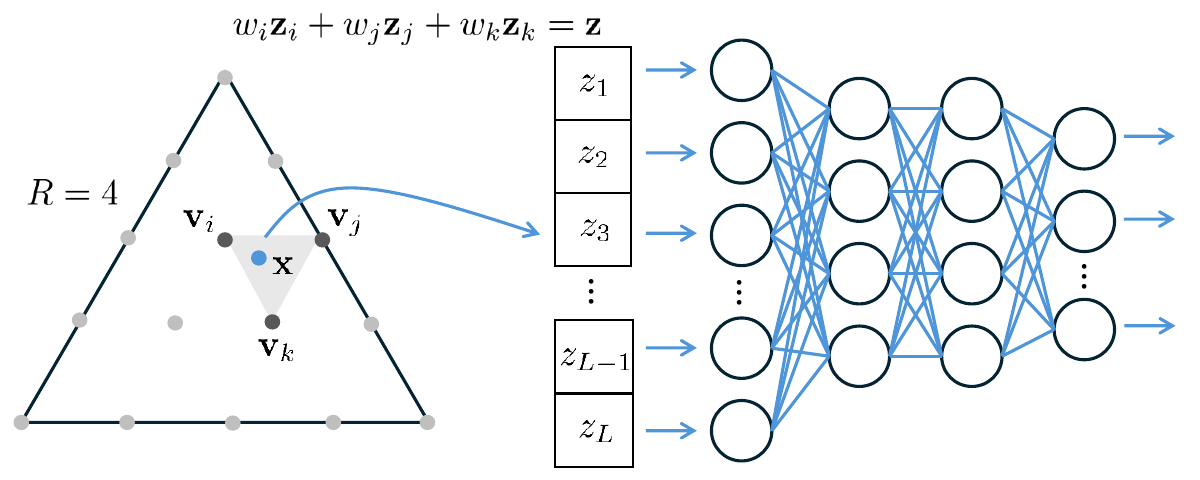}
\caption{An overview of our method. Latent feature vectors are evenly distributed on triangles using mesh colors~\cite{yuksel2010mesh} with resolution $R$. We find a virtual triangle formed by vertices $(\mathbf{v}_i, \mathbf{v}_j, \mathbf{v}_k)$ where the input point $\mathbf{x}$ belongs to. We use barycentric coordinates within this virtual triangle $(w_i, w_j, w_k)$ to interpolate of three feature vectors $(\mathbf{z}_i, \mathbf{z}_j, \mathbf{z}_k)$ associated with the vertices to get the encoded vector $\mathbf{z} = (z_1, \dots, z_L)$, which is subsequently fed to an MLP.}
\label{fig:overview}
\end{figure}

In this section, we present our \emph{geometry-aware trained encoding} (GATE), a parametric positional encoding for points located on triangular meshes, storing latent feature vectors directly on triangles, such that each input point is uniquely specified by the triangle ID, mesh ID, and barycentric coordinates with respect to a given triangle. To infer the feature vector of a point, we locate the three closest feature vectors distributed on the surface of given triangle with mesh colors using barycentric coordinates, and then interpolate them (see Section~\ref{sec:mesh_colors}). The length of a feature vector $L$ is a hyperparameter (typically between $1$ and $8$). All feature vectors are learnable parameters and optimized through backpropagation. An overview of the GATE method is illustrated in Figure~\ref{fig:overview}.

Compared to other competing methods based on grids and/or hashing, our GATE approach has several advantages. Optimizing the training loss will result in feature vectors optimized specifically for every given triangle. This is in contrast to grid-based methods, where feature vectors are stored in the corners of a cell and tri-linearly interpolated to encode arbitrary positions inside the cell (i.e., surface vs. volume). The locations of feature vectors in our approach naturally adapt to the scene geometry, and we also do not need to explicitly normalize the input values (e.g., the world-space position normalized to the scene bounds) due to interpolation based on barycentric coordinates. Since feature vectors are stored consecutively in memory, our approach has a more cache-friendly memory access pattern with no hashing collisions unlike hash-based grid methods. For those grid-based methods, feature vectors belonging to the same cell are scattered in memory and collisions may occur.

\subsection{Distribution of Latent Feature Vectors}
\label{sec:mesh_colors}

We employ \emph{mesh colors}~\cite{yuksel2010mesh} to distribute latent feature vectors across the surface of triangles. The original algorithm was designed to extend vertex colors, where additional colors are distributed on edges and faces based on a triangle given resolution $R$, which is a hyperparameter ($R = 1$ corresponds to when the feature vectors are stored only in the triangle vertices). We employ the same algorithm to distribute feature vectors across a triangle. The idea is to virtually tessellate a given triangle into smaller triangles such that feature vectors are stored at the vertices of virtual triangles. The total number of feature vectors per triangle (including edges and vertices) is $\frac{(R+1)(R+2)}{2}$.

Inferring the feature vector of an input point, we need to fetch the three closest feature vectors (with respect to the virtual triangle) and use the barycentric coordinates (with respect to the virtual triangle) for interpolation. For a virtual triangle formed by vertices $(\mathbf{v}_i, \mathbf{v}_j, \mathbf{v}_k)$, the latent feature vectors are interpolated by:
\begin{equation}
    \mathbf{z} = w_i \mathbf{z}_i + w_j \mathbf{z}_j + w_k \mathbf{z}_k,
    \label{eq:interpolation}
\end{equation}
where $w_t$ are barycentric weights and $\mathbf{z}_t$ are feature vectors (associated with vertex $\mathbf{v}_t$) and $\mathbf{z}$ is the resulting feature vector. The indices of the closest three feature vectors and the barycentric weights (coordinates) within the virtual triangle are provided by mesh colors, based on the integral part and fractional parts, respectively, of the barycentric coordinates scaled by $R$ (see details in~\cite{yuksel2010mesh}).

In addition to the feature vectors at the original vertices, mesh colors create additional feature vectors within a triangle and on its edges. We use information from the mesh's index buffer to share feature vectors at vertices and on the edges of neighboring triangles. This ensures seamless encoding at shared triangle edges of a mesh, assuming a fixed $R$ per mesh. With a fixed $R$ per mesh, our implementation stores $\frac{(R+1)(R+2)}{2}$ feature vectors per triangle consecutively in memory. 

Given a triangle index and barycentric coordinates of a point on the triangle, obtaining the final interpolated feature vector involves multiple steps. First, we compute the start offset of the vertex, edge, and surface feature vectors for the given triangle and mesh. This offset points into a large continuous buffer that holds all feature vectors. These offsets are pre-calculated and stored into an auxiliary buffer for the first triangle of each mesh. Because we use a fixed mesh colors resolution $R$ for every mesh, there is the same amount of feature vectors for every triangle per mesh. The offset where the feature vectors start for a given triangle can be calculated as $ Instance_{Offset} + Triangle_{ID} \cdot FeatureVectorsPerTriangle$. Second, given the barycentric coordinates of the base triangle,  the three closet vertices and barycentric weights of the virtual triangle are provided by mesh colors. The feature vectors of the three closest vertices are loaded from the feature vector buffer and interpolated using the barycentric weights as shown in Equation~\ref{eq:interpolation}, resulting in the final (encoded) feature vector. Note that in case of double-sided triangles, feature vectors need to be stored for each side separately. 

\begin{figure}
\centering
\includegraphics[width=0.475\textwidth]{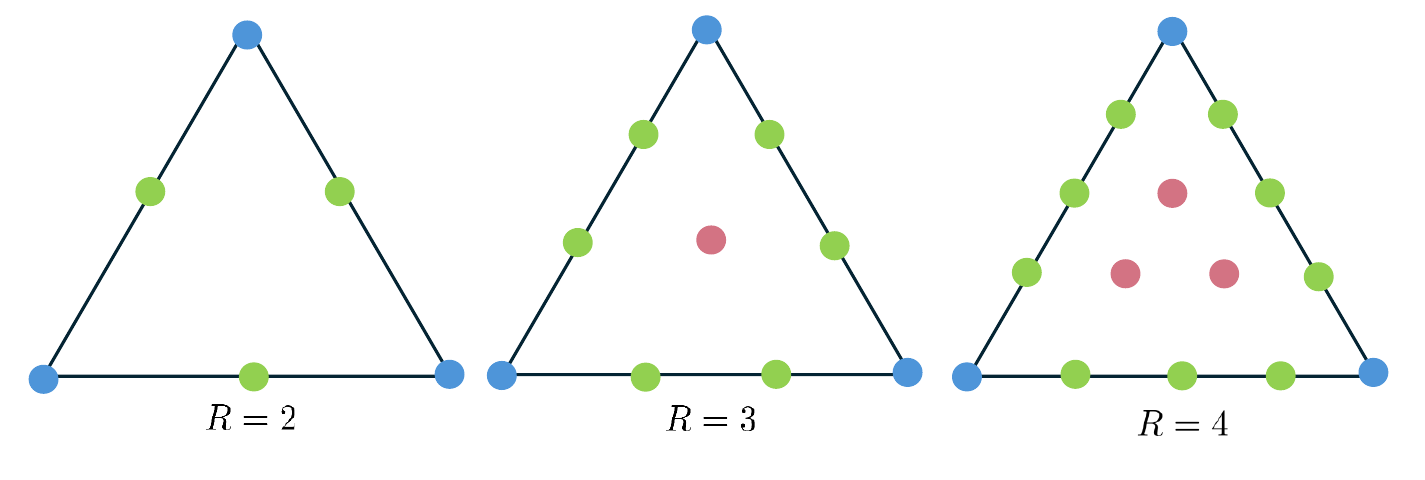}
\caption{An example of mesh colors~\cite{yuksel2010mesh} that we use to distribute feature vectors on triangles using different resolutions. The feature vectors from multiple resolutions can be concatenated to improve the approximation quality.}
\label{fig:resolution}
\end{figure}

The per-triangle resolution~$R$ can be set to achieve a uniform density of feature vectors (in world-space) throughout the scene, regardless of individual triangle sizes. This allows for increasing the feature vector resolution in areas of high interest (e.g., close to the camera position, within a viewing frustum, or in areas of high path tracing sample density). To avoid inconsistencies on edges, we opt to set $R$ per mesh (or instance in the case of instancing) based on triangle area (in world space), with larger triangles assigned higher resolutions. We compute the average normalized triangle area $A_\mathcal{I}$ of mesh/instance $\mathcal{I}$. Then, the adaptive triangle resolution $R_\mathcal{I}$ is given as the following clamped quadratic function: 
\begin{equation}
    R_\mathcal{I} = \clamp{(32 \cdot {A_\mathcal{I}}^2 \cdot R_{scale}, 1, 32)},
\end{equation}

where $R_{scale}$ is a scene-specific parameter for fine-tuning the density of mesh colors. Similarly to grid-based approaches that use multiple resolutions, GATE can hierarchically stack multiple resolutions provided by mesh colors (see Figure~\ref{fig:resolution}). The interpolated feature vectors from different resolutions are then concatenated and fed to the MLP in a similar manner as for multi-resolutions hash-grids, where the number of resolutions is a hyperparameter. 

\begin{table}[!h]
\smallskip{}
    \footnotesize
    \centering
    \begin{tabular}{ccccc}
                   &  Bistro (Ext) &  Bistro (Int) & Bathroom & Kitchen\\
    \hline                                                                 
                   \#Triangles & 2837K &  1020K & 39K & 610K\\
       \vspace{1pt}\\                   
    \hline                                              
    \hline        
                \multicolumn{5}{c}{Neural Ambient Occlusion (NAO)}\\

    \hline        
    \hline                                                  
                   \multicolumn{5}{c}{Training [ms]}\\
    \hline 
       GATE (ours) & 0.45 & 0.38 & 0.38 & 0.40\\
       Hash-Grid   & 25.32 & 3.41 & 1.25 & 2.40 \\
       \textbf{Speedup}     & $\mathbf{56.27 \times}$ & $\mathbf{8.97 \times}$ & $\mathbf{3.29 \times}$ & $\mathbf{6.0 \times}$\\
    \hline
                  \multicolumn{5}{c}{Inference [ms]}\\
    \hline                 
       GATE (ours) & 0.70 & 0.70 & 0.66 & 0.70\\
       Hash-Grid   & 2.05 & 1.91 & 1.65 & 1.92\\
       \textbf{Speedup}     & $\mathbf{2.93 \times}$ & $\mathbf{2.73 \times}$ & $\mathbf{2.5 \times}$ & $\mathbf{2.74 \times}$\\
    \hline
                   \multicolumn{5}{c}{Memory Consumption [MB]}\\
    \hline                 
       GATE (ours)  & 177.09 & 16.00 & 0.67 & 9.92\\
       Hash-Grid    & 189.59 & 17.55 & 0.66 & 9.37\\
       Ratio (G/H)  & $1.07$ & $1.10$ & $1.02$ & $0.95$\\
    \hline  
                   \multicolumn{5}{c}{FLIP Error}\\
    \hline                 
       GATE (ours)  & 0.040 & 0.038 & 0.021 & 0.033\\
       Hash-Grid    & 0.065 & 0.041 & 0.021 & 0.034\\
       Ratio (G/H) & $0.62$ & $0.95$ & $1.00$ & $0.98$\\
    \hline
           \vspace{1pt}\\
    \hline
    \hline
                   \multicolumn{5}{c}{Neural Radiance Caching (NRC)}\\                   
    \hline                           
    \hline        
                   \multicolumn{5}{c}{Training [ms]}\\
    \hline 
       GATE (ours) & 1.43 & 1.37 & 1.45 & 1.41\\
       Hash-Grid   & 41.56 &6.97 & 1.86  & 4.87 \\
       \textbf{Speedup}     & $\mathbf{29.06 \times}$ & $\mathbf{5.09 \times}$ & $\mathbf{1.28 \times}$ & $\mathbf{3.45 \times}$\\
    \hline
                  \multicolumn{5}{c}{Inference [ms]}\\
    \hline                 
       GATE (ours)  & 1.25 & 1.24 & 0.88 & 0.91 \\
       Hash-Grid    & 2.53 & 2.11 & 1.76 & 1.83\\
       \textbf{Speedup} & $\mathbf{2.02 \times}$ & $\mathbf{1.7 \times}$ & $\mathbf{2.00 \times}$ & $\mathbf{2.01 \times}$\\
    \hline
                   \multicolumn{5}{c}{Memory Consumption [MB]}\\
    \hline                 
       GATE (ours)  & 365.46 & 50.00 & 2.47 & 29.76\\
       Hash-Grid    & 328.85 & 46.22 & 2.45 & 29.84\\
       Ratio (G/H)  & $1.11$ & $1.04$ & $1.00$ & $1.00$\\
    \hline
                   \multicolumn{5}{c}{FLIP Error}\\
    \hline                 
       GATE (ours)  & 0.323  & 0.854 & 0.414 & 0.333\\
       Hash-Grid    & 0.382  & 0.985 & 0.409 & 0.348\\
       Ratio (G/H)  & $0.84$ & $0.86$ & $1.01$ & $0.95$\\
    \hline                                                  

    \end{tabular}
    \caption{Timings for NAO (top) / NRC (bottom) training and inference times per frame, and the FLIP error metric~\cite{Andersson2021b} (128 training iterations)  using a variable number of feature vectors per triangle (fixed per mesh). Size of the hash-grid is chosen so that both approaches consume a comparable amount of memory. GATE outperforms hash-grid in terms of training and inference performance, while at the same time providing higher quality (equal or lower FLIP error). Note that for hash-grid training we follow the reference implementation~\cite{muller2022instant} which trains every feature vector in the grid. This leads to a significant increase in training time for larger scenes. For a comparison to smaller hash-grid sizes see Table~\ref{tab:training_inference_comparison_varying_hash_grid_size}. }
    \label{tab:training_inference_comparison}
\end{table}

\subsection{Training}

We train the feature vectors together with the MLP using backpropagation. First, all feature vectors are initialized to random values centered around zero with magnitude up to $10^{-4}$ (similar to \cite{muller2022instant}). During backpropagation, we calculate the partial derivative of each feature vector element (gradient) with respect to a training loss $\mathcal{L}$:
\begin{equation}
    \frac{\partial \mathcal{L}}{\partial \mathbf{z}_t} = \frac{\partial \mathcal{L}}{\partial \mathbf{z}} \frac{\partial \mathbf{z}}{\mathbf{z}_t} = w_t\frac{\partial \mathcal{L}}{\partial \mathbf{z}}.
\end{equation}
Using these gradients, the feature vectors are optimized with the Adam optimizer\cite{diederik2014adam}. 

A typical GPU implementation of the gradient descent-based optimization distributes the learnable parameters (i.e., weights or feature vectors) uniformly among GPU threads, updating all the parameters. This would be rather inefficient for our GATE encoding, since there is a potentially very large number of triangles and their feature vectors, while only a small fraction of them are trained in a single step. Instead, we modify backpropagation to create a record for every feature vector that was updated in a dedicated buffer. The buffer size is chosen to support the maximum training batch size, and an atomic counter keeps track of the number of records actually created. The optimization phase (Adam) reads this buffer and only updates the feature vectors for which the gradient was calculated. 

Another observation is that when training GATE on noisy data, such as radiance, it pays off to reduce input noise to a minimum for more stable results and to learn high-frequency details faster. When training samples are generated, it is more efficient to use a smaller number of triangles with more samples, instead of a larger number of triangles with fewer samples. In this way, the available training budget is used more effectively.  

\newcommand{\imgwidth}{0.19}
\newcommand{\cellwidth}{3cm}

\newcommand{\rowlabel}[1]{%
\begin{minipage}{\imgwidth\textwidth}
    \rotatebox{90}{#1}
\end{minipage}
}

\newcommand{\includeimg}[1]{%
  \begin{minipage}{\imgwidth\textwidth}
    \centering
    \begin{tikzpicture}
        \node[anchor=south west, inner sep=0] (image) at (0,0)
            {\includegraphics[width=\linewidth]{#1}};
    \end{tikzpicture}
\end{minipage}
}

\newcommand{\includeimgcrop}[1]{\includegraphics[height=0.08\textheight, trim=90pt 0 90pt 0, clip]{#1}}

\newcommand{\includeerr}[2]{%
  \begin{minipage}{\imgwidth\textwidth}
    \centering
    \begin{tikzpicture}
        \node[anchor=south west, inner sep=0] (image) at (0,0)
            {\includegraphics[width=\linewidth]{#1}};
        \node[anchor=south east, xshift=-2mm, yshift=2mm, fill=white, opacity=0.7, text opacity=1, font=\scriptsize] 
            at (image.south east) {FLIP #2};
    \end{tikzpicture}
\end{minipage}
}

\begin{figure*}[!t]
\centering

\renewcommand{\arraystretch}{1.5}

\begin{tabular}{>{\centering\arraybackslash}m{0.15cm} m{\cellwidth} m{\cellwidth} m{\cellwidth} m{\cellwidth} m{\cellwidth} c}

& \centering Hash-Grid & \centering GATE (ours) & \centering Hash-Grid & \centering GATE (ours) & \centering Ground Truth & \\

\rowlabel{Bistro (Ext)} &
\includeerr{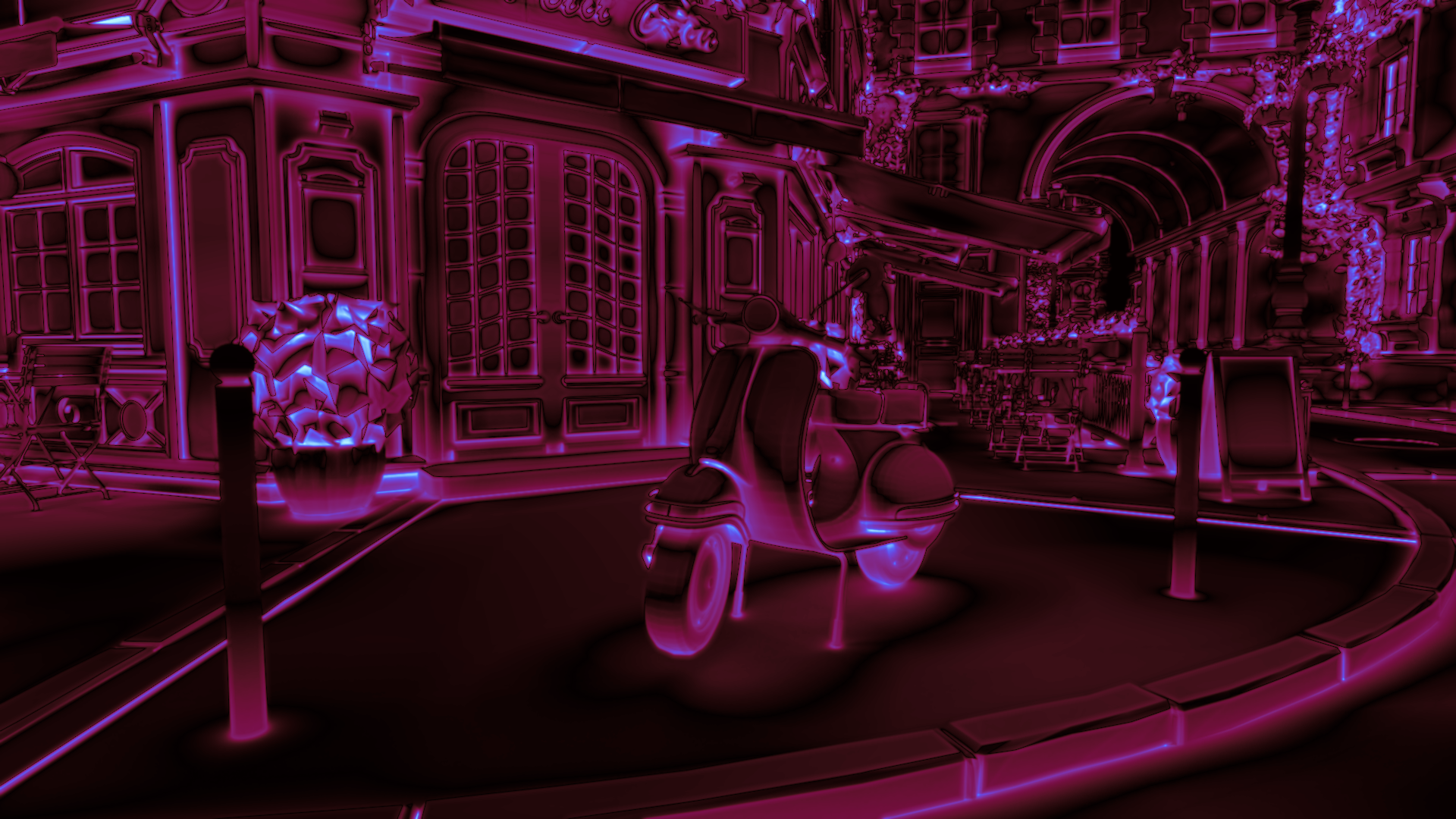}{0.121} &
\includeerr{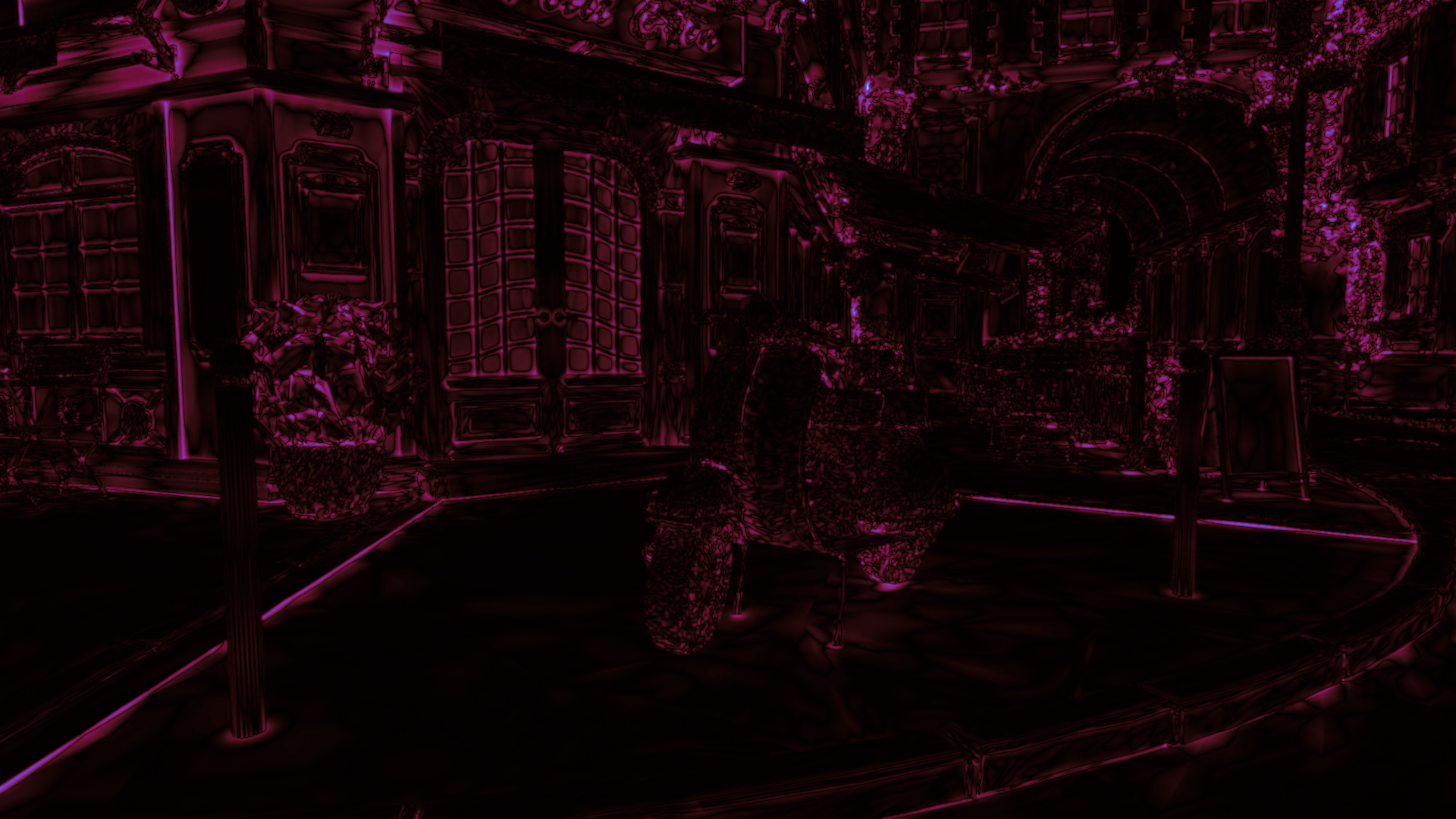}{0.038} &
\includeimg{2_bistro_exterior-AO-HashGrid-R-4-F-2-frames-128.png} &
\includeimg{2_bistro_exterior-AO-GATE-R-4-F-2-frames-128.png} &
\includeimg{2_bistro_exterior-AO-R-4-F-2-GROUND_TRUTH-frames-16384.png}\\

\rowlabel{Bistro (Int)} &
\includeerr{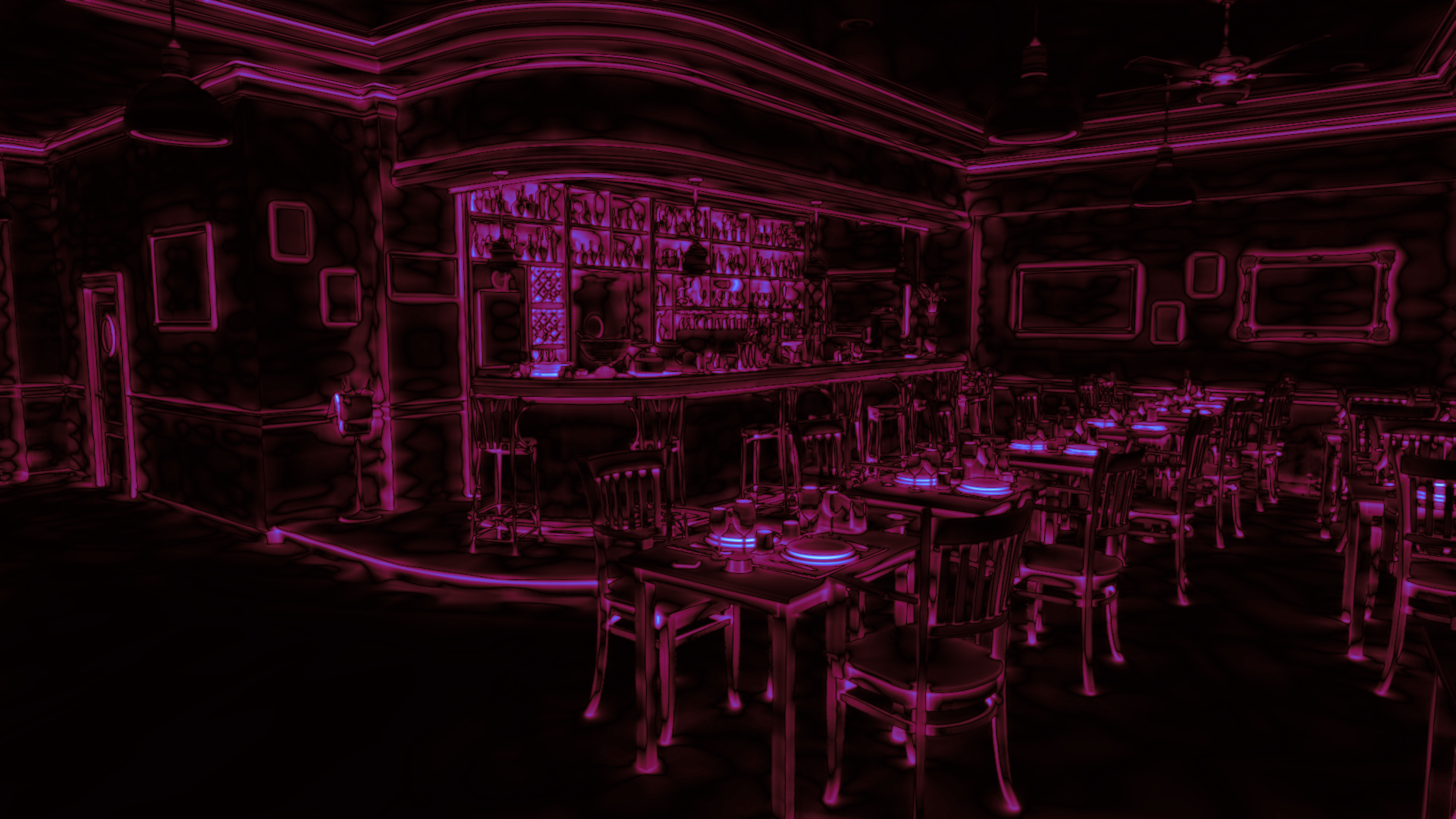}{0.058} &
\includeerr{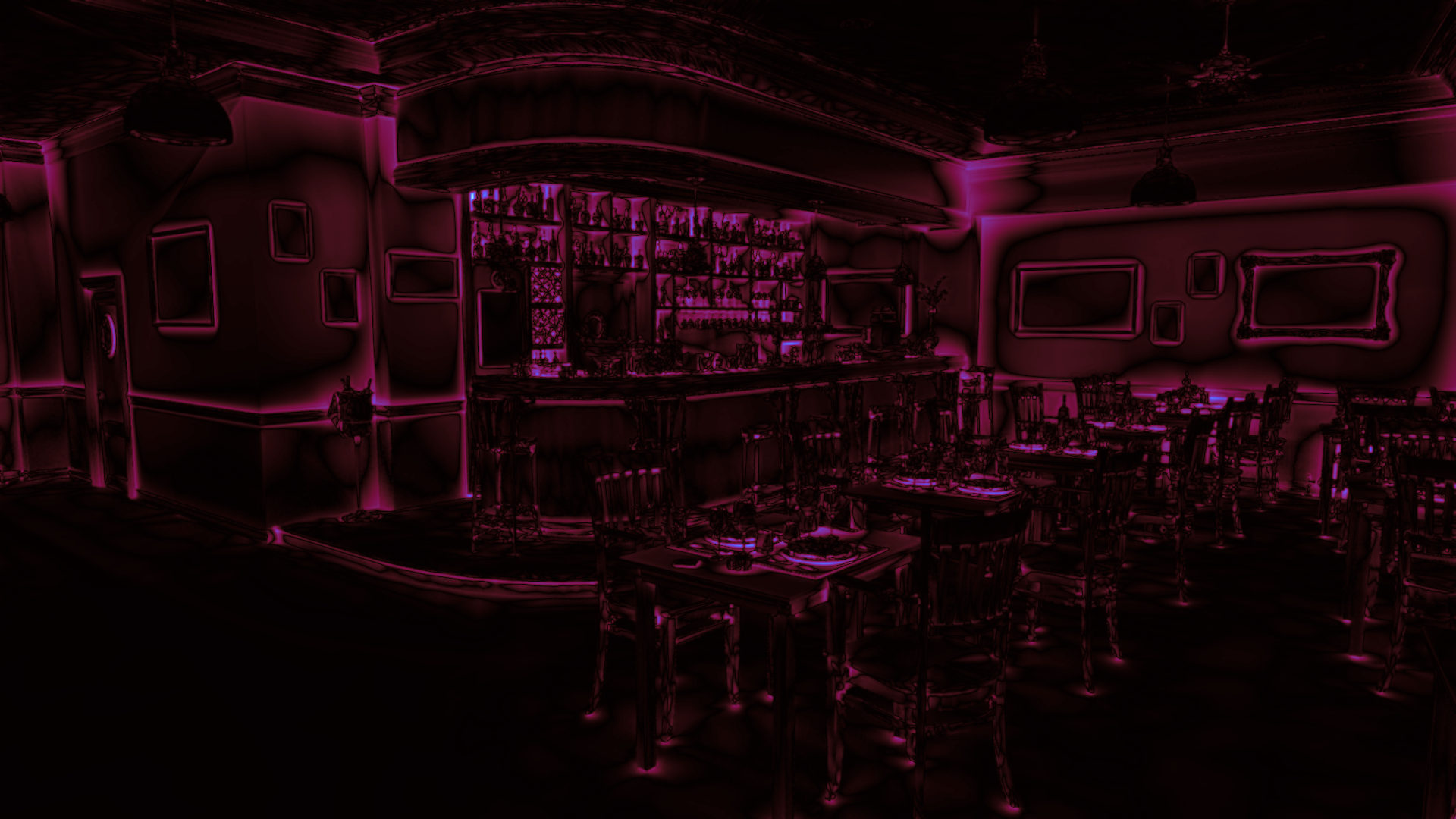}{0.043} &
\includeimg{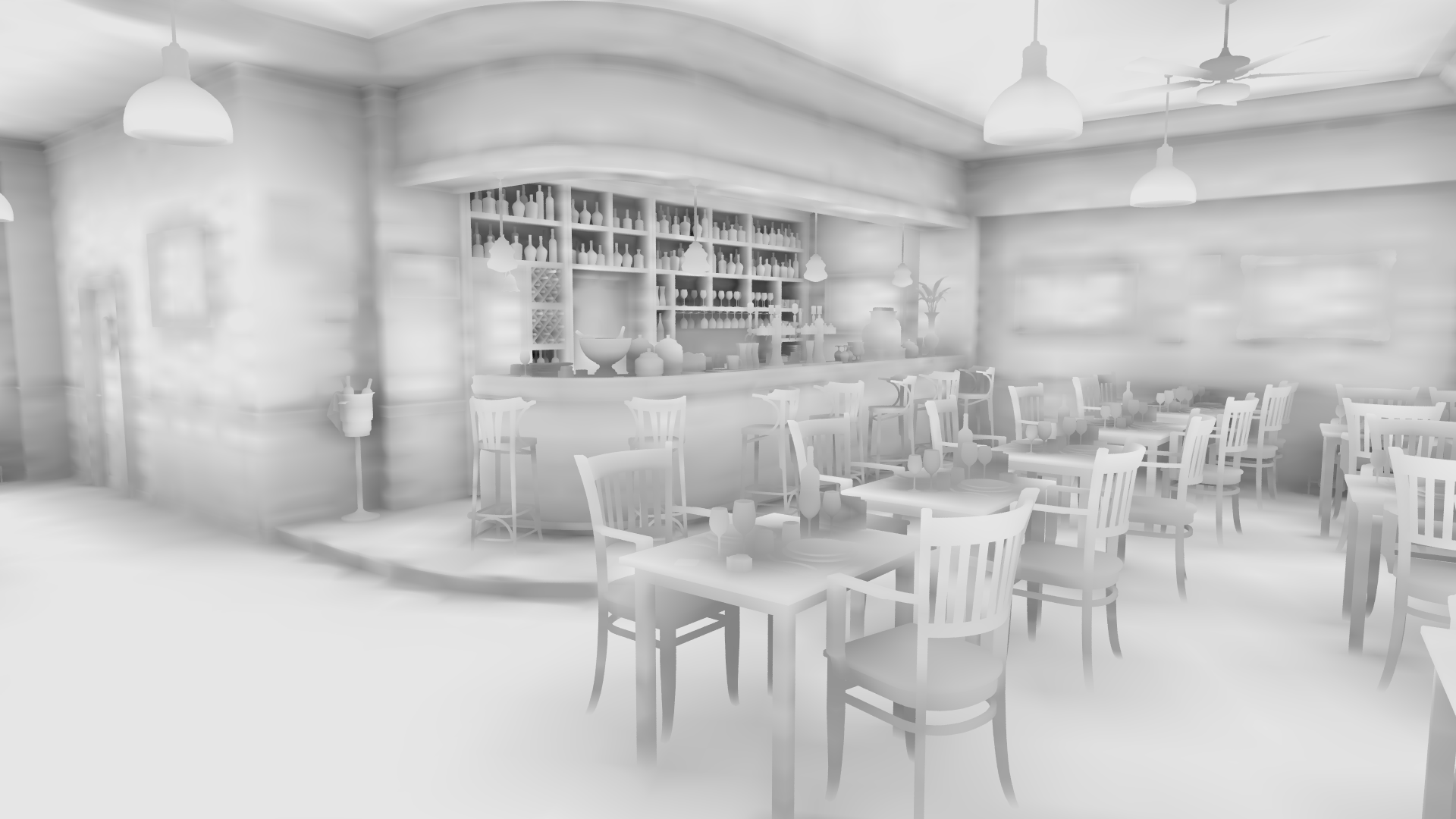} &
\includeimg{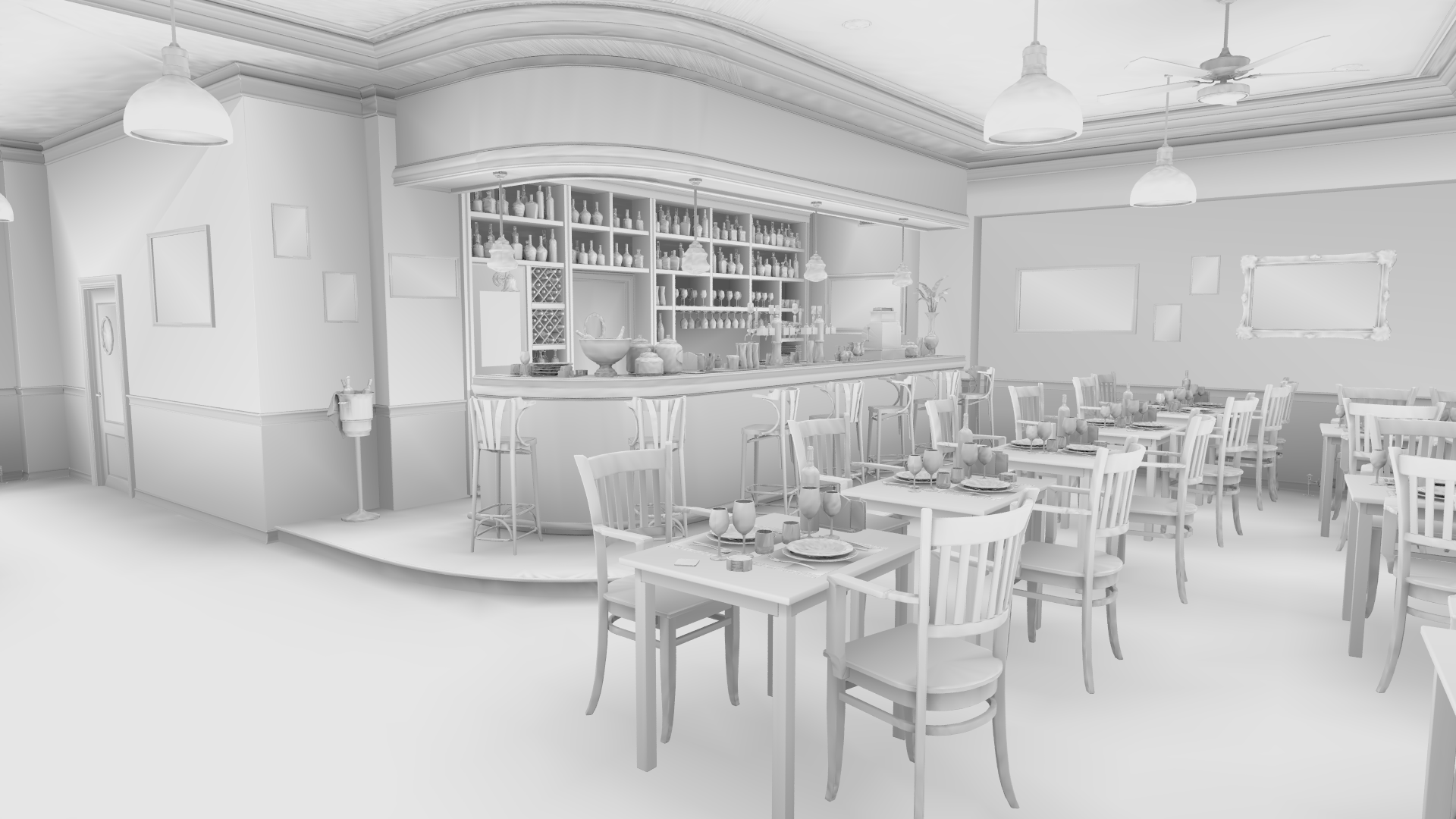} &
\includeimg{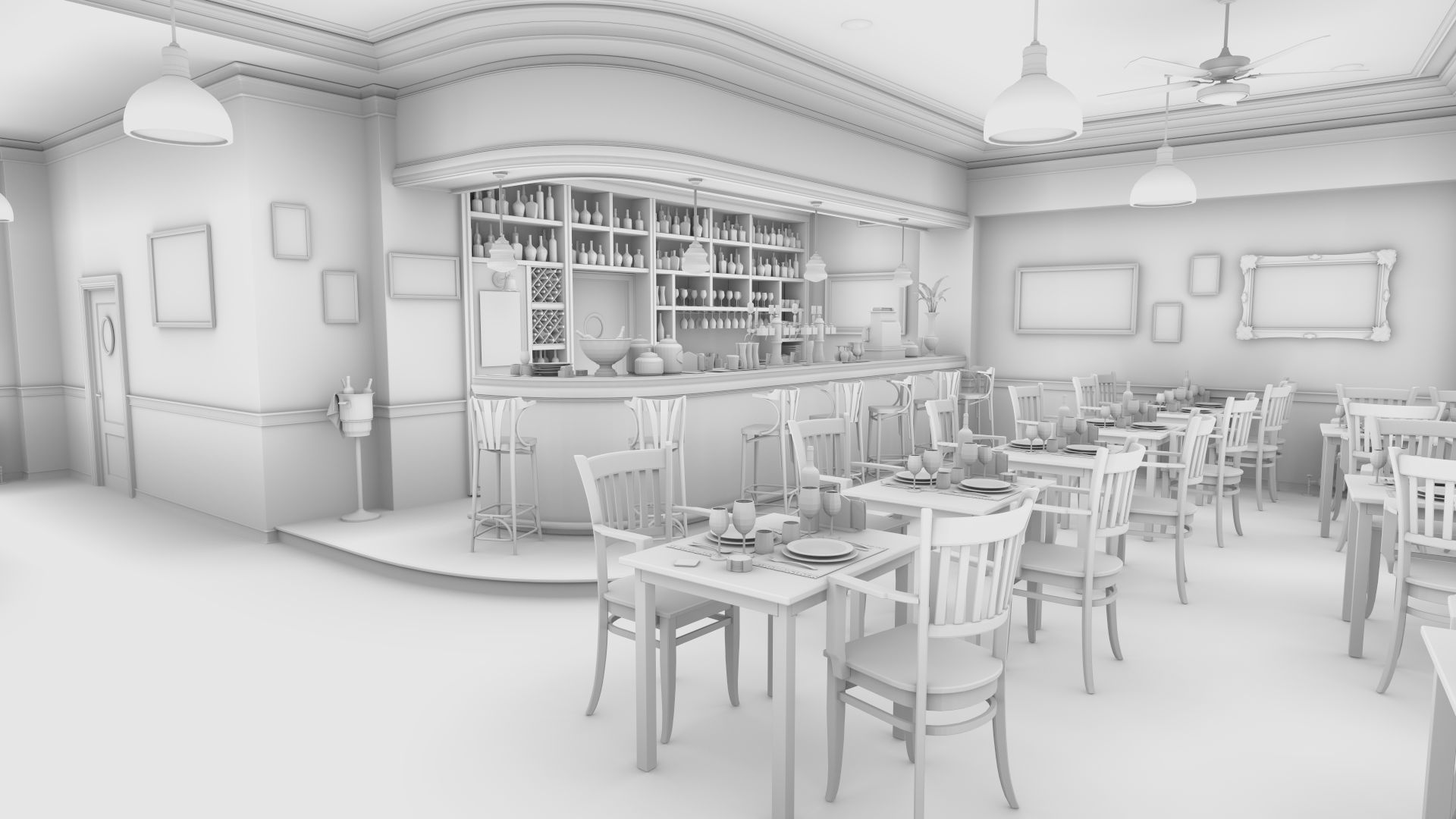}\\

\rowlabel{Bathroom} &
\includeerr{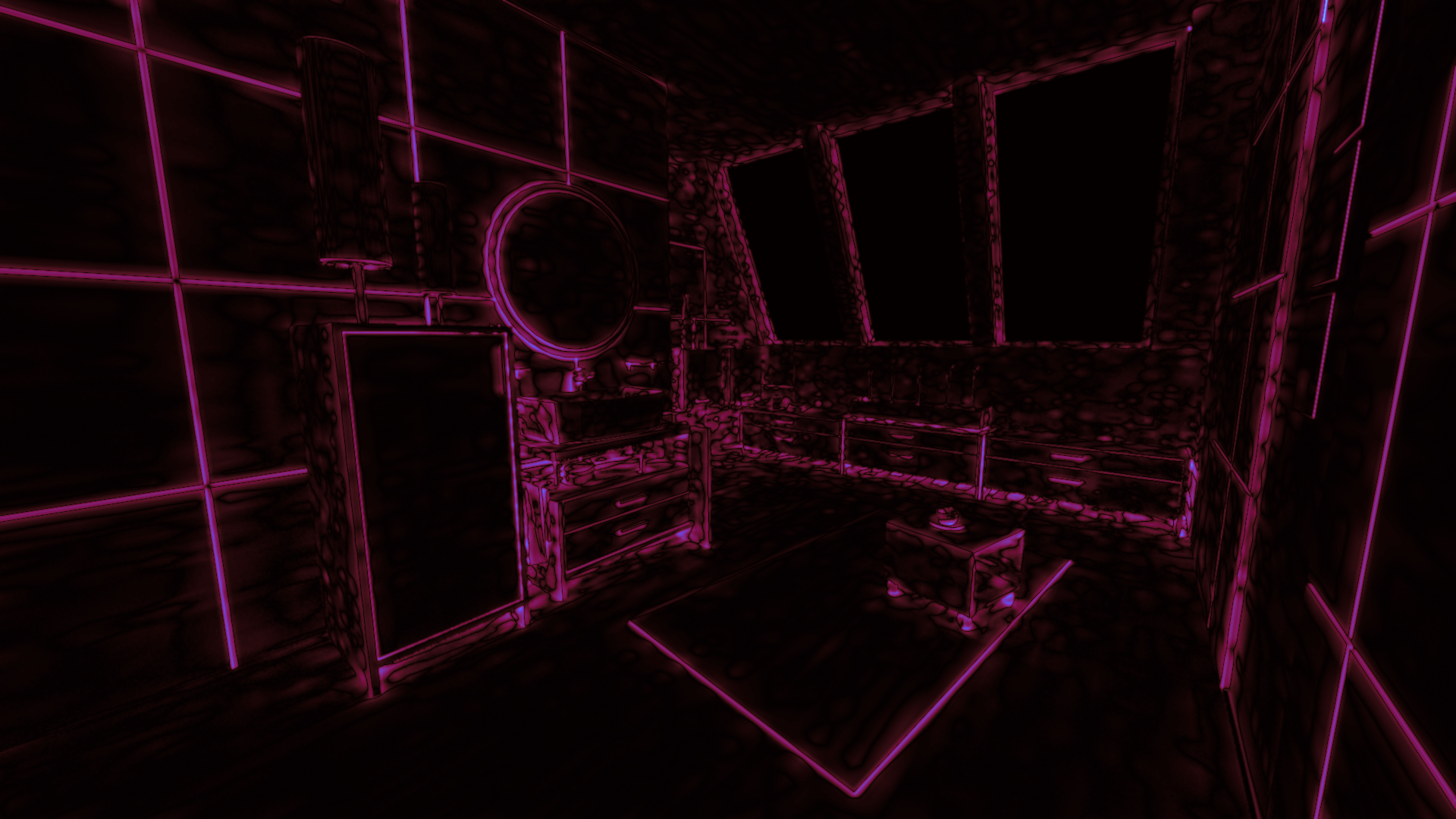}{0.037} &
\includeerr{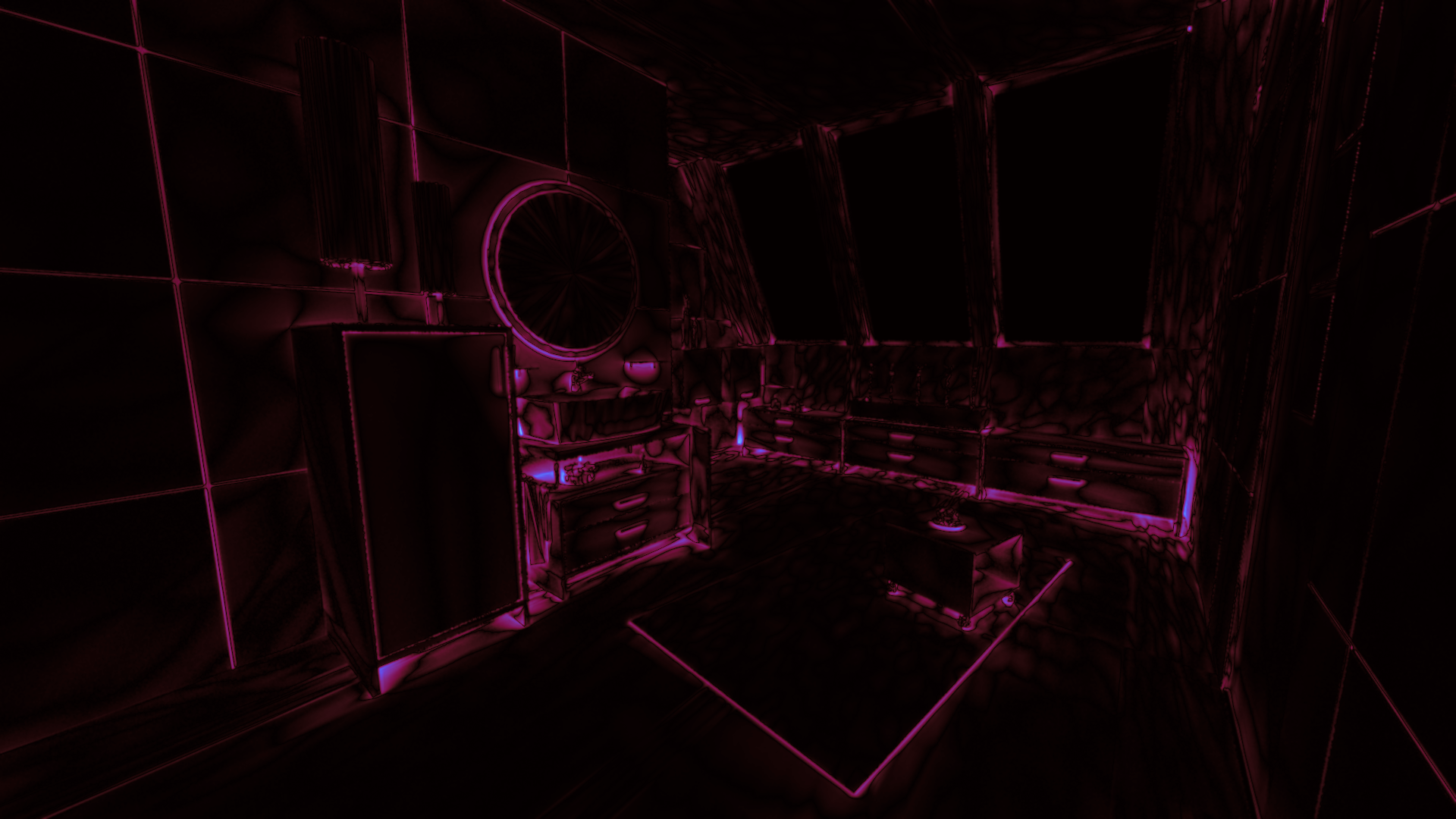}{0.024} &
\includeimg{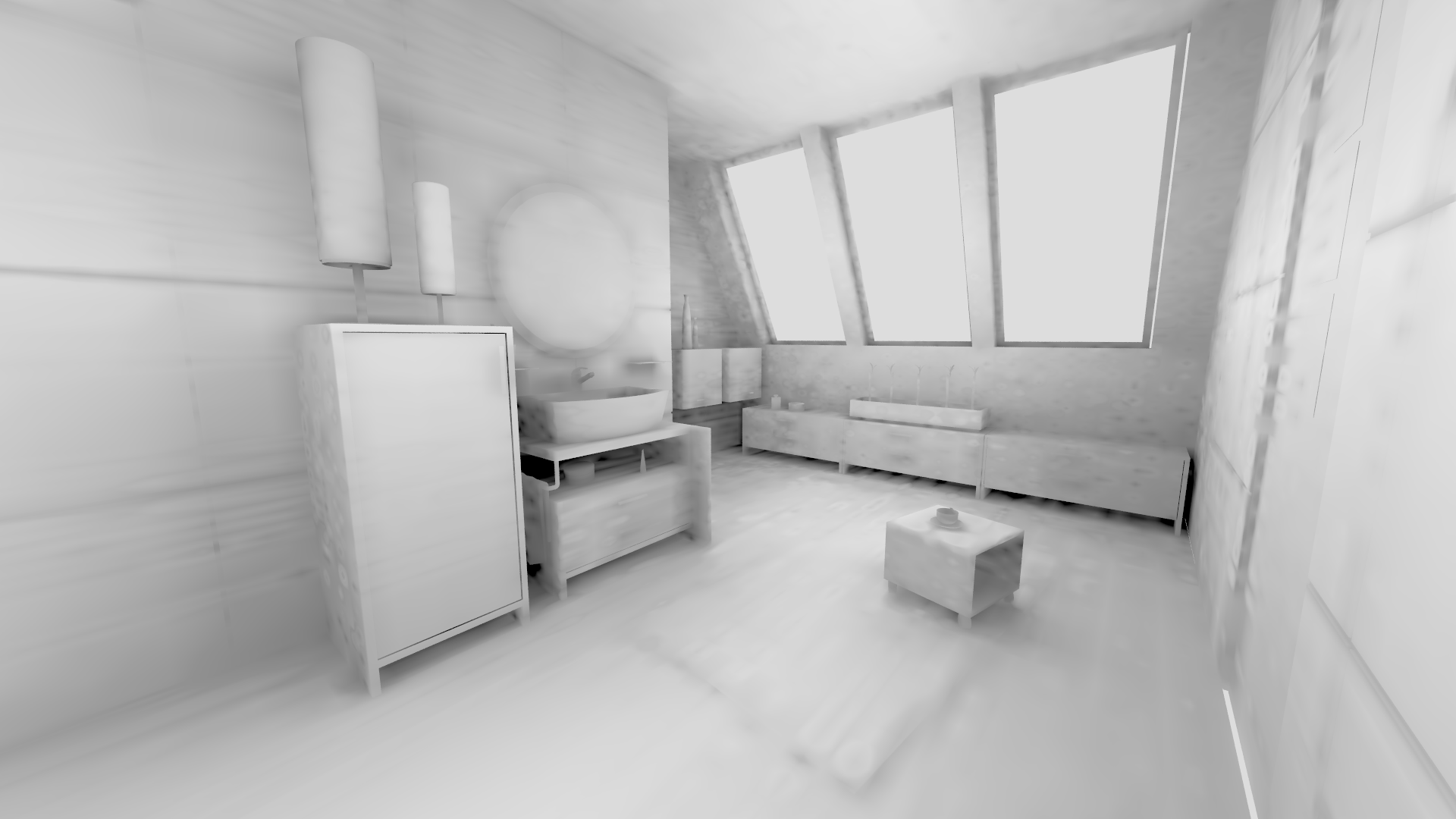} &
\includeimg{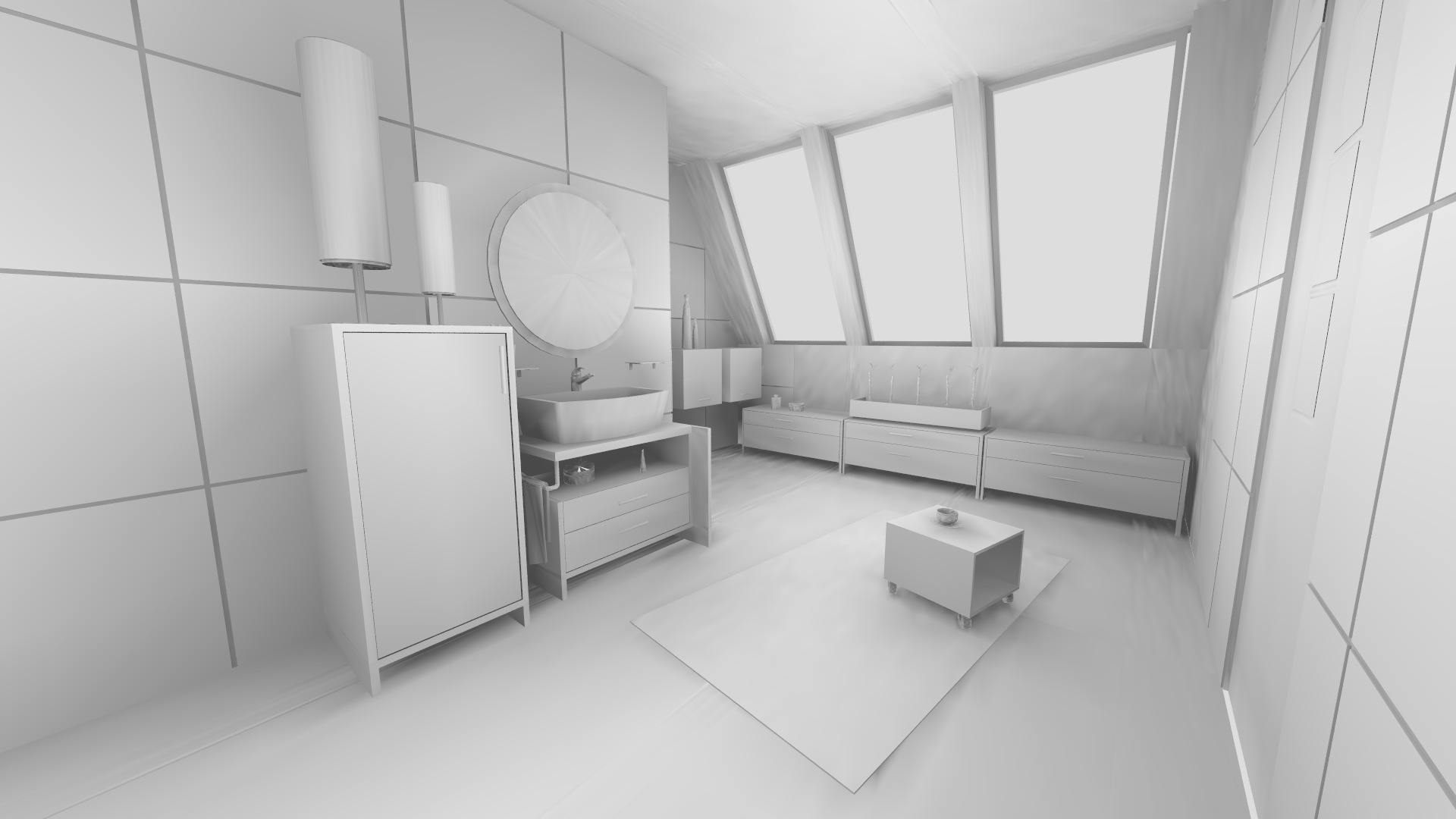} &
\includeimg{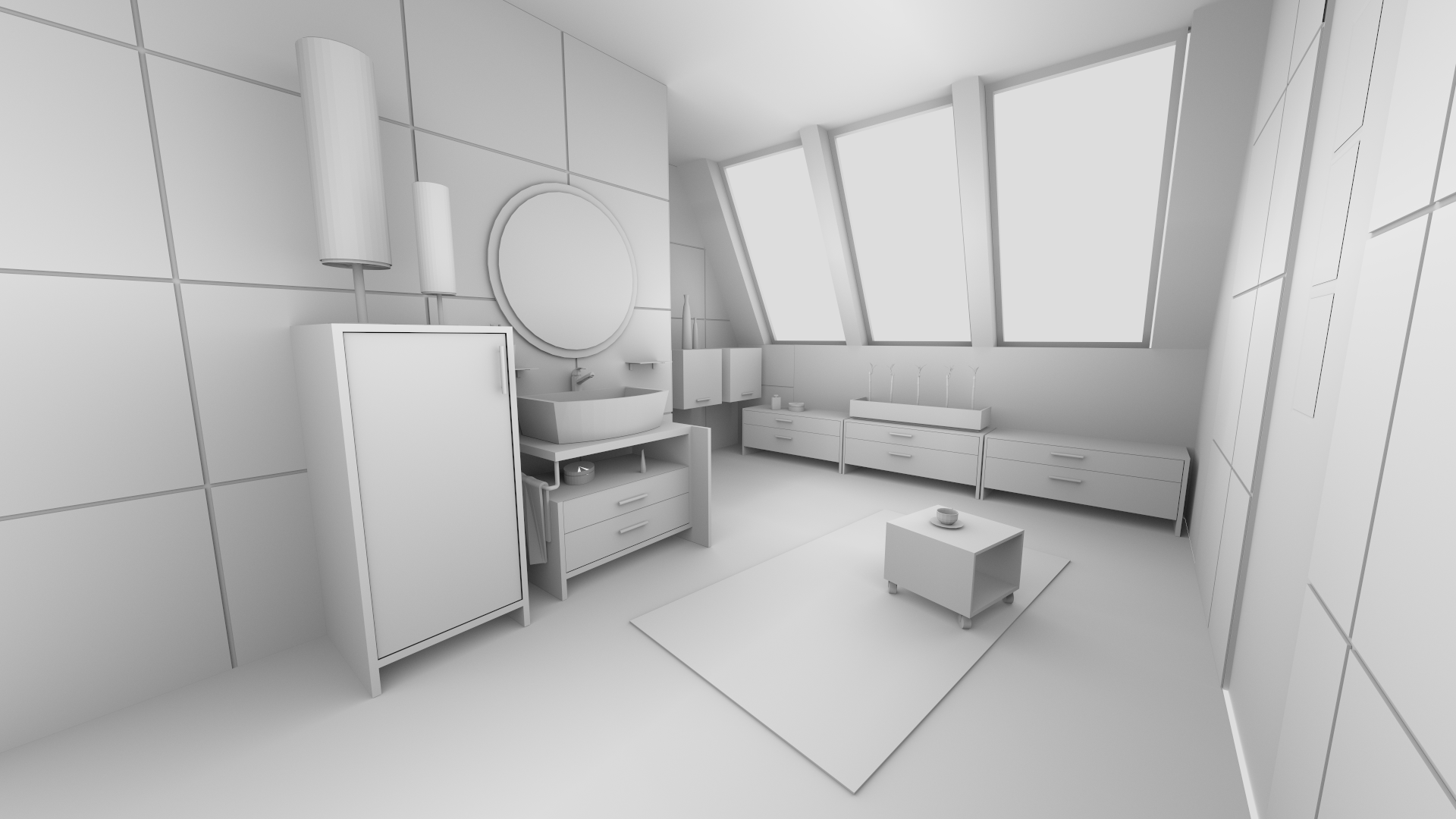}\\

\rowlabel{Kitchen} &
\includeerr{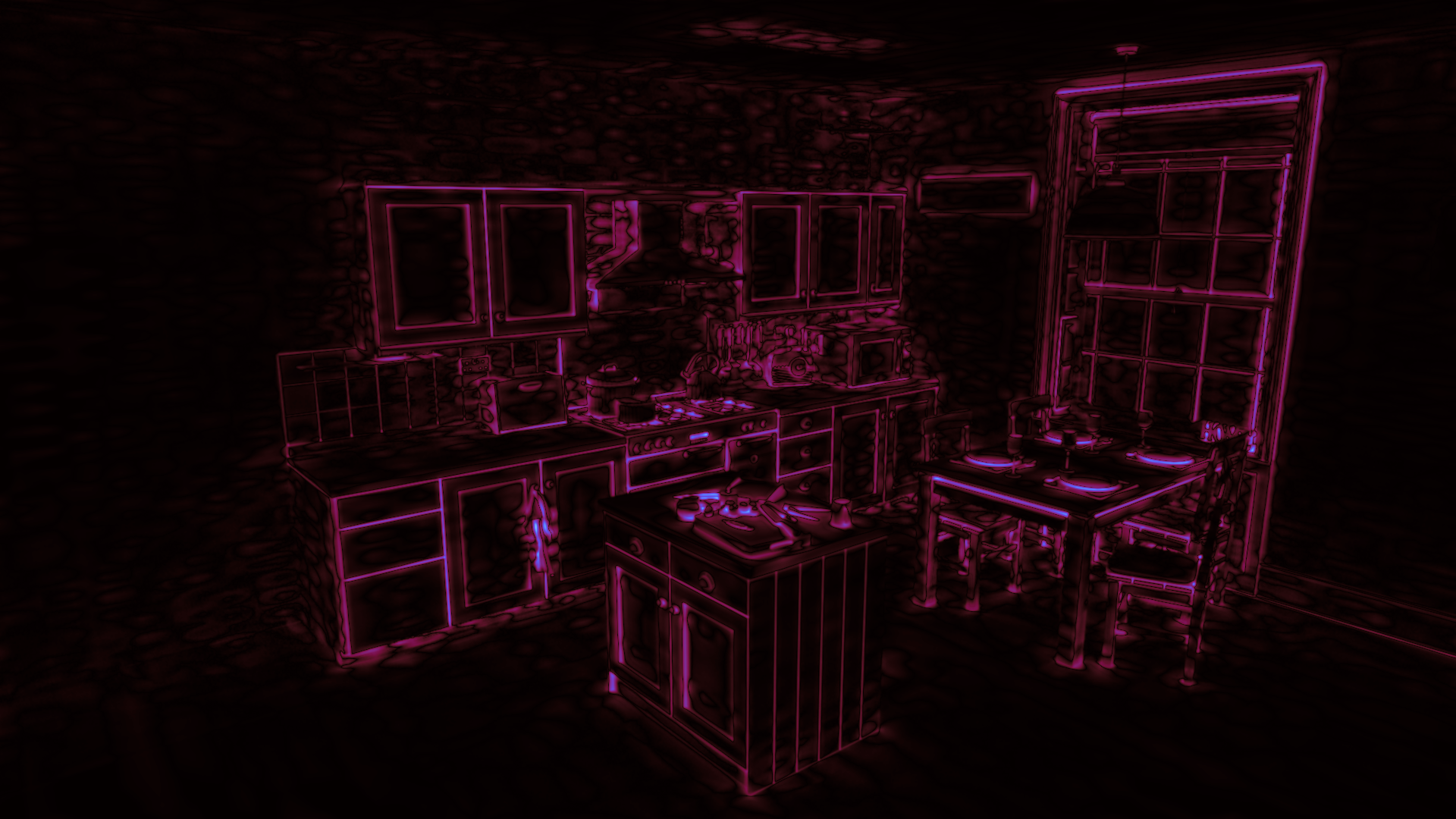}{0.038} &
\includeerr{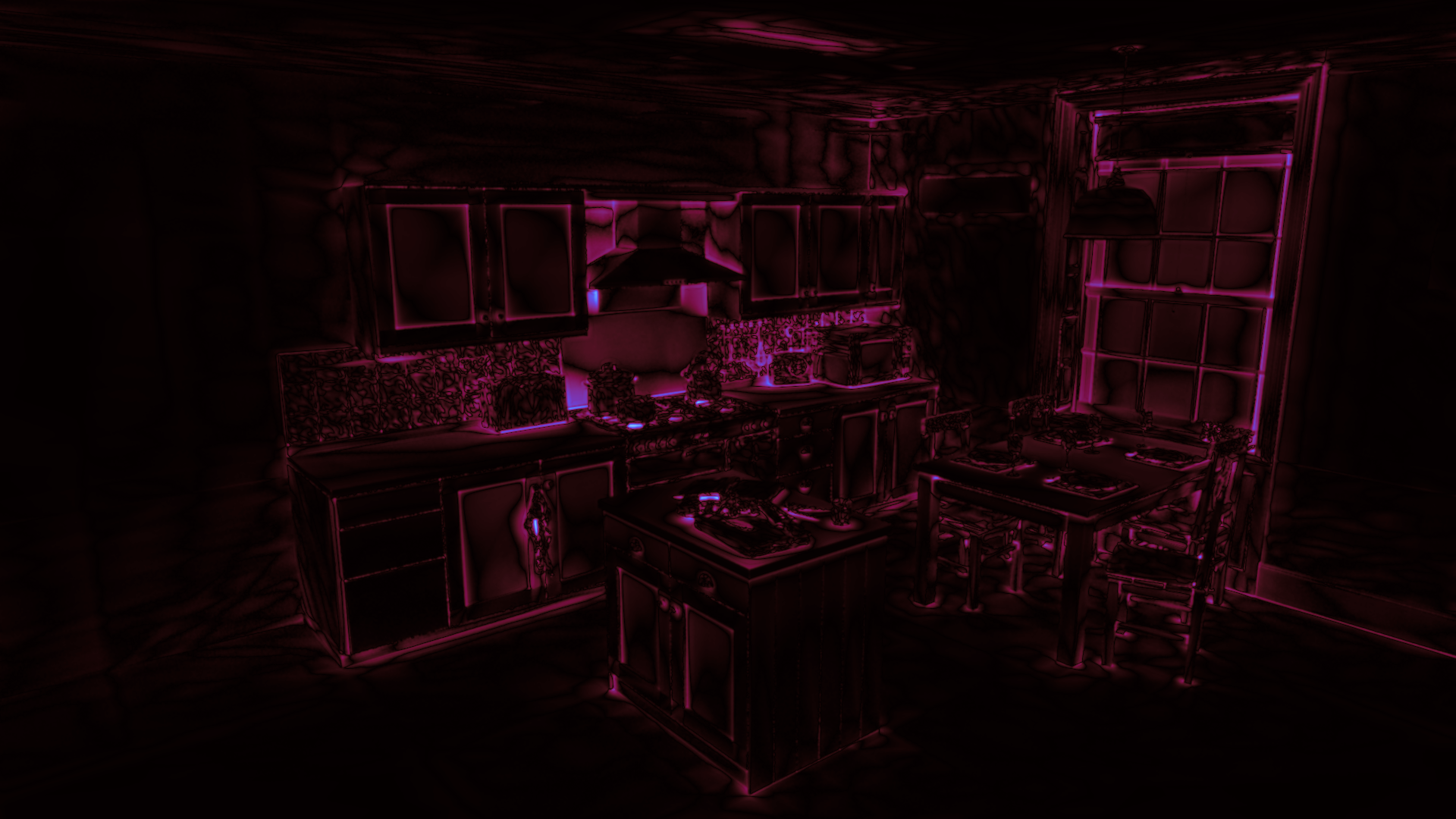}{0.030} &
\includeimg{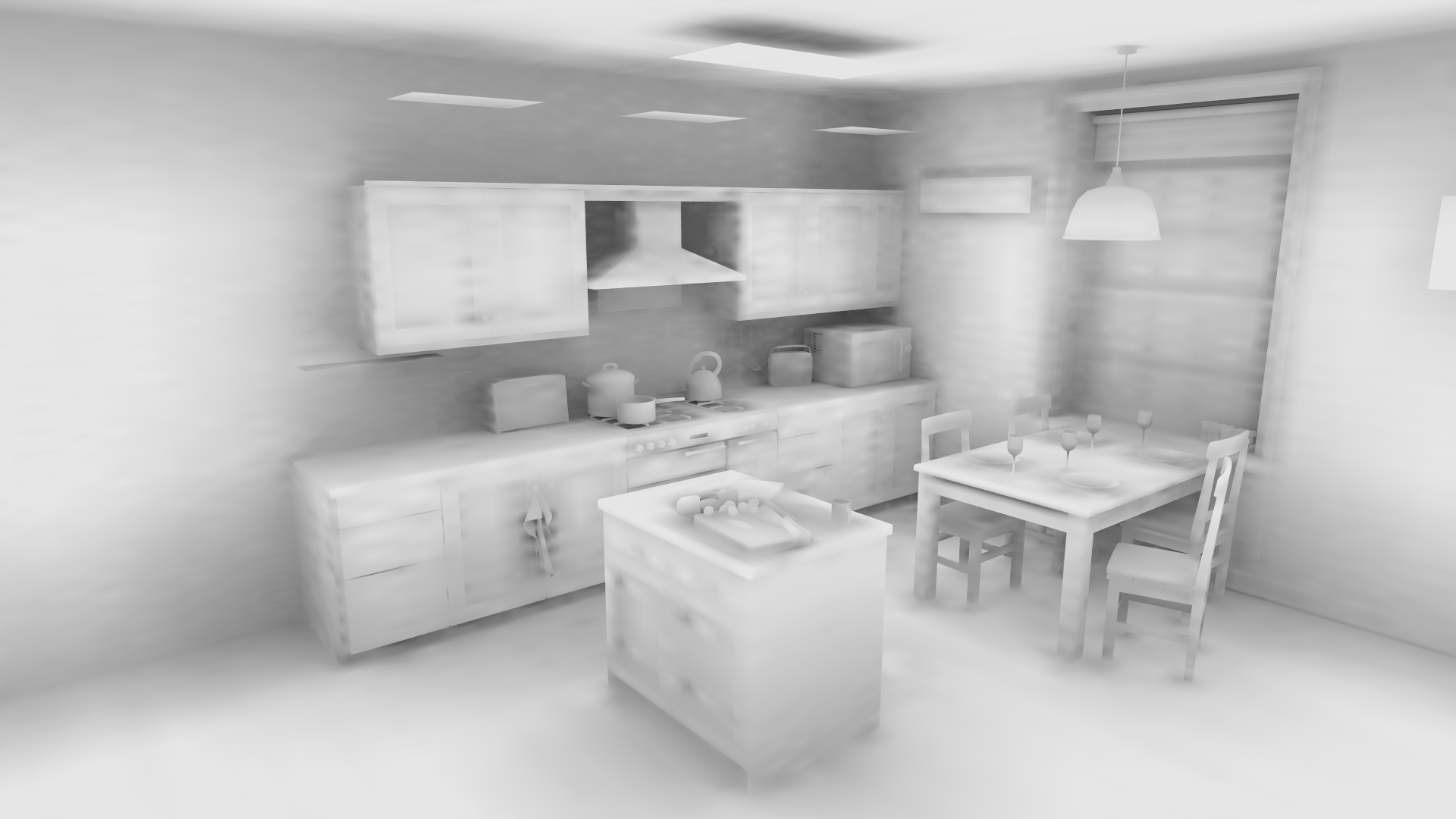} &
\includeimg{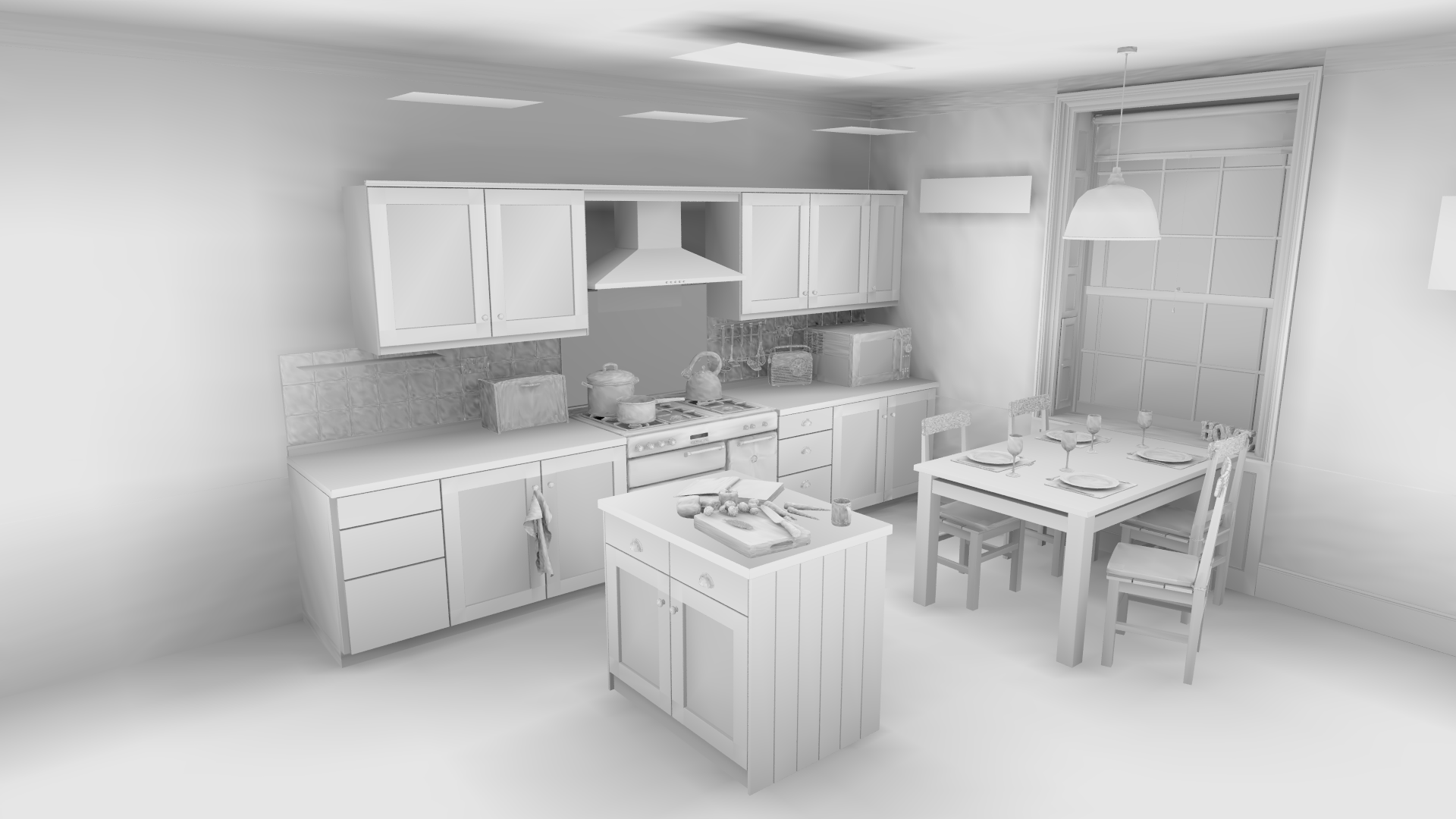} &
\includeimg{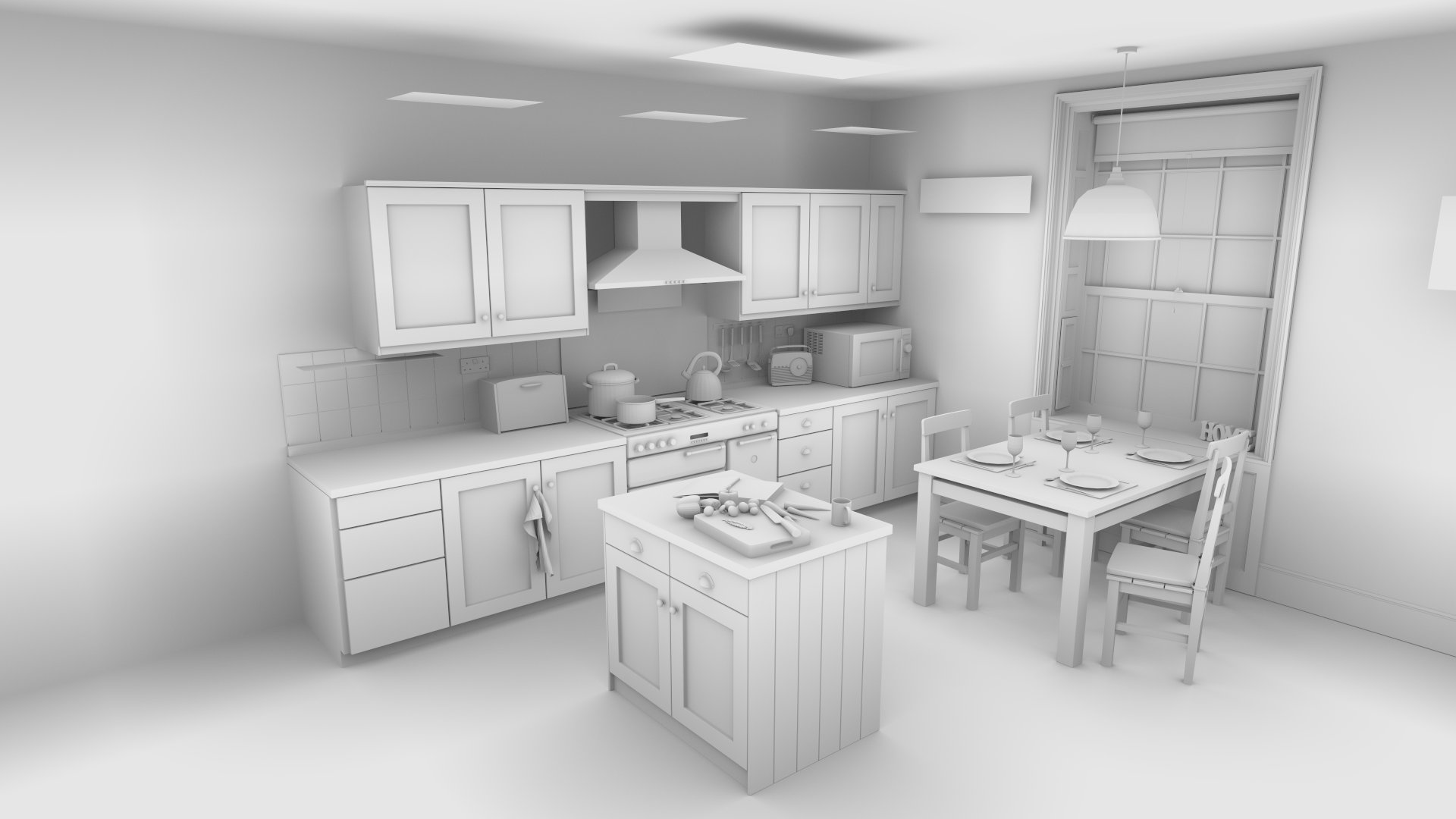}\\

\end{tabular}

\caption{Neural ambient occlusion (NAO). Image quality after 128 training iterations: hash-grid (left), GATE (middle), and ground truth (right). Size of hash-grid $< 1 MB$. GATE provides higher image quality (lower FLIP error) than hash-grid, in particular for the two larger scenes.}
\label{fig:NAO}
\end{figure*}

\begin{figure*}[t]
\centering

\renewcommand{\arraystretch}{1.5}

\begin{tabular}{>{\centering\arraybackslash}m{0.15cm} m{\cellwidth} m{\cellwidth} m{\cellwidth} m{\cellwidth} m{\cellwidth} c}

& \centering Hash-Grid & \centering GATE (ours) & \centering Hash-Grid & \centering GATE (ours) & \centering Ground Truth & \\

\rowlabel{Bistro (Ext)} &
\includeerr{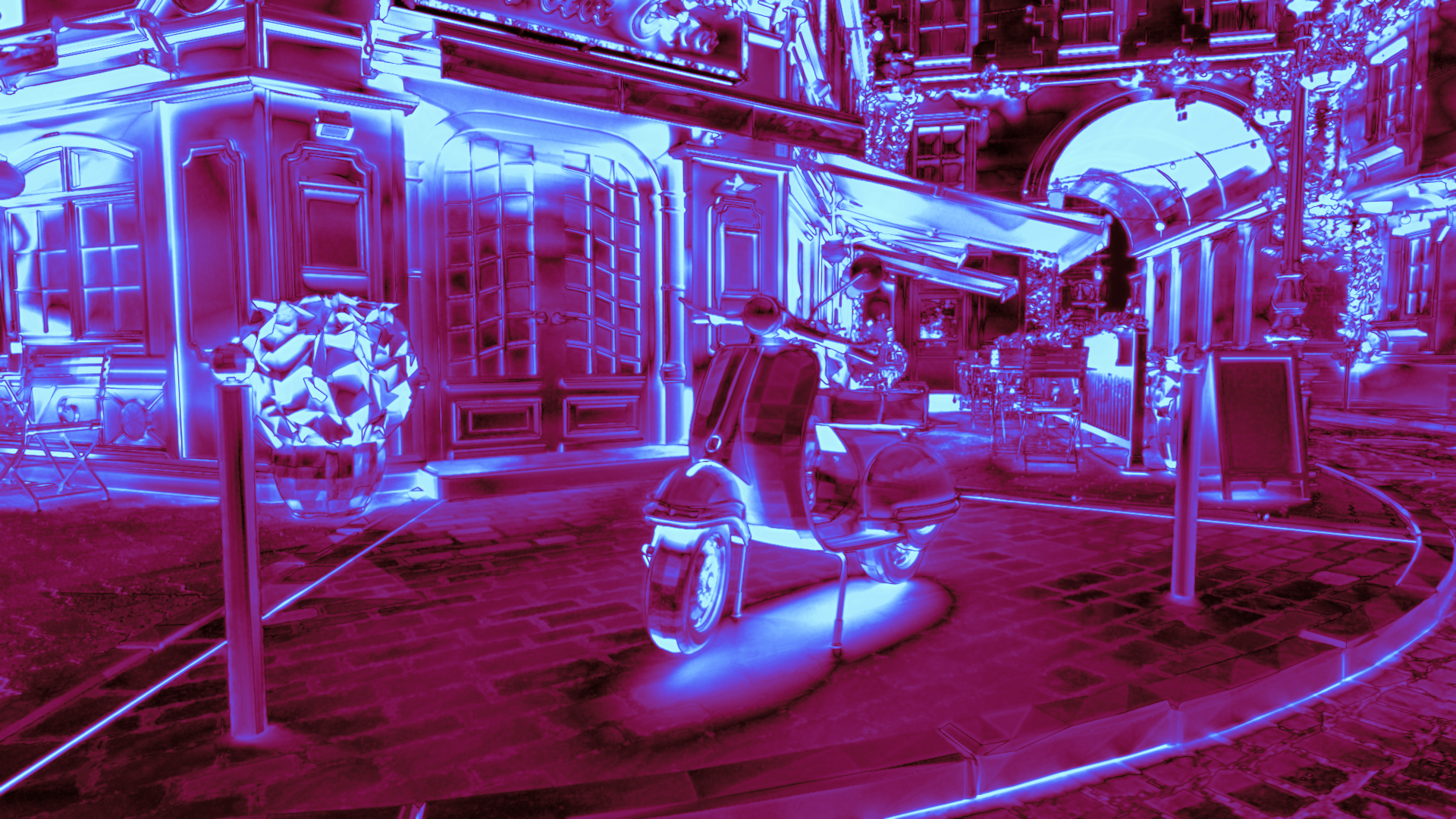}{0.410} &
\includeerr{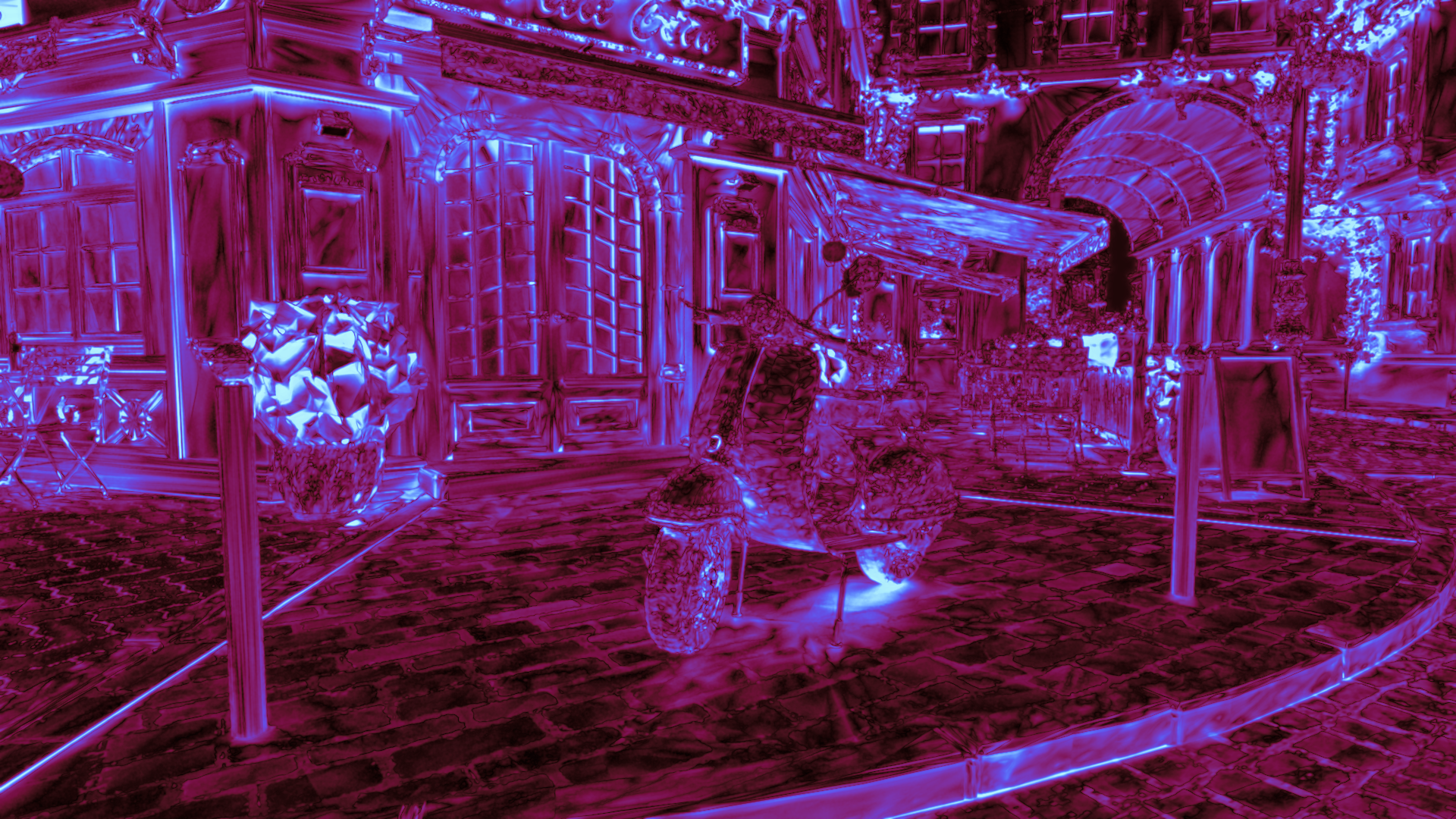}{0.292} &
\includeimg{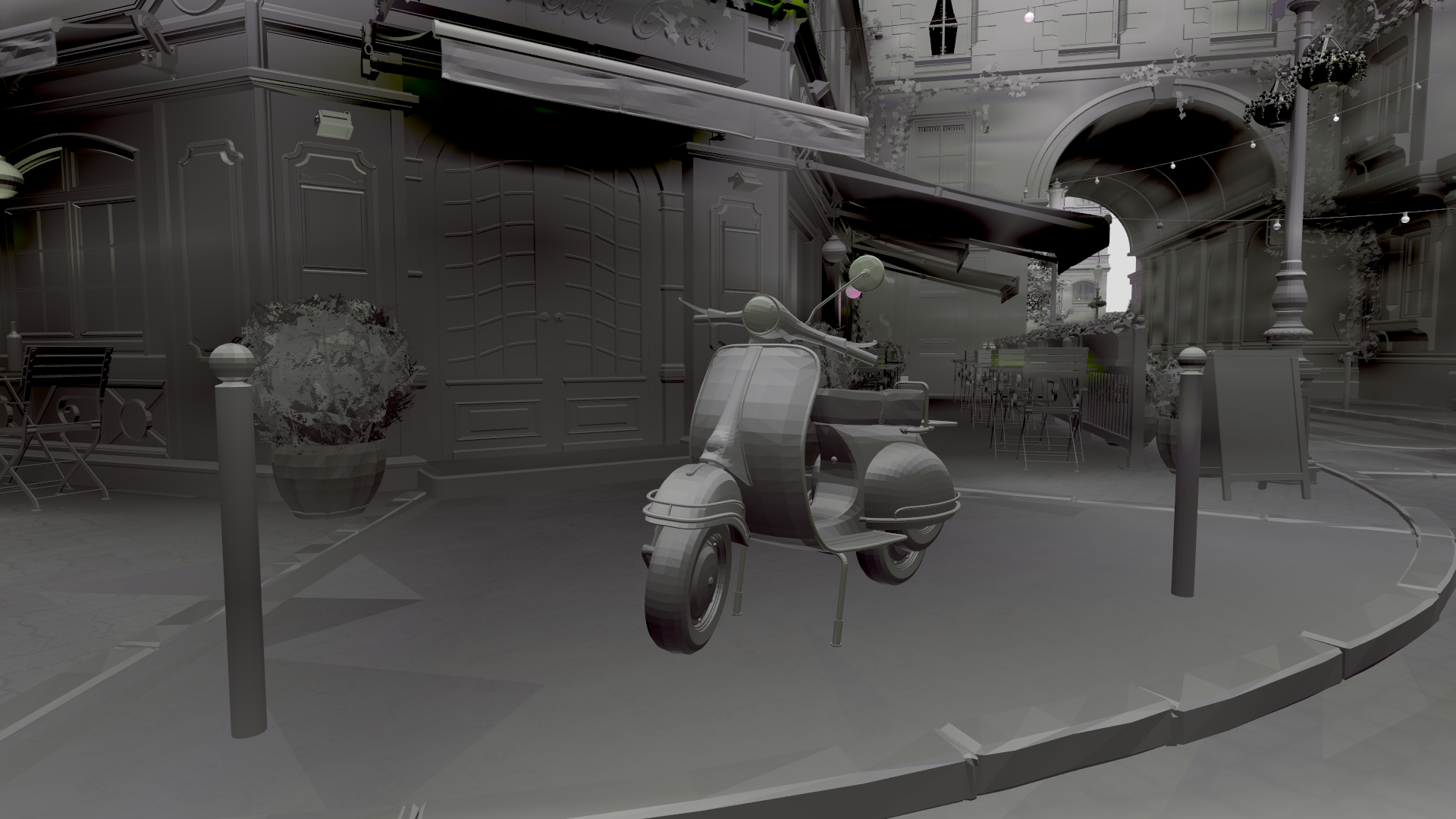} &
\includeimg{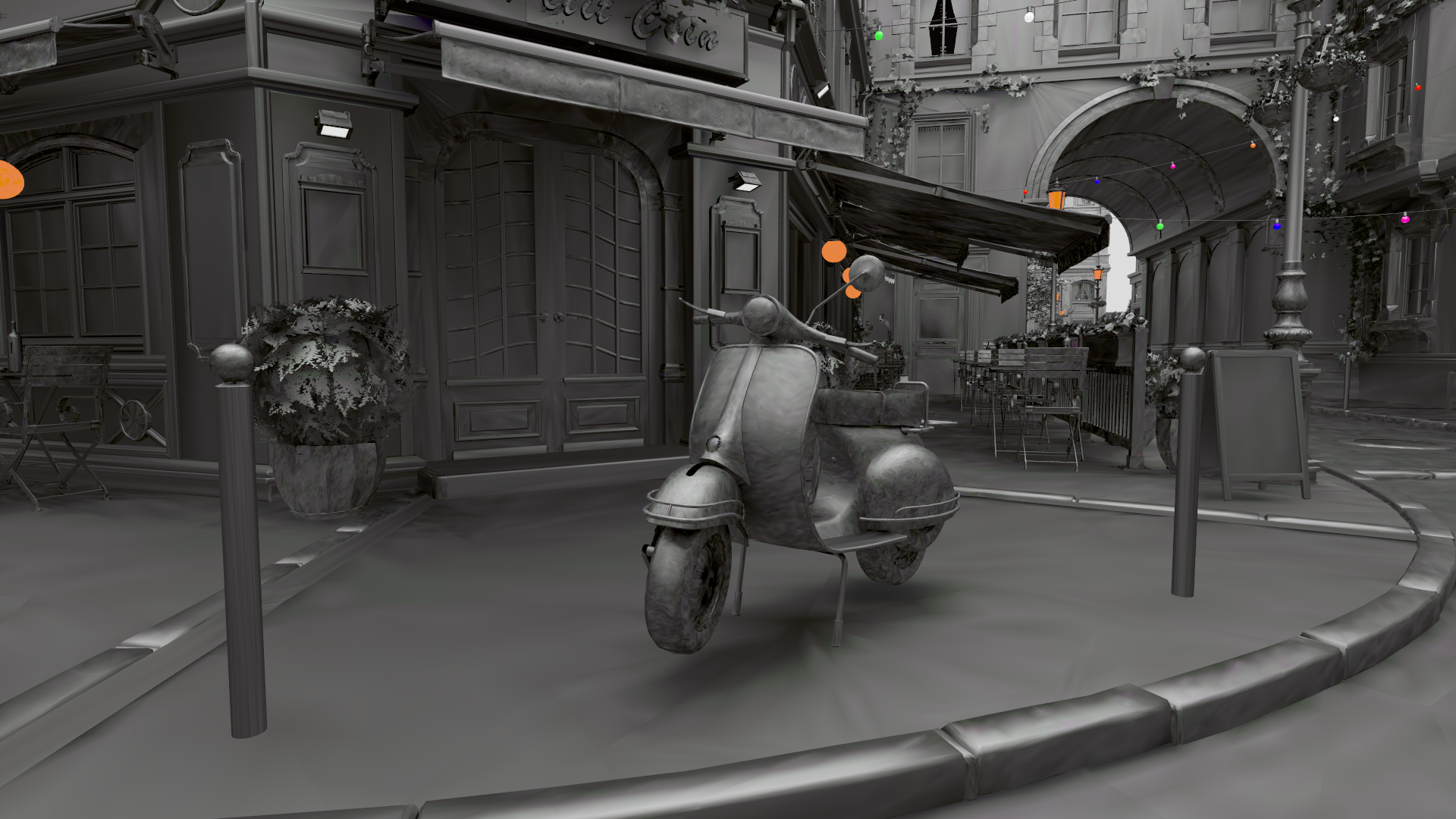} &
\includeimg{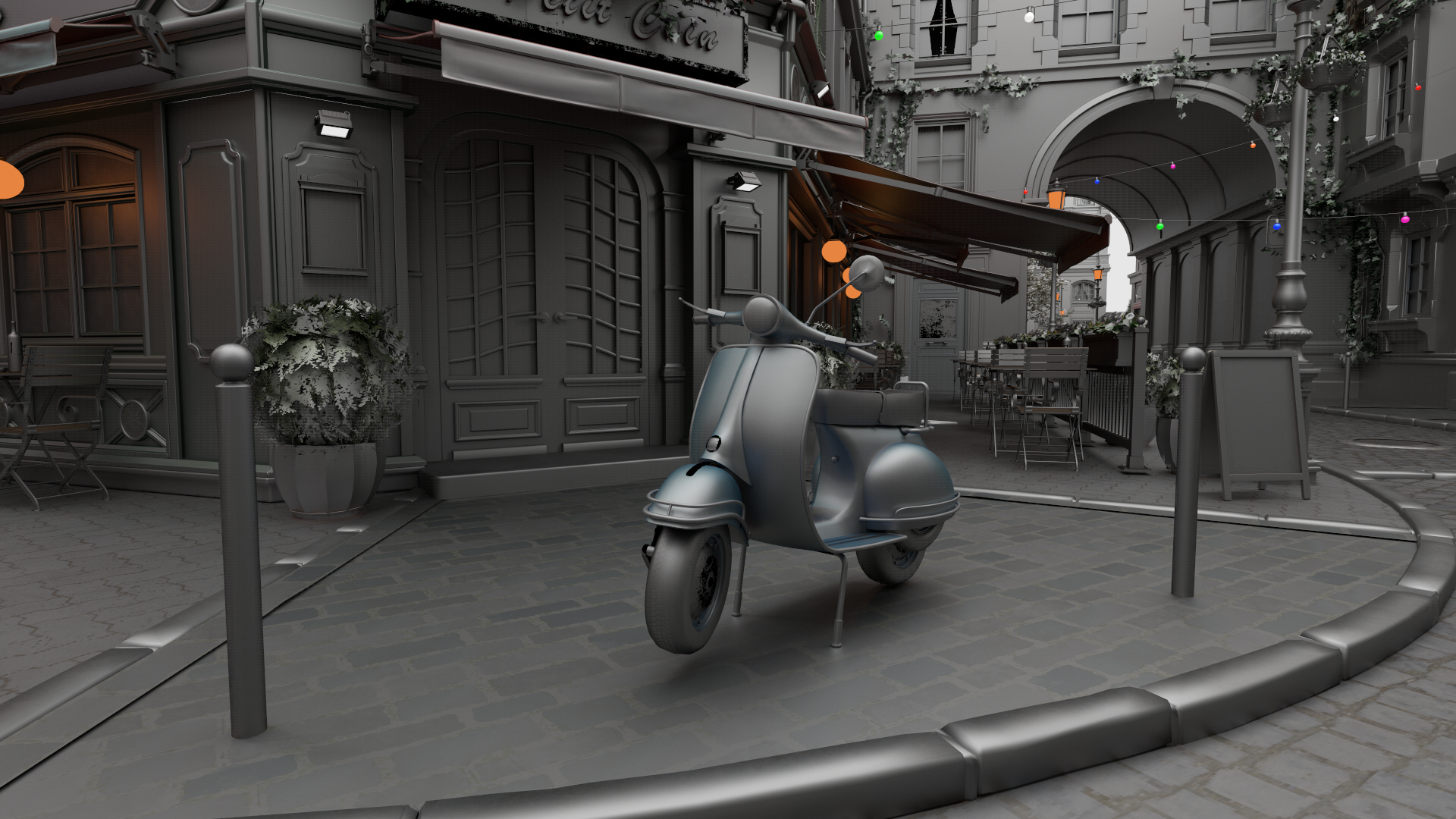}\\

\rowlabel{Bistro (Int)} &
\includeerr{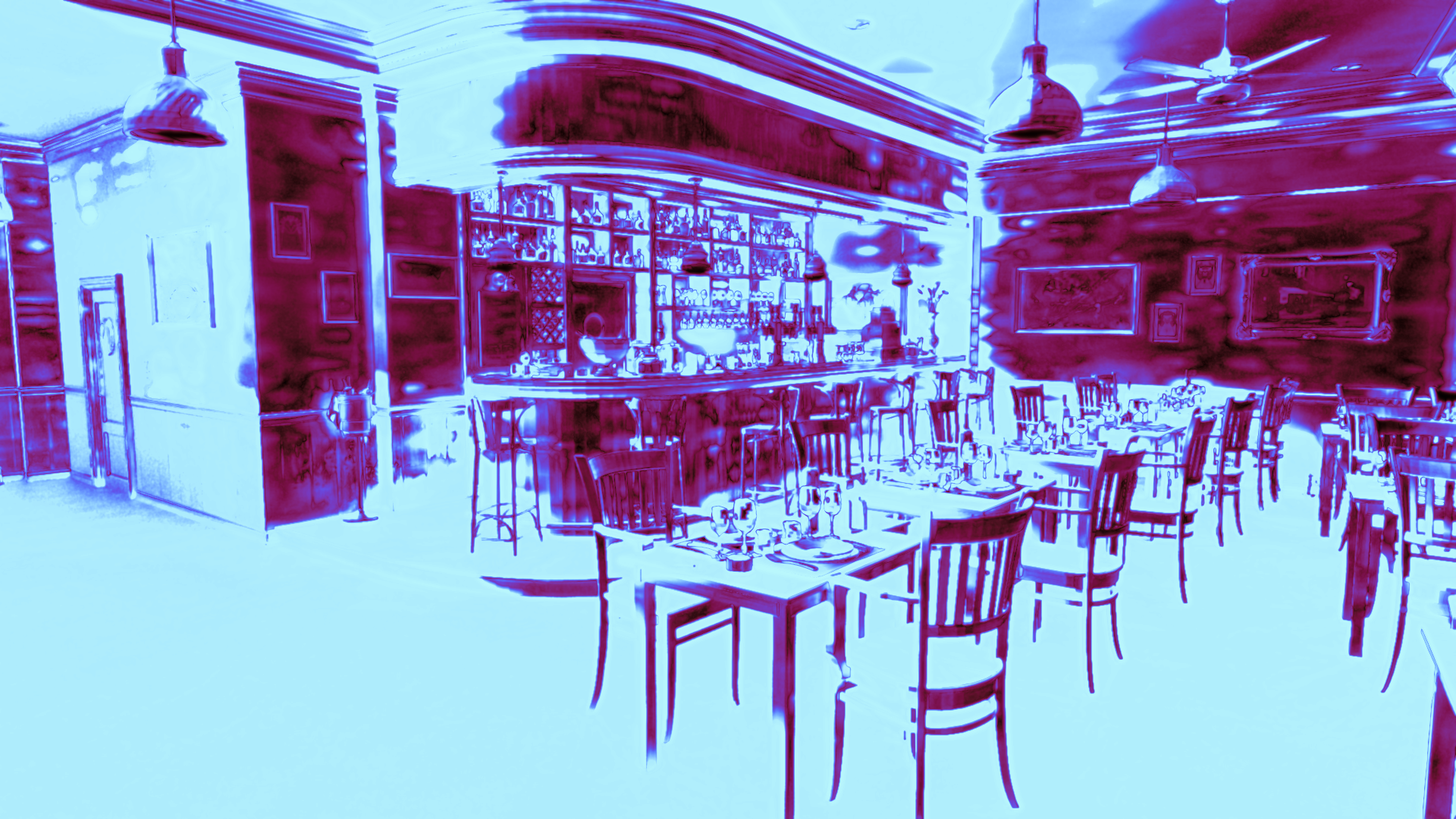}{0.706} &
\includeerr{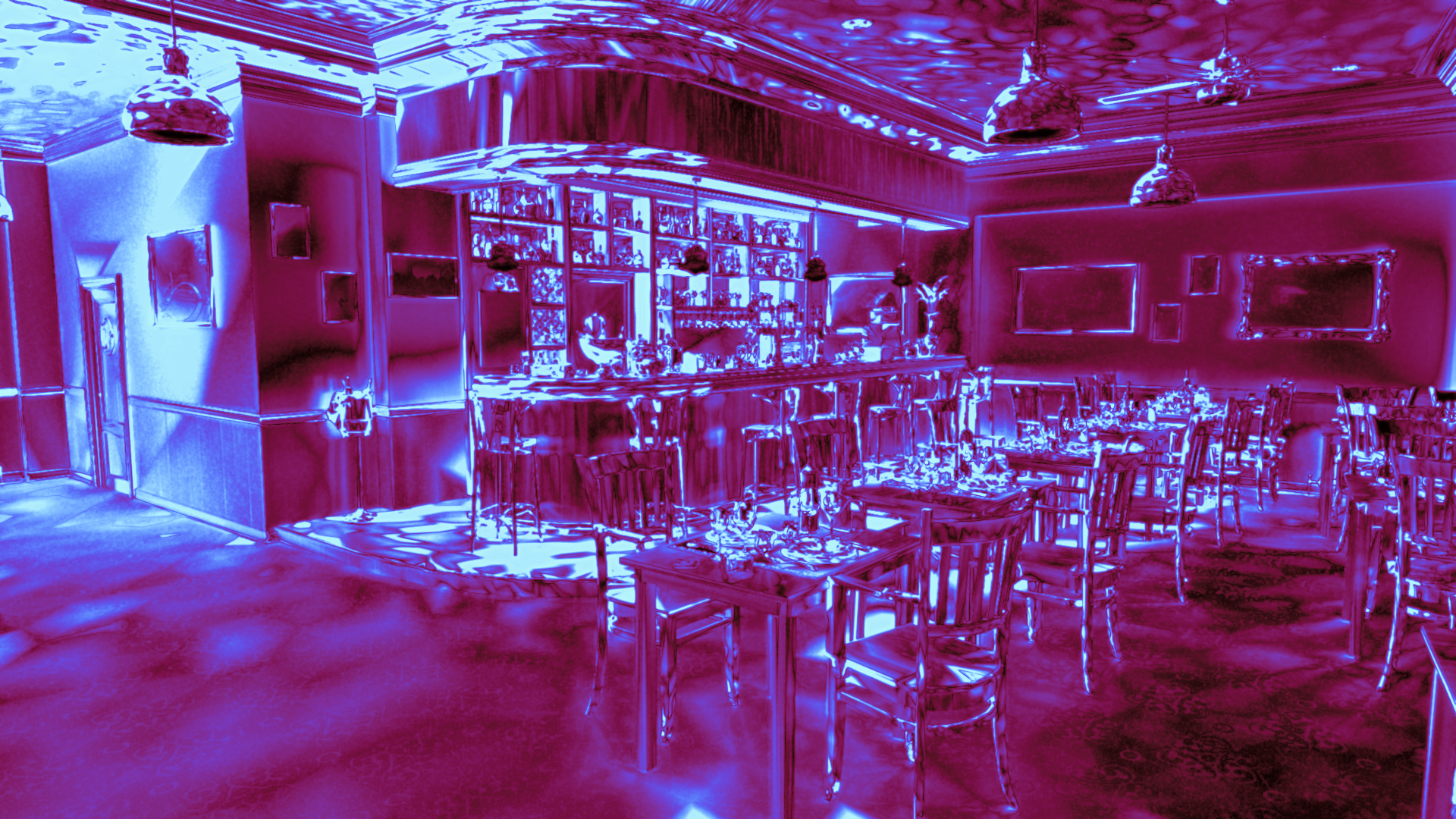}{0.414} &
\includeimg{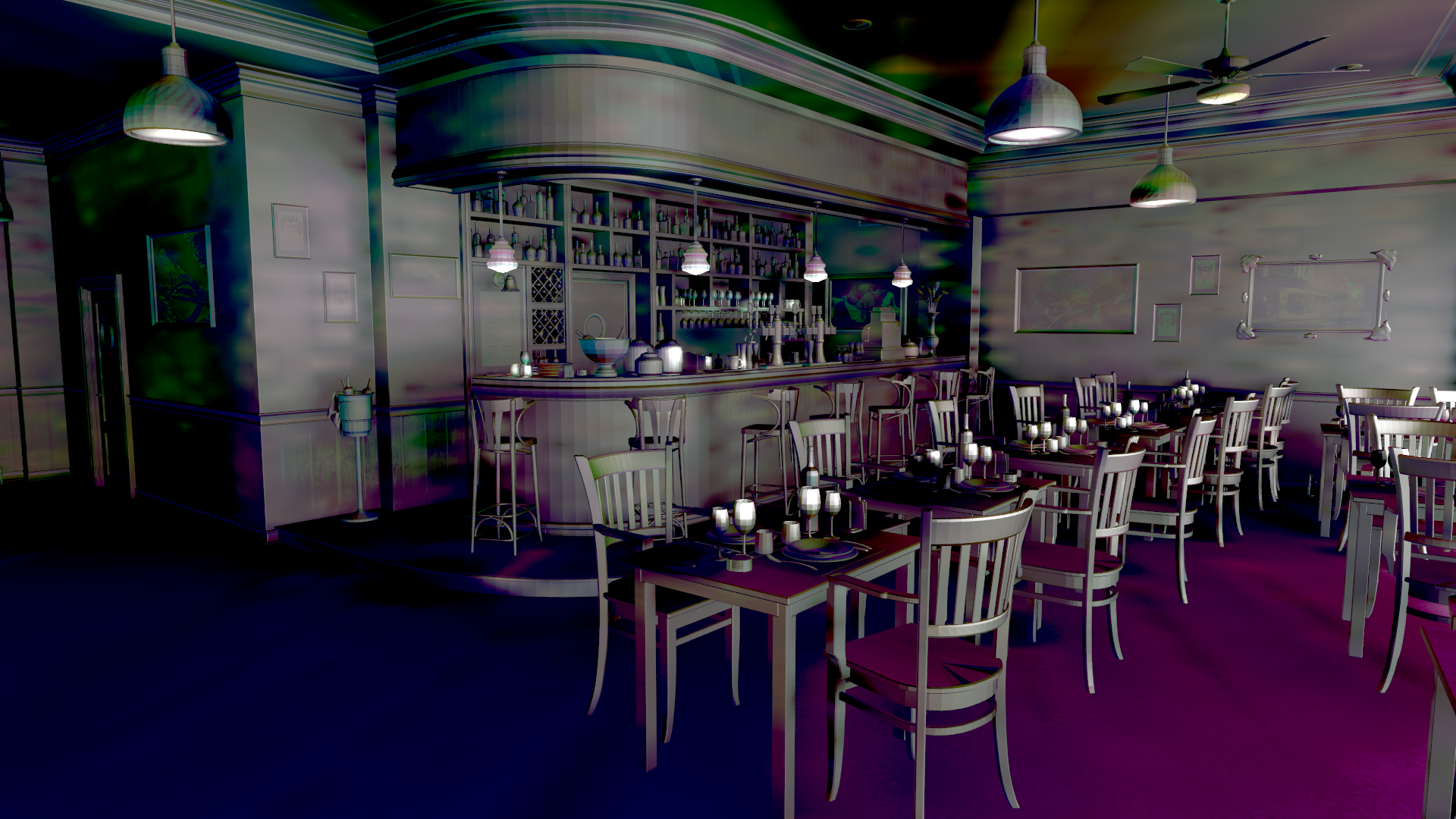} &
\includeimg{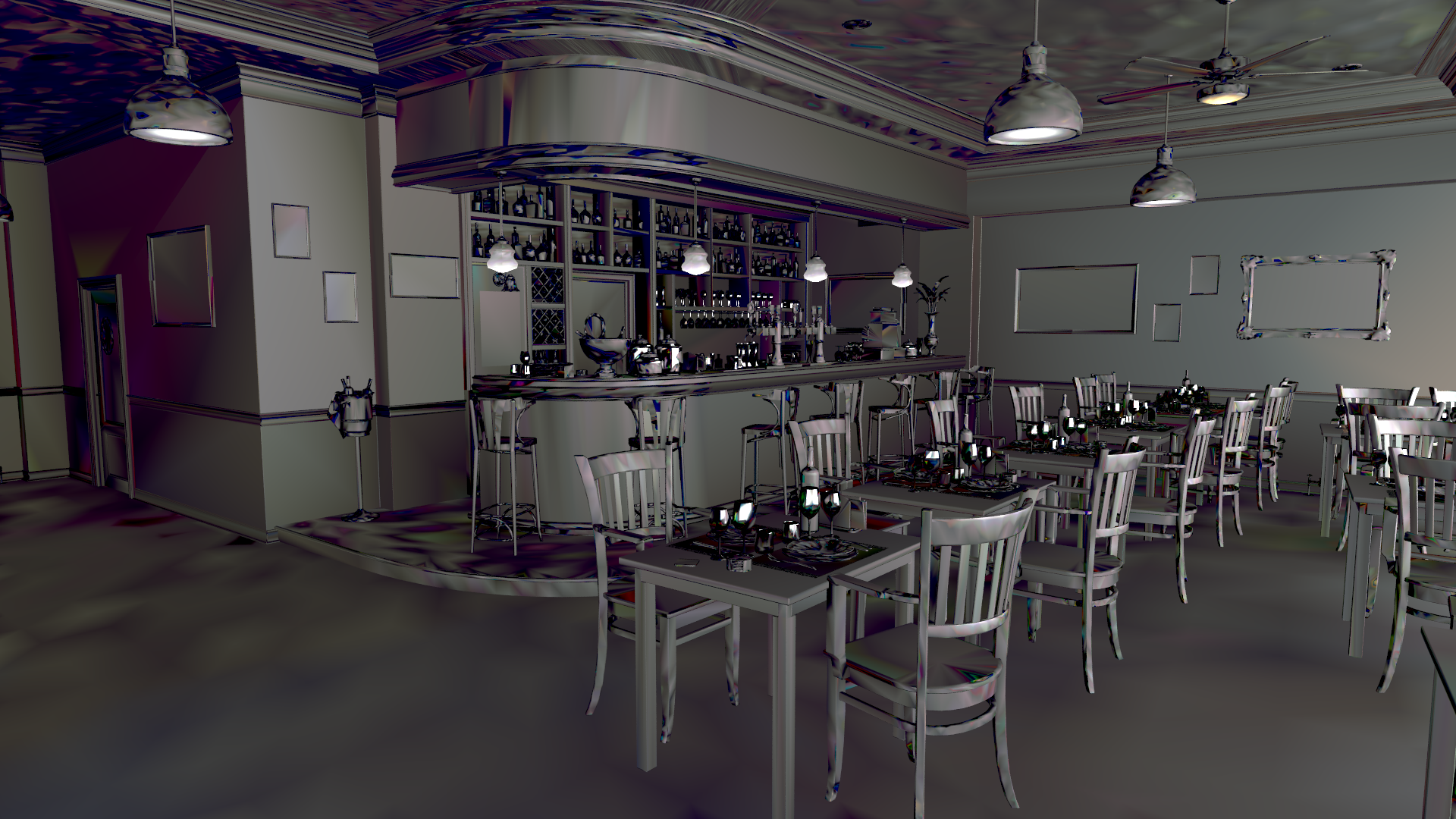} &
\includeimg{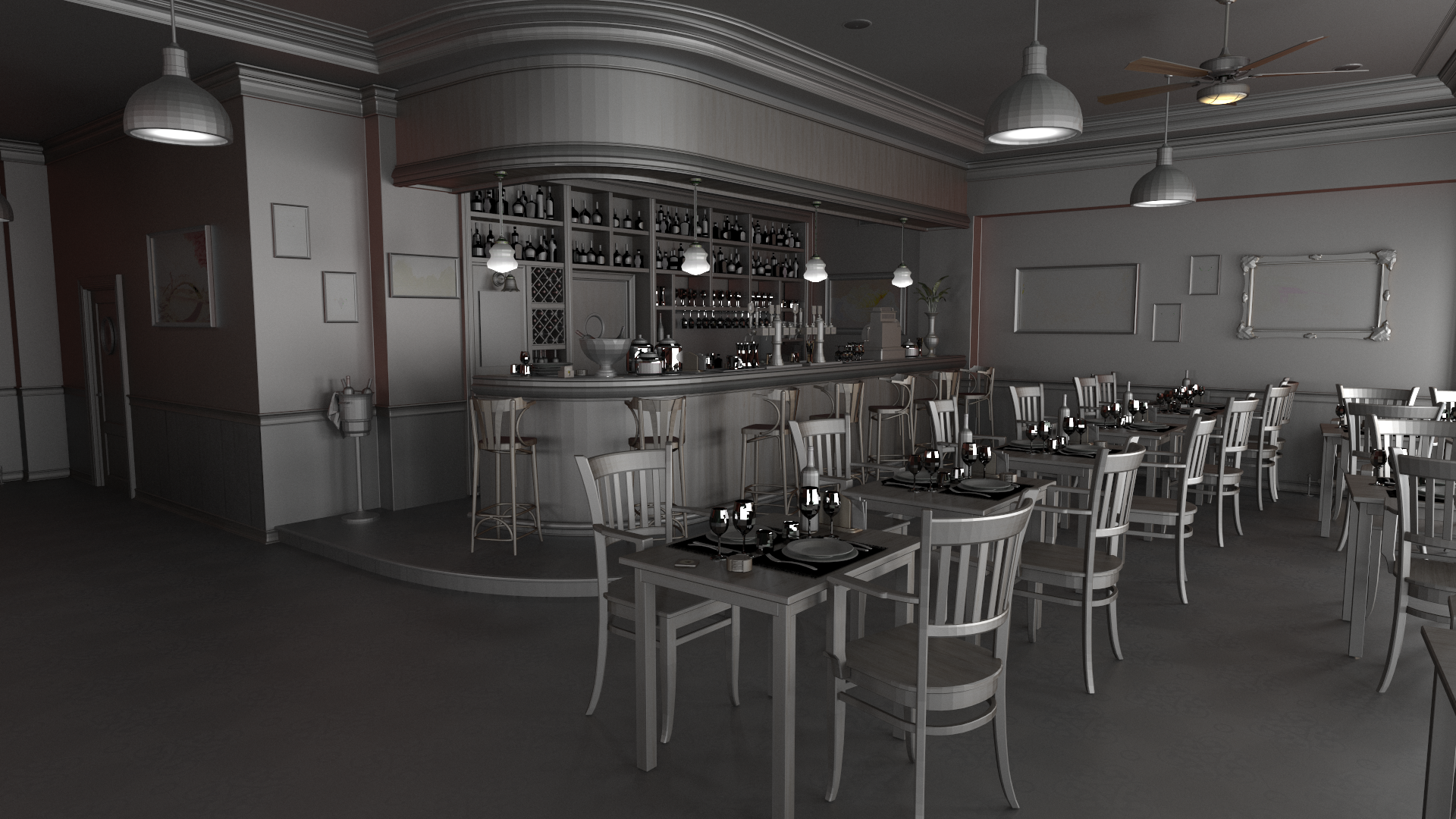}\\

\rowlabel{Bathroom} &
\includeerr{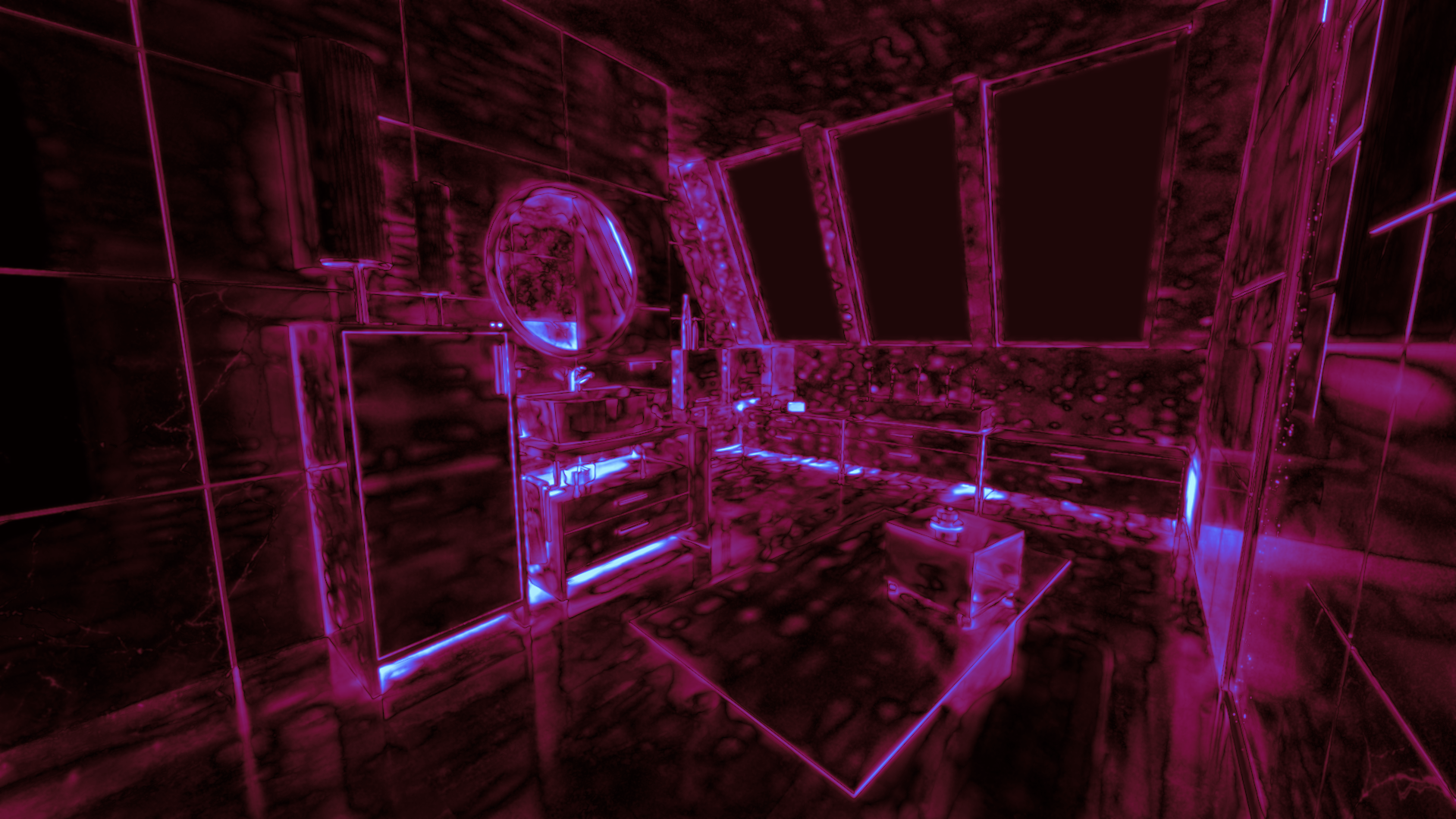}{0.149} &
\includeerr{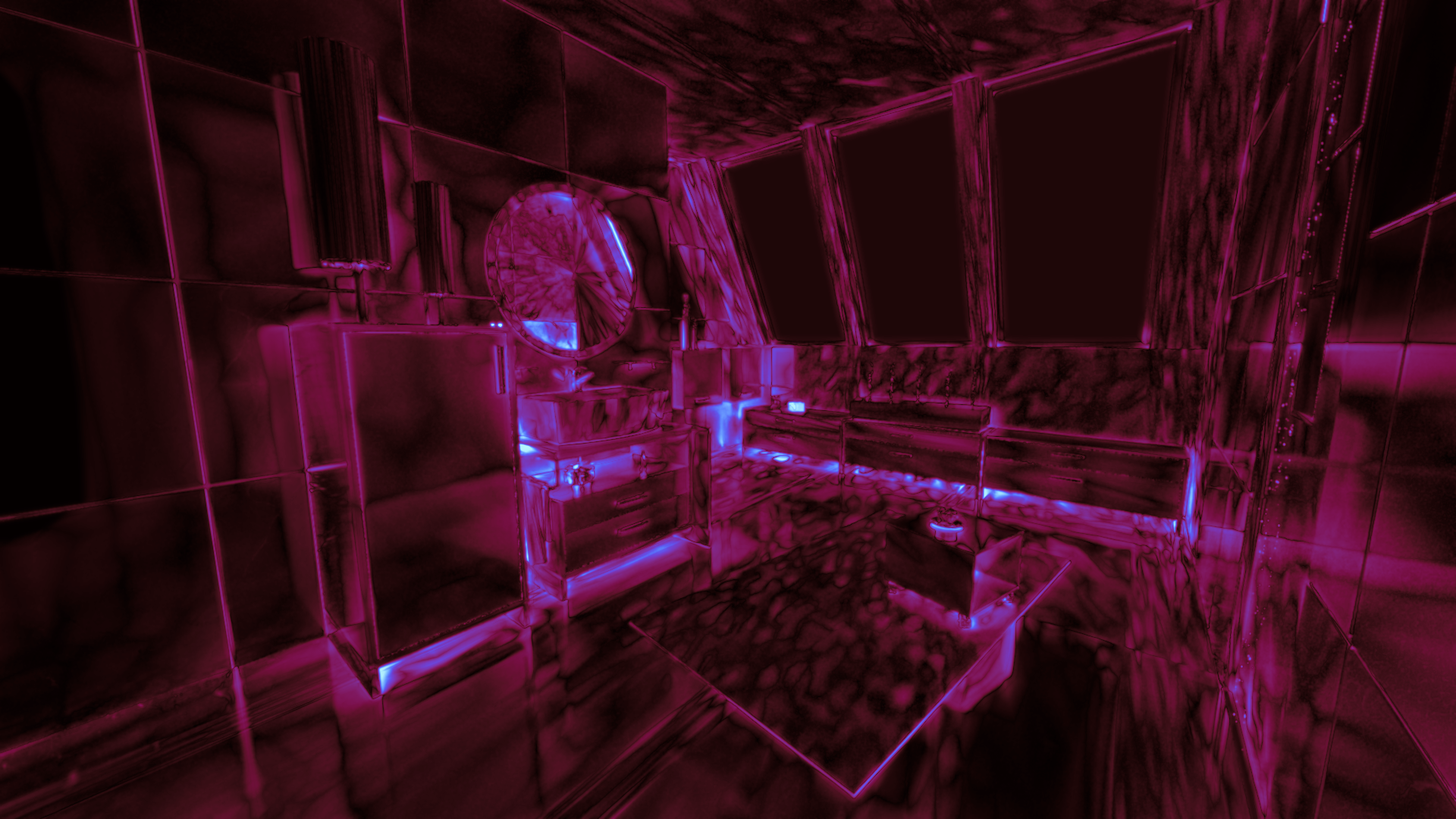}{0.130} &
\includeimg{4_bathroom-NRC-HashGrid-AdaptiveR-F-8-frames-1024.png} &
\includeimg{4_bathroom-NRC-GATE-AdaptiveR-F-8-frames-1024.png} &
\includeimg{4_bathroom-NRC-AdaptiveR-F-8-GROUND_TRUTH-frames-16384.png}\\

\rowlabel{Kitchen} &
\includeerr{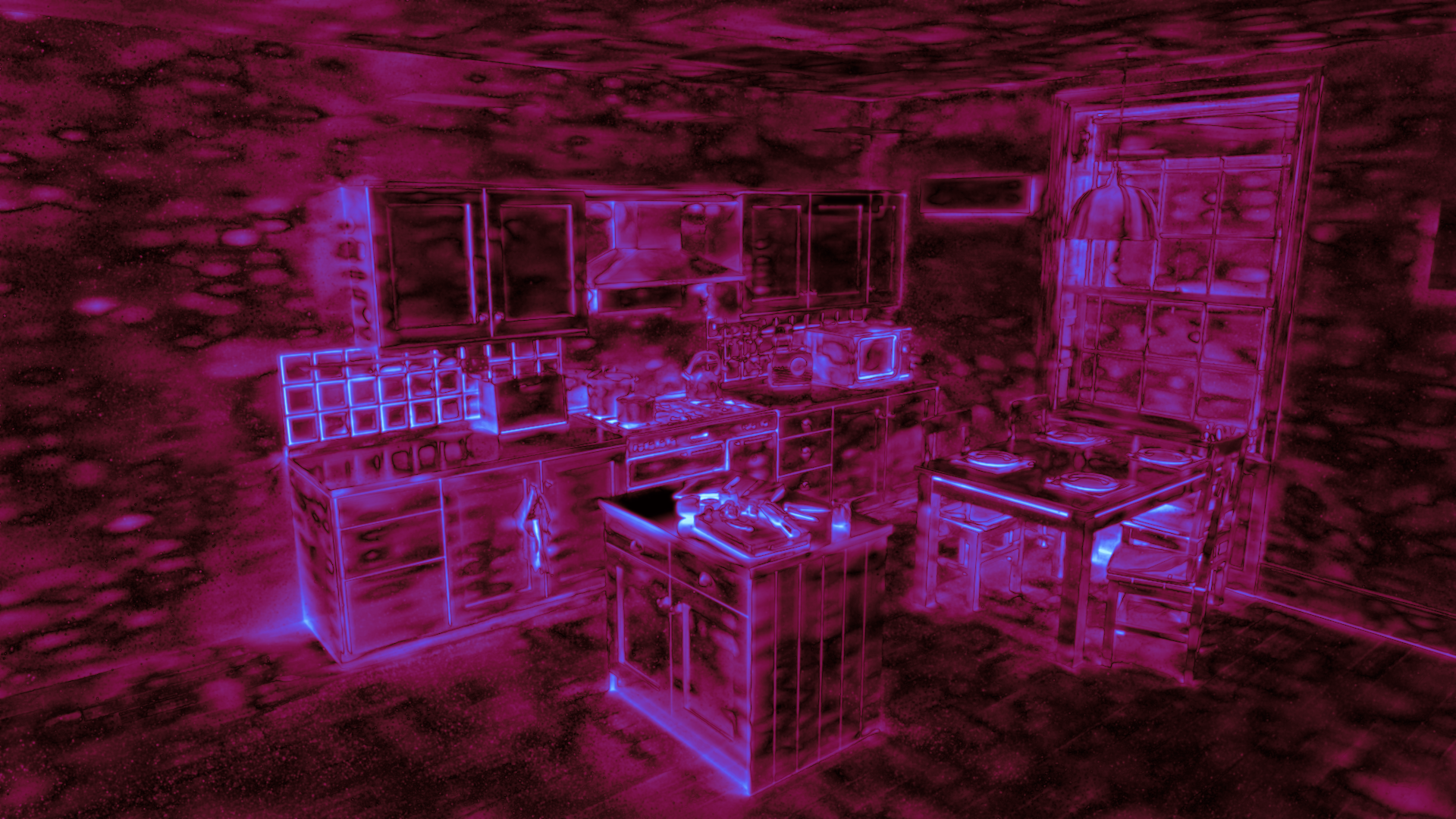}{0.173} &
\includeerr{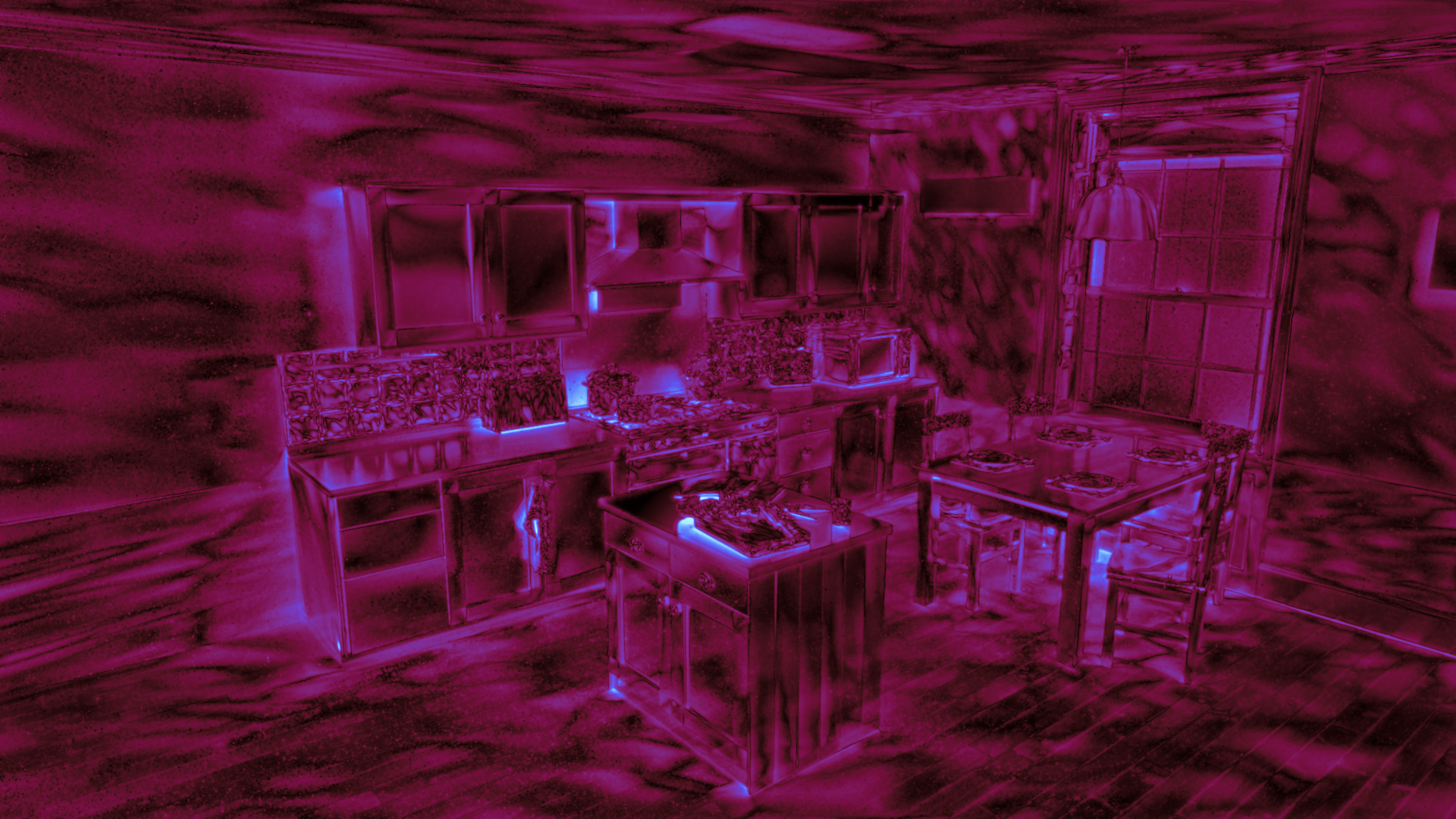}{0.175} &
\includeimg{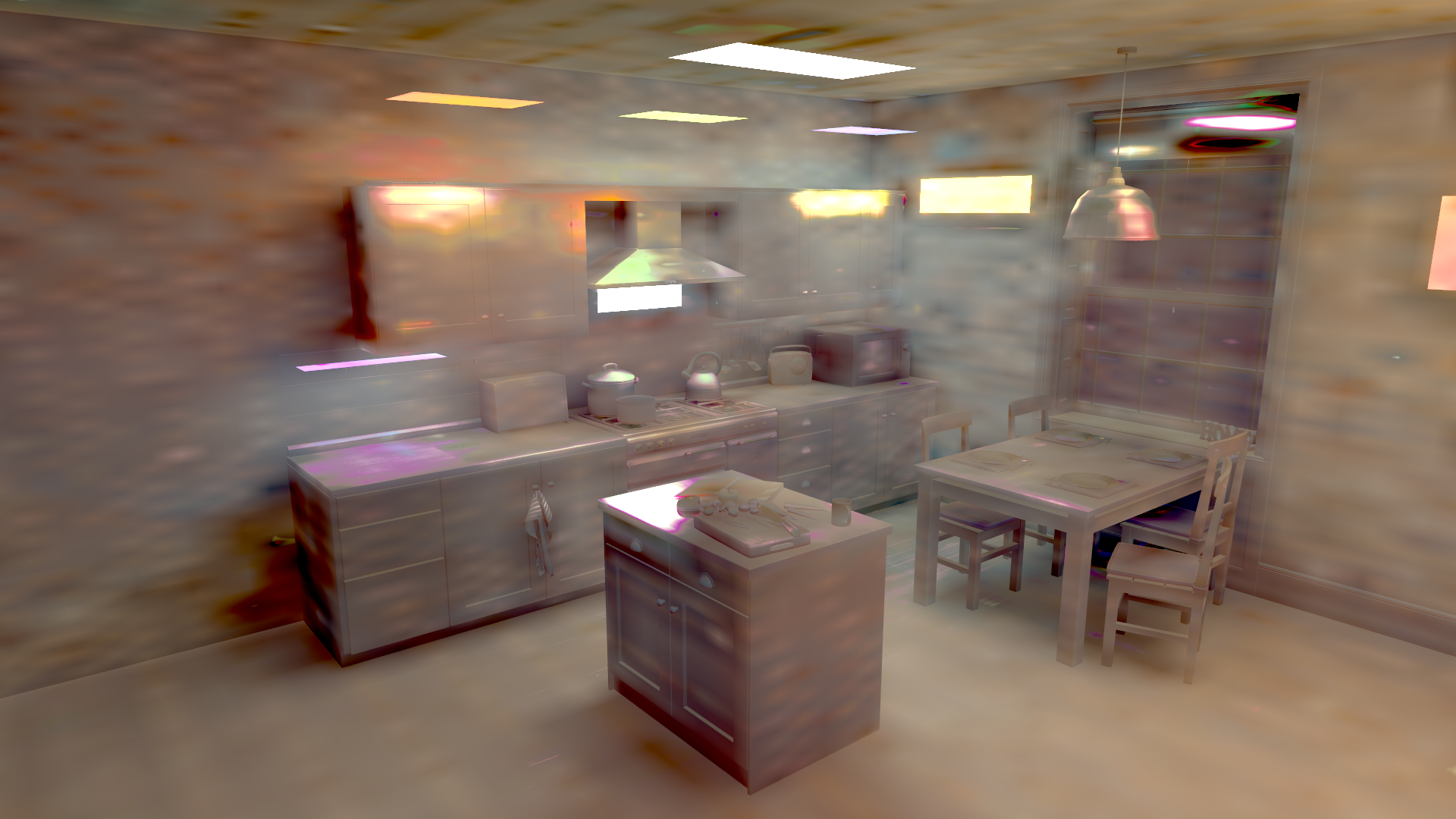} &
\includeimg{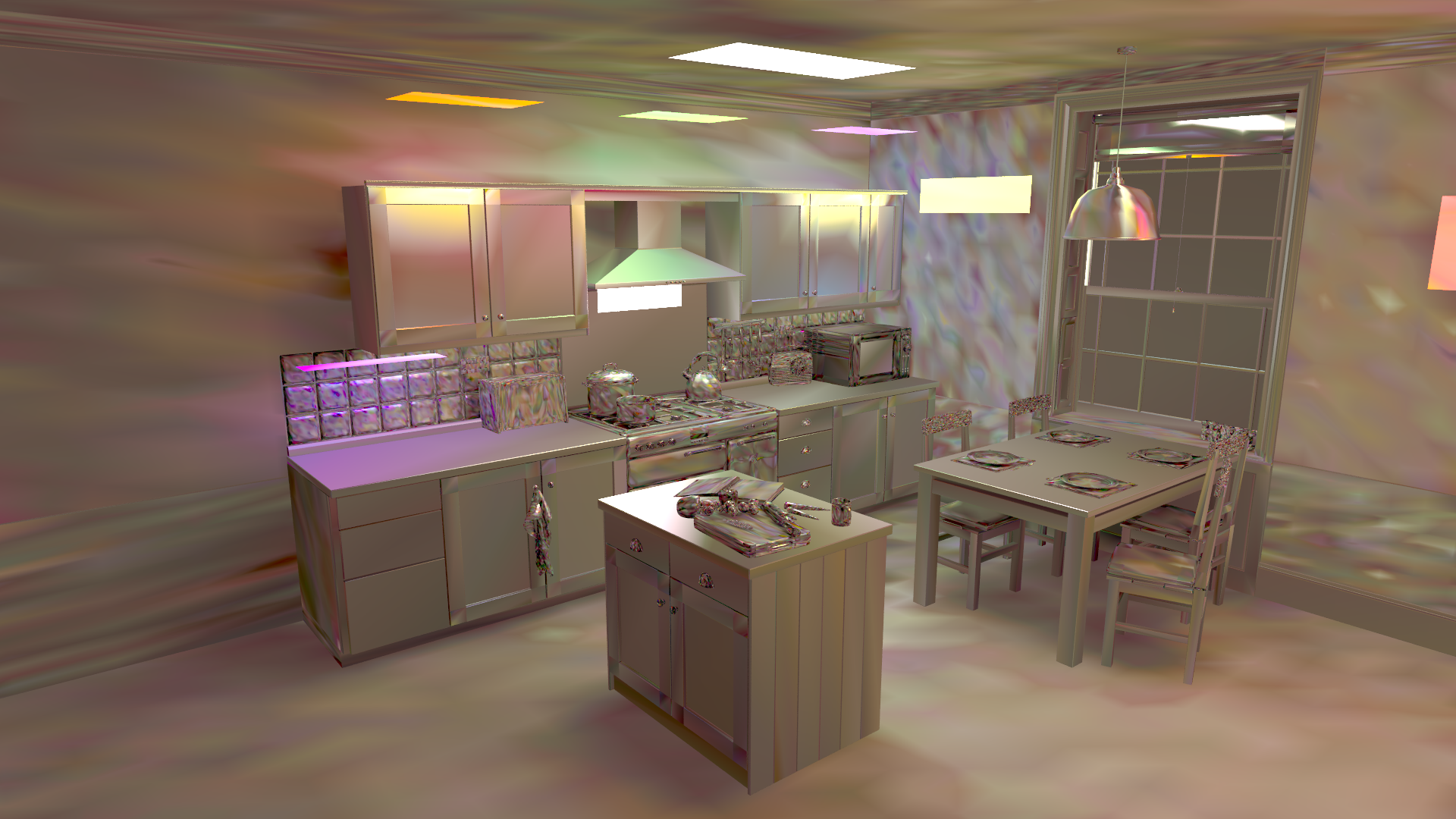} &
\includeimg{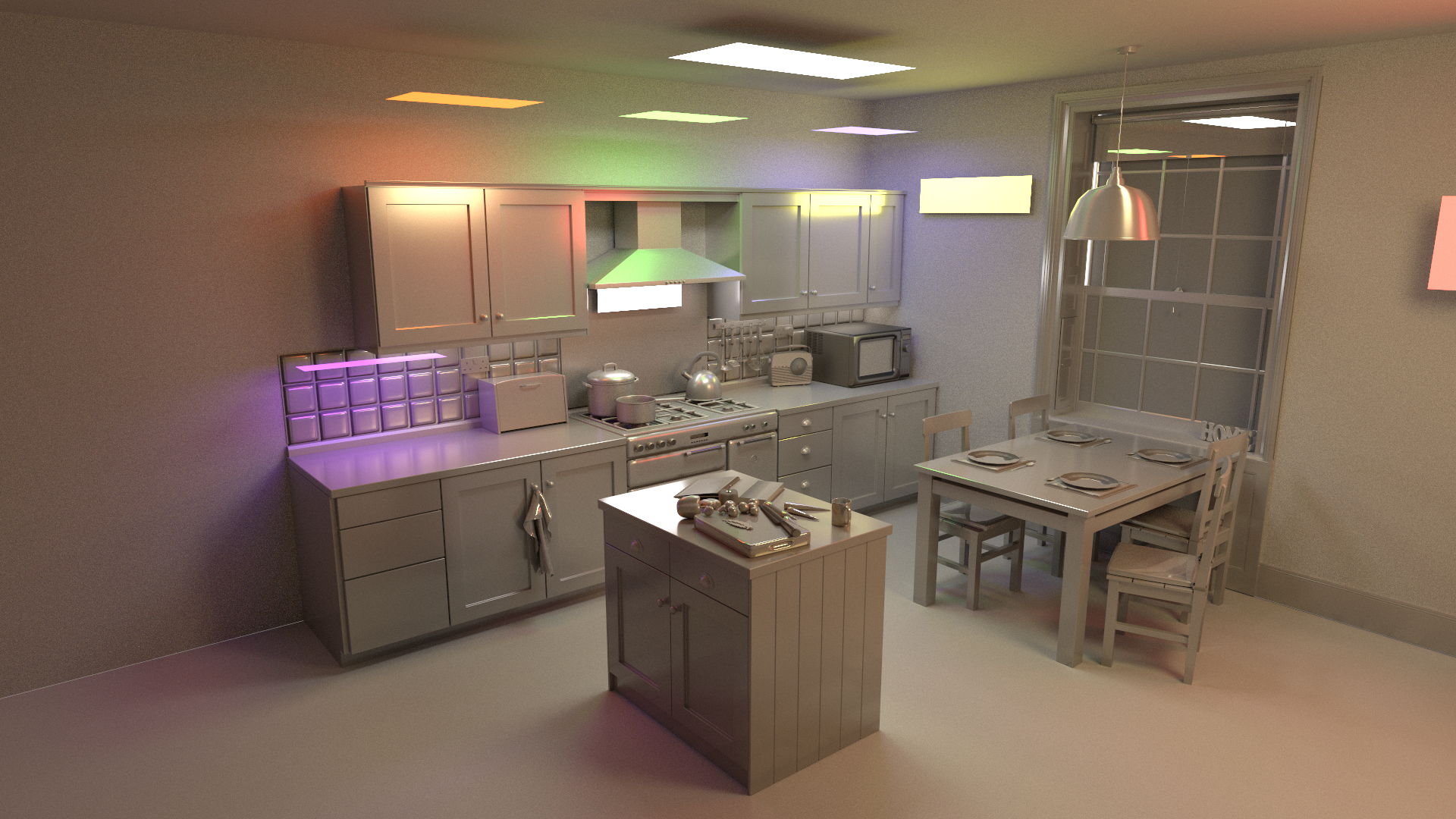}\\

\end{tabular}

\caption{Neural radiance caching (NRC). Image quality after 1024 training iterations: hash-grid (left), GATE (middle), and ground truth (right). Size of hash-grid $<1$ MB. GATE is able to match or exceed the image quality of hash-grid, except for the \emph{Kitchen} scene. Even though GATE has a FLIP higher error, it preserves sharp edges and fine details (e.g., Kitchen scene).}
\label{fig:NRC}
\end{figure*}

\subsubsection{Per-Triangle Learning Rate}
\label{sec:learning_rate}

An advantage of GATE is the link between triangles and feature vectors. The link allows for a per-triangle learning rate, which is in contrast to grid-based approaches, where multiple triangles may occupy the same grid cell or a single triangle spans multiple grid cells. For efficient training, we maintain a per-triangle buffer that stores the number of training steps $t_\mathcal{T}$ applied to triangle $\mathcal{T}$. Using this information, we can fine-tune the training routine in the following ways.

First, we use a per-triangle learning rate for the Adam optimizer, where instead of a global number of training steps, we use per-triangle $t_\mathcal{T}$ for the Adam parameters $\beta_1^{t_\mathcal{T}}$ and $\beta_2^{t_\mathcal{T}}$. We expect only a fraction of triangles to be trained in a single step. Adjusting the parameters $\beta_1^t$ and $\beta_2^t$ only by the number of previous global training iterations would slow down the training of all triangles equally over time, even for those not actually trained. Our modification of tracking the number of training steps per triangle, allows for adjusting the Adam parameters per training step, which ensures a consistent training quality and speed for all triangles.

Second, we can generate training data taking into account which triangles have been trained before and how many times. When generating training sample positions, we prioritize triangles with a smaller training count to quickly distribute more samples to untrained areas. Specifically, we randomly generate $M$ (e.g., $M=16$) samples $x_{\mathcal{T}_1}, \dots, x_{\mathcal{T}_M}$ on triangles $\mathcal{T}_1, \dots, \mathcal{T}_M$, such that we select candidate $x_{\mathcal{T}_i}$ the following probability:
\begin{equation}
    P_{\mathcal{T}_i} = \frac{\weight(x_{\mathcal{T}_i})}{\sum_{j=1}^M \weight(x_{\mathcal{T}_j})},
\end{equation}
where $\weight(x)$ is a weight of sample $x$. To prioritize triangles with fewer training steps, we define the weights as $\weight(x_{\mathcal{T}_i}) = \frac{1}{\min(512, t_{\mathcal{T}_i})}$. To correctly calculate $t_{\mathcal{T}_i}$, we maintain an atomic counter per triangle. This counter is incremented each time a training sample is assigned. We also need to make sure to increment the counter only once per training iteration, even when multiple samples are assigned to the same triangle. Therefore, we track when $\mathcal{T}_i$ was trained the last time and only increment $t_{\mathcal{T}_i}$ if it has not been modified in the current iteration.

\section{Results and Discussion}
\label{sec:results}

For our evaluation, we implemented a minimal MLP framework, written in DirectX 12 and HLSL, to compare GATE to multi-resolution hash-grid encoding~\cite{muller2022instant}. We use an MLP composed of~$2$ hidden layers with~$32$ neurons in each layer, using a leaky \emph{rectified linear unit} (ReLU) as the activation function, with a negative slope of~$0.01$. In our tests, GATE uses two mesh colors resolutions hierarchically: a resolution $R_\mathcal{I}$ calculated adaptively and one with fixed $R=1$. The hash-grid uses~$8$ levels with~$4$ features per level and a base resolution of~$2$. All experiments were conducted on an AMD Radeon 9070 XT (RDNA 4) GPU, running Windows 10 as the operating system. We use the FLIP error~\cite{Andersson2021b} as an image quality metric.

We implemented two neural rendering algorithms on top of our MLP framework: \emph{neural ambient occlusion} (NAO) and \emph{neural radiance caching} (NRC)~\cite{muller2021real}. The implementation of both algorithms encodes a 3D position and feeds the result to the MLP. For NRC, both encoding methods use the view direction as input, as the radiance is view-dependent. For the hash-grid, we use surface normal and albedo as additional input. In the case of GATE, normal and albedo attributes are inherent to the triangles and are therefore omitted. Both view directions and normal are encoded by the one-blob encoding using spherical coordinates~\cite{muller2019neural,muller2021real}. To visualize the approximate radiance field, we evaluate NRC on primary hit points and also remove color data from primary hits to emphasize global illumination effects. The length of training paths is~{4}. Both NAO and NRC use~{49152} training samples, performing one training iteration per frame. GATE employs the per-triangle learning rate (see Section~\ref{sec:learning_rate}). We use the L2 loss function, and the batch size is equal to the number of training samples. The feature vectors for both GATE and hash-grid are stored as single precision floating point values (FP32), using $2$ (NAO) and $8$ (NRC) features at each resolution. For GATE using $2$ stacked resolutions, this leads to $2L$ number of neurons in the input layer, where $L$ is the length of the feature vector, resulting in $4$ input neurons for NAO, and $16$ for NRC. Note that the MLP inference in our implementation is not limited to primary hits but can be done at any stage in the shader code, including at deeper bounces. All images were rendered at a resolution of 1920$\times$1080.

\begin{figure*}[!h]
\begin{center}
\scalebox{1.0}{
\begin{minipage}{\textwidth}
\centering

\begin{minipage}[t]{0.6\textwidth}
\centering
\setlength{\tabcolsep}{1pt}
\renewcommand{\arraystretch}{0.5}
\begin{tabular}{cccccc}
& 8 & 16 & 32 & 64 & 128 \\

\rotatebox{90}{Hash-Grid} &
\includeimgcrop{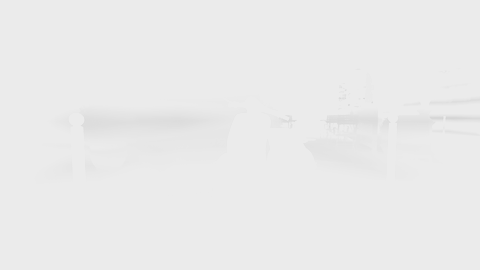} &
\includeimgcrop{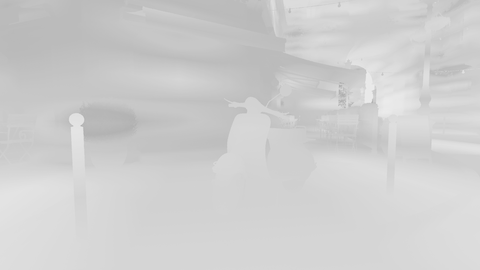} &
\includeimgcrop{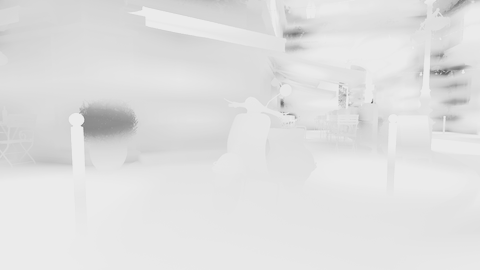} &
\includeimgcrop{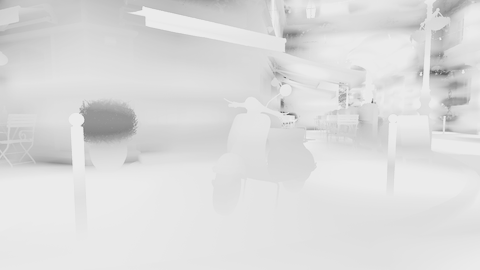} &
\includeimgcrop{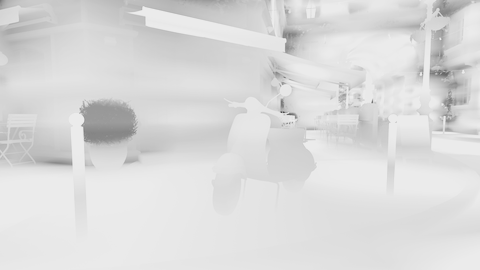} \\
&
\includeimgcrop{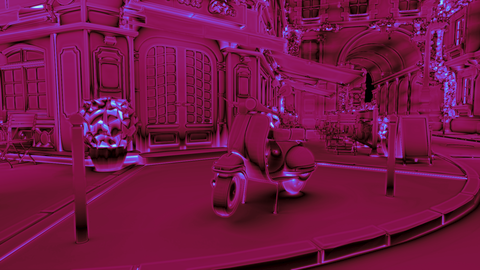} &
\includeimgcrop{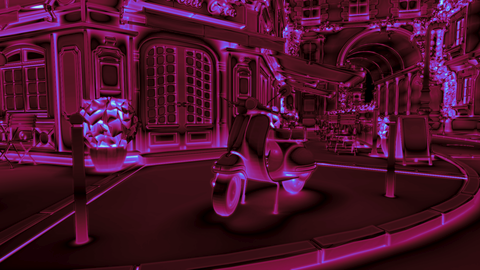} &
\includeimgcrop{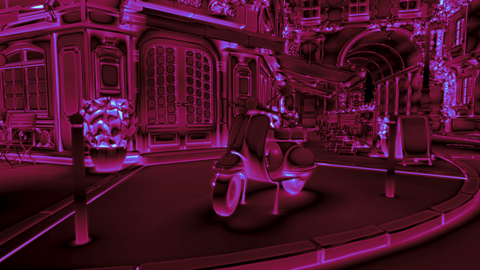} &
\includeimgcrop{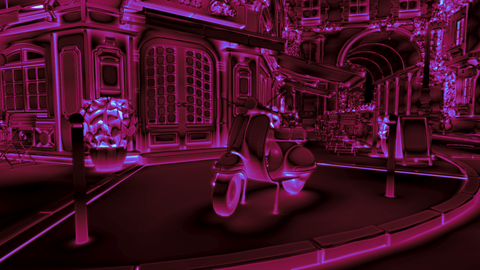} &
\includeimgcrop{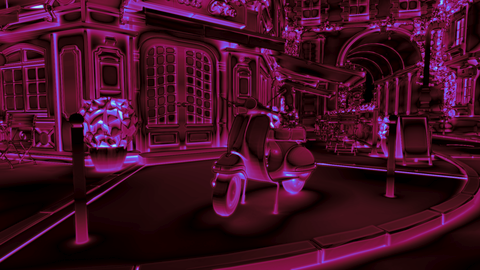} \\
\rotatebox{90}{GATE (ours)} &
\includeimgcrop{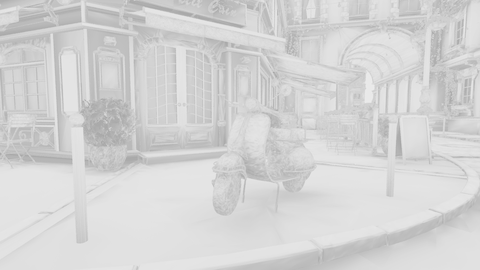} &
\includeimgcrop{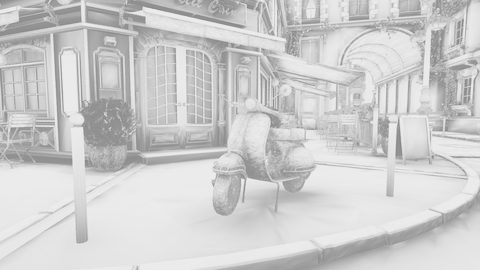} &
\includeimgcrop{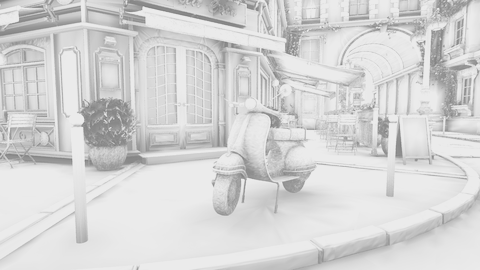} &
\includeimgcrop{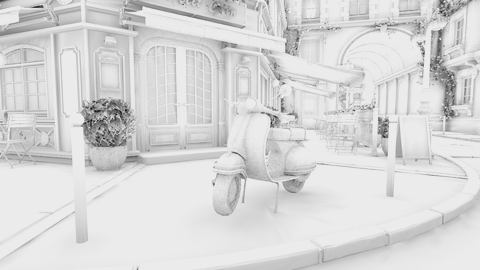} &
\includeimgcrop{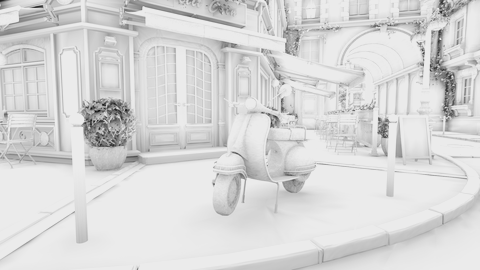} \\
&
\includeimgcrop{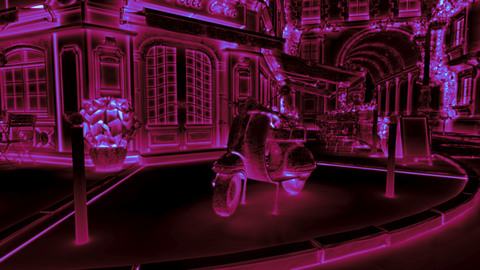} &
\includeimgcrop{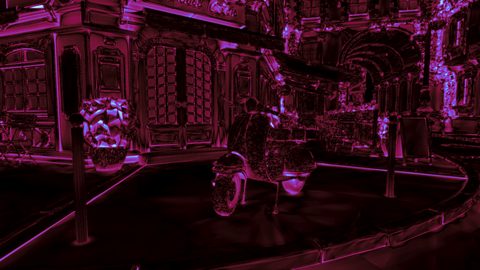} &
\includeimgcrop{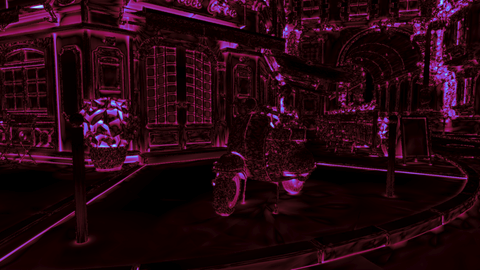} &
\includeimgcrop{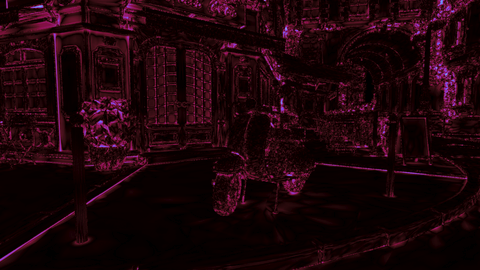} &
\includeimgcrop{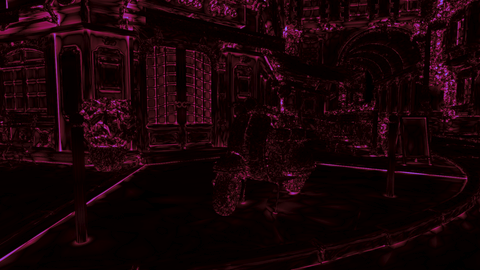} \\
\end{tabular}
\end{minipage}
\hspace{-0.25cm}
\begin{minipage}[t]{0.35\textwidth}
\centering
\begin{tabular}{c}
\\
\includegraphics[height=0.16\textheight]{2_bistro_exterior-AO-R-4-F-2-GROUND_TRUTH-frames-16384.png} \\
\includegraphics[height=0.16\textheight]{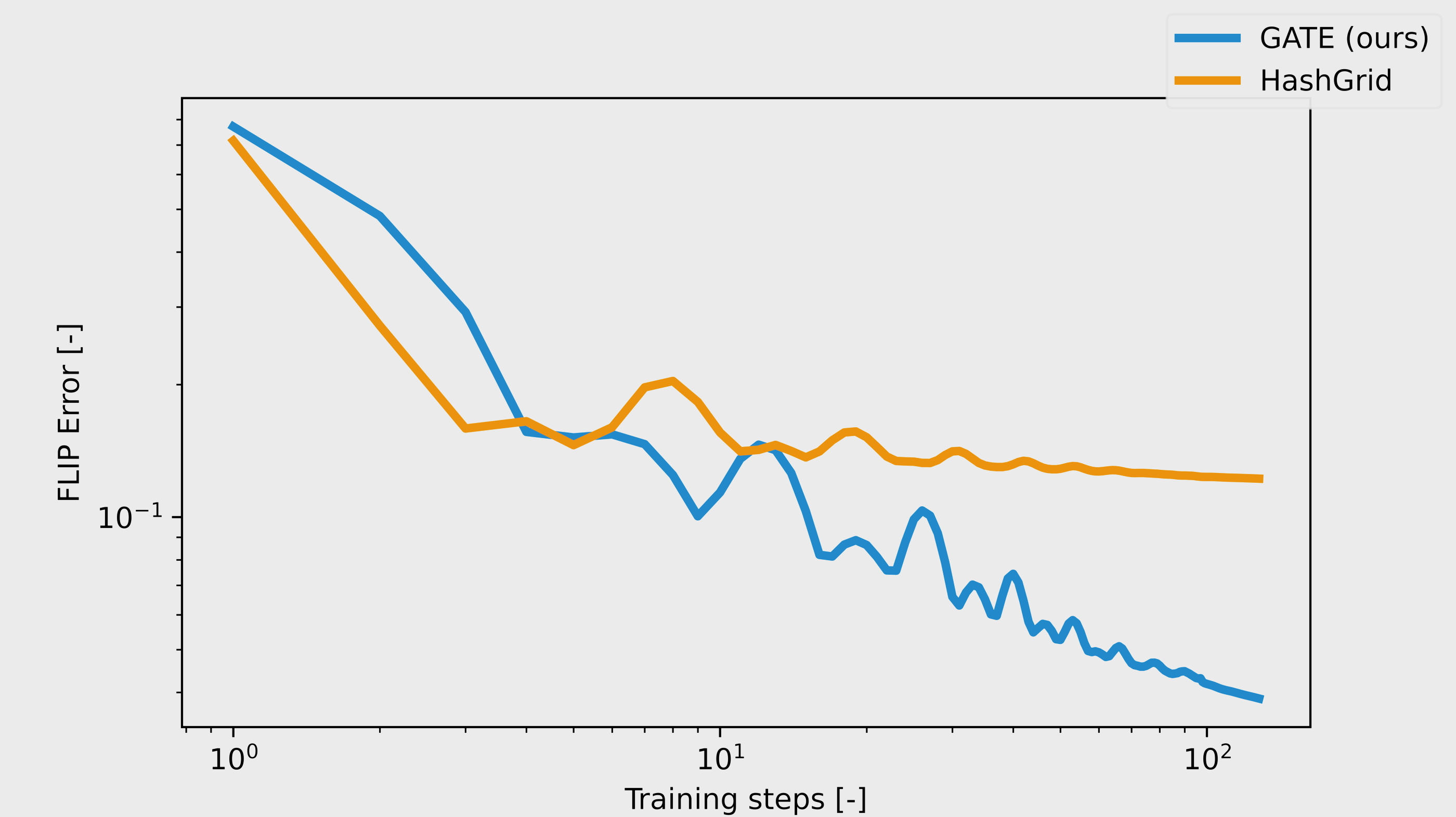}
\end{tabular}
\end{minipage}

\end{minipage}
}

\end{center}
\caption{A comparison of the multi-resolution hash-grid encoding (top) and our GATE (bottom) for the Bistro (Ext) scene rendered with the neural ambient occlusion (NAO) with 1 sample per pixel, showing how the FLIP error decreases with an increasing number of training steps (up to 128 training steps). GATE benefits from an increased number of training steps, while hash-grid encoding shows only small quality improvements. }
\label{fig:training_steps}
\end{figure*}

In the context of real-time applications, our goal is to support continuous training over multiple frames. Figure~\ref{fig:NAO} (NAO) and Figure~\ref{fig:NRC} (NRC) visualize the quality difference for four example scenes with varying complexity after 128 (NAO) and 1024 (NRC) training steps. For all example scenes, GATE is able to deliver a higher image quality (lower FLIP error) compared to hash-grid-based encoding. The higher quality is due to the feature vector density being aligned to the actual geometry, while a hash-grid distributes the feature vectors evenly over the volume of the scene. If a part of the scene needs a high feature vector density on specific parts of the geometry (the teapot-in-the-stadium problem), a distribution over the scene volume does not provide enough resolution. GATE, on the other hand, is able to adapt to varying density requirements on the actual geometry. Figure~\ref{fig:training_steps} shows how quickly the neural network learns with increasing number of training steps. The images are rendered using one sample per pixel. Since the neural rendering algorithms operate in world-space (i.e., encoding 3D positions), they amortize multiple training steps through multiple frames. Our experiment with a fixed view represents the ideal case of perfect temporal coherence. This does not prevent our method to be used for dynamic content, as it adapts to changes thanks to continuous online training.

Table~\ref{tab:training_inference_comparison} compares training and inference times, when giving both hash-grid and GATE a comparable memory budget, and using same training data of equal size. GATE achieves a significant speedup for both training and inference over hash-grids. The speedup for training increases with increasing model complexity, as for hash-grid encoding every feature vector needs to be trained (we follow the reference implementation~\cite{muller2022instant}). GATE, on the other hand, only needs to train a subset of feature vectors, which provides a significant cost advantage. The inference speedup of $1.7-3\times$ is mainly due to GATE offering simpler and more efficient access to feature vectors, e.g., by providing a more cache-friendly memory access pattern, as it stores feature vectors in a linear buffer that preserves locality. In contrast, the hash-grid's hashing function causes frequent incoherent memory accesses, as neighboring input values are mapped to vastly different output values.

Table~\ref{tab:training_inference_comparison_varying_hash_grid_size} compares the performance of training and inference with reduced hash-grid sizes. Even with smaller hash-grids, GATE is able to provide superior training and inference performance. Note that smaller hash-grids sizes also reduce image quality (higher FLIP error). GATE provides image quality similar to or greater than any hash-grid size.

\begin{table}[!h]
\smallskip{}
    \footnotesize
    \centering
    \begin{tabular}{ccccc}
                   &  Mem. & Training &  Inference & FLIP \\
                   &  [MB] &   [ms] &    [ms] &  \\                   
    \hline                                                                 
    \hline 
                  \multicolumn{5}{c}{NAO}\\
    \hline                 
    \hline
       GATE (ours)      & 177.09 &  0.45 ($\mathbf{1.0\times}$)  & 0.70 ($\mathbf{1.0\times}$) & 0.040\\
       Hash-Grid (L)    & 189.89 & 25.30 ($\mathbf{56.2\times}$) & 2.05 ($\mathbf{2.9\times}$) & 0.065\\
       Hash-Grid (M)    & 83.09  & 11.66 ($\mathbf{25.9\times}$) & 2.02 ($\mathbf{2.9\times}$) & 0.074\\
       Hash-Grid (S)    & 0.91   &  1.66 ($\mathbf{3.7\times}$)  & 1.92 ($\mathbf{2.7\times}$) & 0.122\\    
    \hline
    \hline                                                                 
                  \multicolumn{5}{c}{NRC}\\
    \hline                 
    \hline                     
       GATE (ours)     & 365.56 & 1.43  ($\mathbf{1.0\times}$)   & 1.23 ($\mathbf{1.0\times}$) & 0.403 \\
       Hash-Grid (L)   & 328.85 & 41.54 ($\mathbf{29.0\times}$)  & 2.53 ($\mathbf{2.1\times}$) & 0.407 \\
       Hash-Grid (M)   & 132.24 & 17.63 ($\mathbf{12.3\times}$)  & 2.54 ($\mathbf{2.1\times}$) & 0.465\\
       Hash-Grid (S)   & 0.91   & 1.98  ($\mathbf{1.4\times}$)   & 2.03 ($\mathbf{1.7\times}$) & 0.461\\
    \hline
    \end{tabular}
    \caption{
Quality, training and inference times of varying hash-grid sizes compared to GATE for the \emph{Bistro Exterior} scene. (\emph{L}) comparable memory budget to GATE, (\emph{M}) medium size, (\emph{S}) small size, below 1 MB. For all hash-grid sizes, GATE is able to provide higher training (up to $56.2\times$) and inference (up to $2.9\times$) performance and lower FLIP error.} 
    \label{tab:training_inference_comparison_varying_hash_grid_size}
\end{table}

\subsection{Ablation Study}

\begin{figure}
\centering
\setlength{\tabcolsep}{1pt}
\begin{tabular}{cc}
$R=4$ & Adaptive $R$\\
\includegraphics[width=0.225\textwidth]{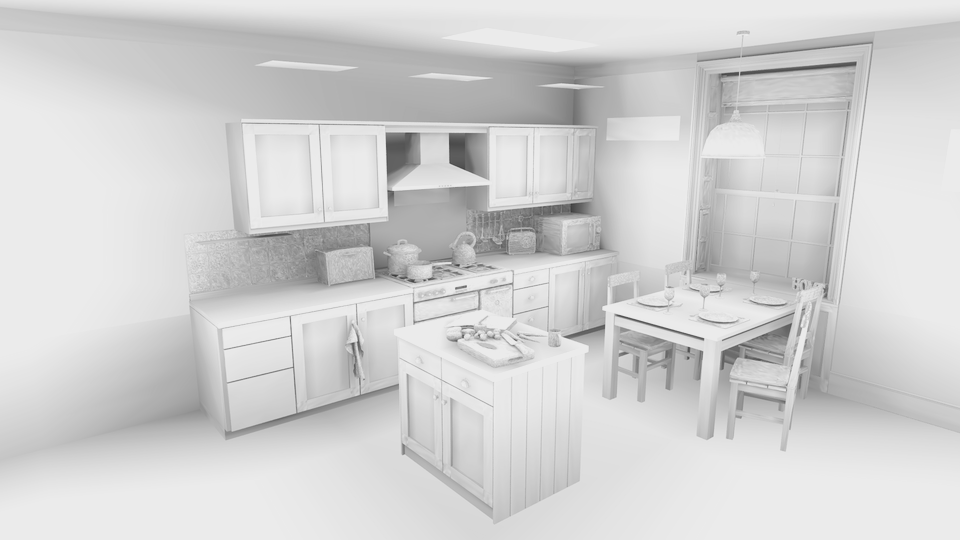} & 
\includegraphics[width=0.225\textwidth]{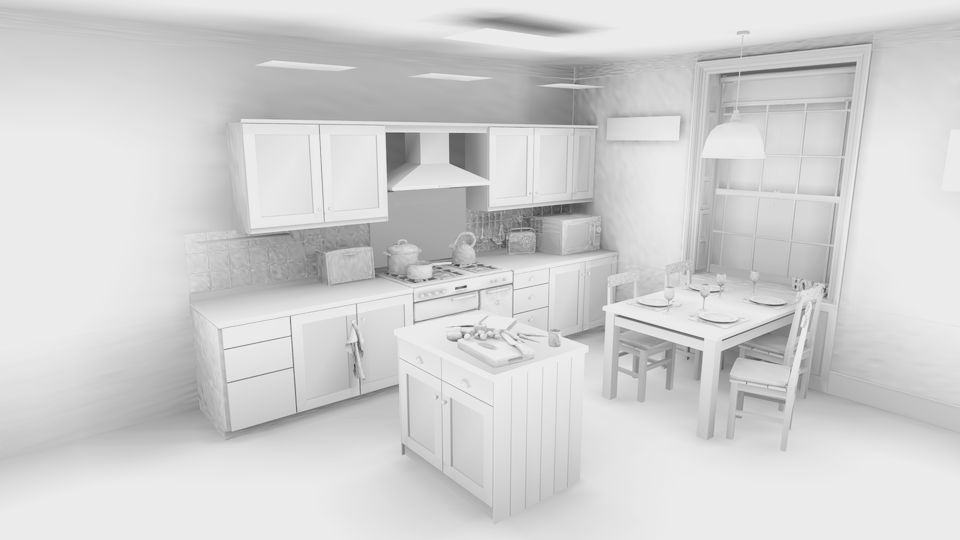} \\
\includegraphics[width=0.225\textwidth]{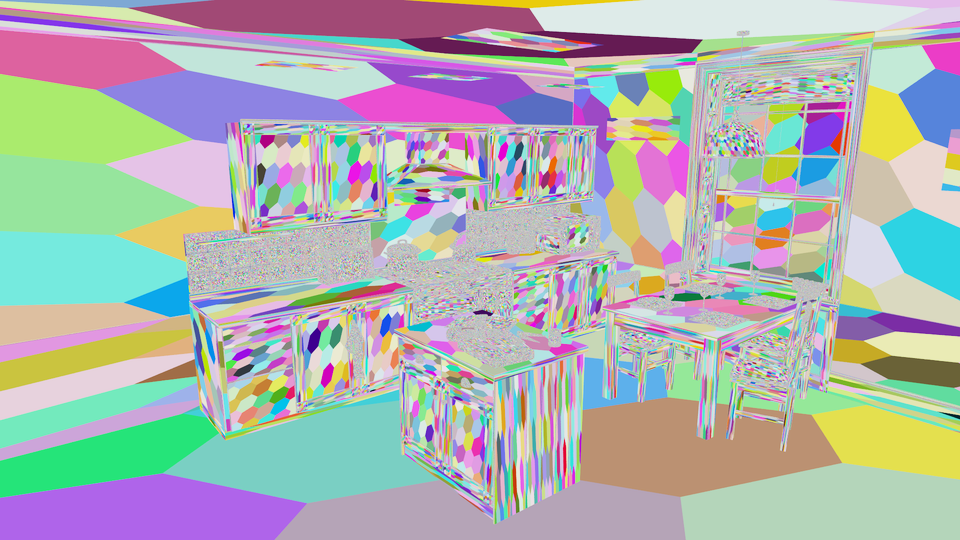} & 
\includegraphics[width=0.225\textwidth]{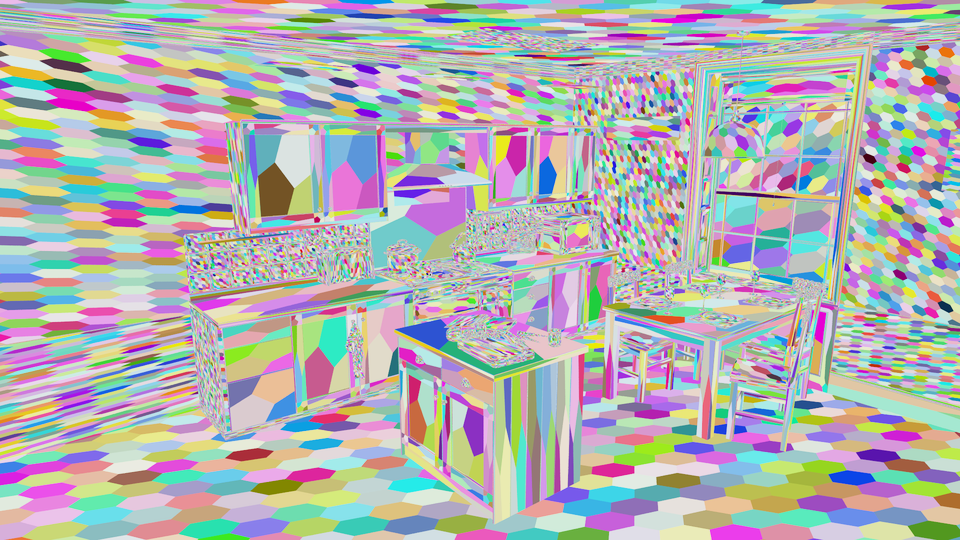} \\
75 MB & 20 MB\\
\end{tabular}
\caption{A comparison of GATE with a fixed resolution (left) and the adaptive one (right) in the \emph{Kitchen} scene. Voronoi diagram (bottom) visualizes the closest feature vector location, using the same color for points belonging to the same feature vector. Notice how the adaptive version tessellates the large triangles more than smaller ones, achieving a similar quality with significantly less memory usage.}
\label{fig:adaptive_R}
\end{figure}

Figure~\ref{fig:adaptive_R} illustrates the advantage of supporting an adaptive resolution per triangle when dealing with an extreme variation in triangle sizes. In this case, applying a fixed resolution to all triangles does not provide enough density on large triangles while at the same time assigning an unnecessarily high density to small triangles. An adaptive per-triangle resolution fixes the density issue, which leads to significantly reduced memory requirements.

\begin{figure}
\centering
\includegraphics[width=0.235\textwidth]{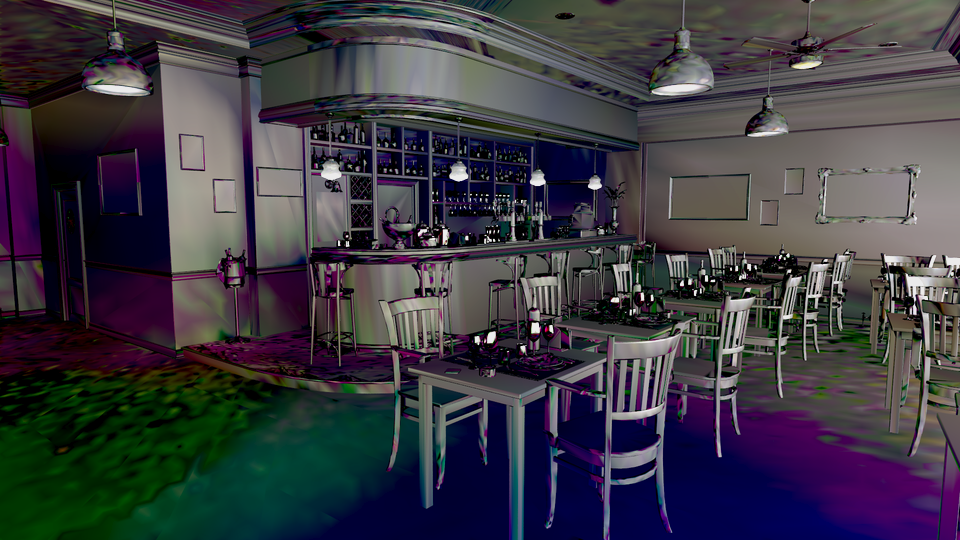}
\includegraphics[width=0.235\textwidth]{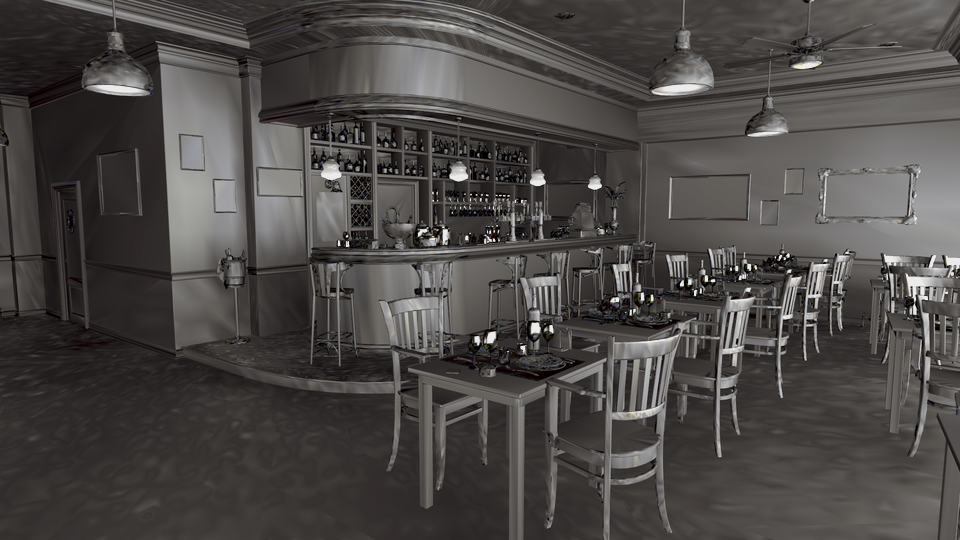} 
\caption{A comparison of NRC + GATE after 1024 training iterations with (left) and without (right) additional G-Buffer inputs such as normal and albedo. Compared to hash-grids, GATE does not benefit from these two additional inputs. With additional inputs, the neural network using GATE requires more training iterations to achieve same quality as without them, as the signal is more complex to learn, but the final quality is not improved. This behavior is different than for the hash-grid, where additional inputs improve quality, however, still requiring more training iterations to achieve it.}
\label{fig:extra_inputs_comparison}
\end{figure}

For neural radiance caching, hash-grids require additional inputs such as normal and albedo to achieve sufficient quality when trained, which is not the case for GATE. Figure~\ref{fig:extra_inputs_comparison} illustrates for GATE that these inputs do not impact convergence speed and image quality, given a fixed set of training iterations.

\section{Conclusion and Future Work}
\label{sec:conclusion}
We presented GATE, a positional encoding tailored for triangular meshes, storing latent feature vectors directly on the triangles. We demonstrated that our method can achieve reconstruction quality higher than multi-resolution hash-grid encoding given a comparable memory budget, while at the same time providing significantly higher training (up to $3.7\times$) and inference (up to $2.7 \times$) performance against a small sized ($< 1$ MB)  hash-grid. GATE also does not require information about scene bounds and effectively handles the teapot-in-the-stadium problem.

There are a couple of interesting directions for future work. Mesh colors can adaptively increase the number of feature vectors on large triangles; however, it is not possible to share feature vectors among multiple (possibly disconnected) triangles, which might pose an issue for finely tessellated meshes with many tiny triangles, introducing significant memory overhead. This problem could be addressed by clustering algorithms. Here, a cluster of locally adjacent triangles could be gathered, which use coarser feature vector resolution compared to the given tessellation. Additionally, we would like to further enhance the heuristic for setting the adaptive resolution based on triangle area, possibly combining it with other metrics such as distance from the camera position. 

We also demonstrated that the use of multiple hierarchically stacked mesh colors resolutions can reduce errors and visual artifacts. Currently, the number of resolutions is a hyperparameter, but the number of resolutions could be adaptively selected similarly to the resolution itself (which might be difficult for grid-based approaches). Furthermore, early tests indicate that feature vectors could be stored with reduced precision without a visible loss of image quality, for example, using FP16 instead of FP32, thereby halving the memory requirements. 

Another possible direction would be to completely replace mesh colors by different mesh parametrization (e.g., the methods used in texture mapping~\cite{yuksel2019star}, or Ptex~\cite{burley_ptex}) to tackle drawbacks of mesh colors. Mesh colors are easy to implement and work well for meshes with near-equilateral triangles of similar sizes, but they do not handle elongated triangles well, leading to either over- or under-sampling. Over-sampling case requires more training and under-sampling case cannot represent high-frequency details. With the industry moving towards meshes represented by micro-triangles (e.g., Nanite), this problem can become less important, especially when we apply clustering of nearby triangles mentioned above.

Finally, we would like to explore the possibility of pre-training. If we think about the MLP part as a \textit{decoder} from the space of latent feature vectors into output representation, then we can theoretically pre-train the decoder for given application and only update feature vectors during the runtime. Such pre-training can be beneficial for several reasons. Consider an algorithm where we generate training samples corresponding to the camera view. Abrupt visibility changes, such as teleportation of the viewer or passing through a wall, can suddenly cause a situation when most feature vectors used for training are untrained and random. This will cause re-training of the MLP to the new situation, changing the \textit{decoder} that was trained for feature vectors in the previous view. Going back will again cause the same situation and another re-training.

\section*{Acknowledgements}
We would like to thank Holger Gr\"un for fruitful discussion and valuable feedback. Model courtesy: Bistro (Amazon Lumberyard), Bathroom (Blendswap/cenobi), Country Kitchen (Blendswap/Jay-Artist).


\section*{Copyright Notice and Trademarks}
©2025 Advanced Micro Devices, Inc. All rights reserved. AMD, Ryzen, Radeon, and combinations thereof are trademarks of Advanced Micro Devices, Inc. Other product names used in this publication are for identification purposes only and may be trademarks of their respective companies.

\bibliographystyle{eg-alpha}

\bibliography{main}

\newcommand{\etalchar}[1]{$^{#1}$}
\begin{thebibliography}{\uppercase{TMND{\etalchar{*}}23}}

\bibitem[ANA21]{Andersson2021b}
\textsc{Andersson P., Nilsson J., Akenine{-}M{\"{o}}ller T.}:
\newblock {Visualizing and Communicating Errors in Rendered Images}.
\newblock In \emph{Ray Tracing Gems II}, Marrs A., Shirley P., Wald I., (Eds.). 2021, ch.~19, pp.~301--320.

\bibitem[BL08]{burley_ptex}
\textsc{Burley B., Lacewell D.}:
\newblock Ptex: Per-face texture mapping for production rendering.
\newblock \emph{Computer Graphics Forum 27}, 4 (2008), 1155--1164.

\bibitem[DKHD25]{Dereviannykh2025}
\textsc{Dereviannykh M., Klepikov D., Hanika J., Dachsbacher C.}:
\newblock Neural two-level monte carlo real-time rendering.
\newblock \emph{Computer Graphics Forum} (2025).

\bibitem[FH24]{fujieda2024neural}
\textsc{Fujieda S., Harada T.}:
\newblock Neural texture block compression.
\newblock \emph{arXiv preprint arXiv:2407.09543} (2024).

\bibitem[FYH23]{Fujieda2023}
\textsc{Fujieda S., Yoshimura A., Harada T.}:
\newblock {Local Positional Encoding for Multi-Layer Perceptrons}.
\newblock In \emph{Pacific Graphics Short Papers and Posters} (2023), Chaine R., Deng Z., Kim M.~H., (Eds.), The Eurographics Association.

\bibitem[GSS{\etalchar{*}}24]{govindarajan2024laghashes}
\textsc{Govindarajan S., Sambugaro Z., Shabhanov A., Takikawa T., Sun Weiweiand~Rebain D., Conci N., Yi K.~M., Tagliasacchi A.}:
\newblock Lagrangian hashing for compressed neural field representations.
\newblock In \emph{ECCV} (2024).

\bibitem[HCZ21]{Hadadan2021}
\textsc{Hadadan S., Chen S., Zwicker M.}:
\newblock Neural radiosity.
\newblock \emph{ACM Trans. Graph. 40}, 6 (Dec. 2021).

\bibitem[KB14]{diederik2014adam}
\textsc{Kingma D., Ba J.}:
\newblock Adam: A method for stochastic optimization.
\newblock \emph{International Conference on Learning Representations} (12 2014).

\bibitem[KMX{\etalchar{*}}21]{Kuznetsov2021}
\textsc{Kuznetsov A., Mullia K., Xu Z., Ha\v{s}an M., Ramamoorthi R.}:
\newblock Neumip: multi-resolution neural materials.
\newblock \emph{ACM Trans. Graph. 40}, 4 (July 2021).

\bibitem[MESK22]{muller2022instant}
\textsc{M{\"u}ller T., Evans A., Schied C., Keller A.}:
\newblock Instant neural graphics primitives with a multiresolution hash encoding.
\newblock \emph{ACM transactions on graphics (TOG) 41}, 4 (2022), 1--15.

\bibitem[MLL{\etalchar{*}}21]{Martell2021}
\textsc{Martel J. N.~P., Lindell D.~B., Lin C.~Z., Chan E.~R., Monteiro M., Wetzstein G.}:
\newblock Acorn: adaptive coordinate networks for neural scene representation.
\newblock \emph{ACM Trans. Graph. 40}, 4 (July 2021).

\bibitem[MMR{\etalchar{*}}19]{muller2019neural}
\textsc{M{\"u}ller T., McWilliams B., Rousselle F., Gross M., Nov{\'a}k J.}:
\newblock Neural importance sampling.
\newblock \emph{ACM Transactions on Graphics (ToG) 38}, 5 (2019), 1--19.

\bibitem[MRNK21]{muller2021real}
\textsc{M\"{u}ller T., Rousselle F., Nov\'{a}k J., Keller A.}:
\newblock Real-time neural radiance caching for path tracing.
\newblock \emph{ACM Trans. Graph. 40}, 4 (July 2021).

\bibitem[MST{\etalchar{*}}20]{mildenhall2020nerf}
\textsc{Mildenhall B., Srinivasan P.~P., Tancik M., Barron J.~T., Ramamoorthi R., Ng R.}:
\newblock Nerf: Representing scenes as neural radiance fields for view synthesis.
\newblock In \emph{ECCV} (2020).

\bibitem[RBA{\etalchar{*}}19]{rahaman2019}
\textsc{Rahaman N., Baratin A., Arpit D., Draxler F., Lin M., Hamprecht F., Bengio Y., Courville A.}:
\newblock On the spectral bias of neural networks.
\newblock In \emph{Proceedings of the 36th International Conference on Machine Learning} (09--15 Jun 2019), Chaudhuri K., Salakhutdinov R., (Eds.), vol.~97 of \emph{Proceedings of Machine Learning Research}, PMLR, pp.~5301--5310.

\bibitem[SLR24]{Sivaram2024}
\textsc{Sivaram V., Li T.-M., Ramamoorthi R.}:
\newblock Neural geometry fields for meshes.
\newblock In \emph{ACM SIGGRAPH 2024 Conference Papers} (New York, NY, USA, 2024), SIGGRAPH '24, Association for Computing Machinery.

\bibitem[TET{\etalchar{*}}22]{Takikawa2022}
\textsc{Takikawa T., Evans A., Tremblay J., M\"{u}ller T., McGuire M., Jacobson A., Fidler S.}:
\newblock Variable bitrate neural fields.
\newblock In \emph{ACM SIGGRAPH 2022 Conference Proceedings} (New York, NY, USA, 2022), SIGGRAPH '22, Association for Computing Machinery.

\bibitem[TLY{\etalchar{*}}21]{Takikawa2021}
\textsc{Takikawa T., Litalien J., Yin K., Kreis K., Loop C., Nowrouzezahrai D., Jacobson A., McGuire M., Fidler S.}:
\newblock Neural geometric level of detail: Real-time rendering with implicit {3D} shapes.

\bibitem[TMND{\etalchar{*}}23]{takikawa2023compact}
\textsc{Takikawa T., M\"{u}ller T., Nimier-David M., Evans A., Fidler S., Jacobson A., Keller A.}:
\newblock Compact neural graphics primitives with learned hash probing.
\newblock In \emph{SIGGRAPH Asia 2023 Conference Papers} (2023).

\bibitem[TZN19]{thies2019}
\textsc{Thies J., Zollh\"{o}fer M., Nie\ss{}ner M.}:
\newblock Deferred neural rendering: image synthesis using neural textures.
\newblock \emph{ACM Trans. Graph. 38}, 4 (July 2019).

\bibitem[VSP{\etalchar{*}}17]{Vaswani2017}
\textsc{Vaswani A., Shazeer N., Parmar N., Uszkoreit J., Jones L., Gomez A.~N., Kaiser L.~u., Polosukhin I.}:
\newblock Attention is all you need.
\newblock In \emph{Advances in Neural Information Processing Systems} (2017), Guyon I., Luxburg U.~V., Bengio S., Wallach H., Fergus R., Vishwanathan S., Garnett R., (Eds.), vol.~30, Curran Associates, Inc.

\bibitem[VSW{\etalchar{*}}23]{vaidyanathan2023random}
\textsc{Vaidyanathan K., Salvi M., Wronski B., Akenine-M{\"o}ller T., Ebelin P., Lefohn A.}:
\newblock Random-access neural compression of material textures.
\newblock \emph{arXiv preprint arXiv:2305.17105} (2023).

\bibitem[WZK{\etalchar{*}}23]{Weier2023}
\textsc{Weier P., Zirr T., Kaplanyan A., Yan L.-Q., Slusallek P.}:
\newblock Neural prefiltering for correlation-aware levels of detail.
\newblock \emph{ACM Trans. Graph. 42}, 4 (July 2023).

\bibitem[YKH10]{yuksel2010mesh}
\textsc{Yuksel C., Keyser J., House D.~H.}:
\newblock Mesh colors.
\newblock \emph{ACM Transactions on Graphics (TOG) 29}, 2 (2010), 1--11.

\bibitem[YLT19]{yuksel2019star}
\textsc{Yuksel C., Lefebvre S., Tarini M.}:
\newblock Rethinking texture mapping.
\newblock \emph{Computer Graphics Forum 38}, 2 (2019), 535--551.

\bibitem[ZRW{\etalchar{*}}24]{zeltner2024real}
\textsc{Zeltner* T., Rousselle* F., Weidlich* A., Clarberg* P., Nov{\'a}k* J., Bitterli* B., Evans A., Davidovi{\v{c}} T., Kallweit S., Lefohn A.}:
\newblock Real-time neural appearance models.
\newblock \emph{ACM Transactions on Graphics 43}, 3 (2024), 1--17.

\end{thebibliography}

\end{document}